\documentclass[epj-spec,final]{svjour} 
\usepackage{graphicx}
\usepackage{dcolumn}
\usepackage{amsmath}
\usepackage{amssymb}

\def\be{\begin{equation}}
\def\ee{\end{equation}}

\def\bea{\begin{eqnarray}}
\def\eea{\end{eqnarray}}

\begin{document}
\title{Testing theories of Gravity and Supergravity with inflation and observations of the cosmic microwave background}
\author{G. K. {Chakravarty}\inst{1} \and {G. Lambiase}\inst{2,3} \and S. {Mohanty}\inst{1}}
\institute{ Physical Research Laboratory, Ahmedabad 380009, India \and Dipartimento di Fisica ``E. R. Caianiello'',
Universit\`a degli
  Studi di Salerno, Via Giovanni Paolo II, Fisciano, Italy \and INFN - Gruppo Collegato di Salerno, Italy.}

\renewcommand{\theequation}{\thesection.\arabic{equation}}
\abstract{Cosmological and astrophysical observations lead to the emerging picture of a universe that is spatially flat and presently undertaking an accelerated expansion. The observations supporting this picture come from a range of measurements encompassing estimates of galaxy cluster masses, the Hubble diagram derived from type-Ia supernovae observations, the measurements of Cosmic Microwave Background radiation anisotropies, etc.  The present accelerated expansion of the universe can be explained by admitting the existence of a cosmic fluid, with negative pressure. In the simplest scenario this unknown component of the universe, the Dark Energy, is represented by the cosmological constant ($\Lambda$), and accounts for about 70\% of the global energy budget of the universe. The remaining 30\% consists of a small fraction of baryons (4\%) with the rest cold Dark Matter (CDM).
The Lambda Cold Dark Matter ($\Lambda$CDM) model, i.e. General Relativity with cosmological constant, is in good agreement with observations. It can be assumed as the first step towards a new standard cosmological model. However, despite the satisfying agreement with observations, the $\Lambda$CDM model presents several lacks of congruence and shortcomings, and therefore theories beyond Einstein’s General Relativity are called for.
Many extensions of Einstein's theory of gravity have been studied and proposed with various motivations like the quest for a quantum theory of gravity to extensions of anomalies in observations at the solar system, galactic and cosmological scales. These extensions include adding higher powers of Ricci curvature $R$, coupling the Ricci curvature with scalar fields and generalized functions of $R$.  In addition when viewed from the perspective of Supergravity (SUGRA) many of these theories may originate from the same SUGRA theory, but interpreted in different frames. SUGRA therefore serves as a good framework for organizing and generalizing theories of gravity beyond General Relativity. All these theories when applied to inflation (a rapid expansion of early Universe in which primordial gravitational waves might be generated and might still be detectable by the imprint they left or by the ripples that persist today) can have distinct signatures in the Cosmic Microwave Background radiation  temperature and polarization anisotropies.
We give a review of $\Lambda$CDM cosmology and survey the theories of gravity beyond Einstein’s General Relativity, specially which arise from SUGRA, and study the consequences of these theories in the context of inflation and put bounds on the theories and the parameters therein from the observational experiments like PLANCK, Keck/BICEP etc. The possibility of testing these theories in the near future in CMB observations and new data coming from colliders like the LHC, provides an unique opportunity for constructing verifiable models of particle physics and General Relativity.}

\maketitle

\section{Introduction}
\setcounter{equation}{0}

The attempt to find a consistent quantum theory of gravity has motivated attempts to find generalizations of Einsteins General Theory of Relativity. The recent cosmological observations, in fact, suggest that the present Universe is in accelerating phase \cite{accUn}.
To explain such a behavior one is forced to introduce the concept of Dark Energy. Moreover, for explaining the rotation curves of Galaxies one needs to introduce Dark Matter. The origin and nature of both Dark Energy (DE) and Dark Matter (DM), that undoubtedly represent a fundamental issue
in particle physics, astrophysics and cosmology, is unknown and although many attempts have been done to explain such {\it dark components}, no final conclusions has been reached, and the question till now is completely open.

Observations of the cosmic microwave background anisotropy motivate the idea that the universe in the past went through a period of accelerated expansion called inflation. The specific model of inflation which fits the temperature and polarization anisotropies is still not settled. It's not established yet whether inflation requires a fundamental scalar field in a particle physics model or whether the scalar degrees of freedom  of the metric in modified gravity theories can give rise to viable inflation. In this review we survey the generailised theories of gravity which can give inflation compatible with observations of the CMB by experiments like Planck \cite{Ade:2015lrj,Planck:2015} and BICEP2/{\it Keck}+Planck~\cite{BKP:2015,bicep2keck2015,Ade:2015xua}.

%


\subsubsection*{\it General features  of metric theories of gravity}

In order that a theory of gravity might be considered a valid theory, it  must fulfill some minimal requirements:
$1)$ It must reproduce the Newtonian dynamics in the
weak-energy limit, which means that it has to pass the classical Solar
System tests which are all experimentally well founded
\cite{Will93}. $2)$ It must reproduce Galactic dynamics in relation to
the observed baryonic constituents (hence luminous components as
stars, or planets, dust and gas), radiation and Newtonian potential which is, by assumption,
extrapolated to Galactic scales. $3)$ It must address the problem of large scale structure
and the cosmological dynamics.
General Relativity (GR) is the simplest theory that satisfies these requirements.
It is based on the assumption that space and
time are entangled into a single space-time structure,
which must reduce to Minkowski's space-time structure in absence of gravitational forces.
GR underlies also on Riemann's ideas according to which the Universe
is a curved manifold \cite{riemann} and that the distribution of matter affects point
by point the local curvature of the space-time structure.

GR is strongly based on some assumptions that the Physics of Gravitation has to
satisfy:

\begin{quote}
The {\it Principle of Relativity}: All frames are good frames for describing Physics. This implies that no preferred
inertial frame should be chosen a priori (if any exists).
\end{quote}
\begin{quote}
The {\it Principle of Equivalence}: Inertial effects are locally indistinguishable from
gravitational effects (i.e. the equivalence between the inertial and the gravitational mass).
\end{quote}
\begin{quote}
The {\it Principle of General Covariance}: Field equations  are generally covariant.
\end{quote}
\begin{quote}
The {\it Principle of Causality}: Each point of space-time should admit a universally valid
notion of past, present and future.
\end{quote}
On the basis of the above principles, Einstein postulates that the
gravitational forces must be related to the curvature of a
metric tensor field $ds^2 = g_{\mu\nu}dx^{\mu}dx^{\nu}$ defined on a
four-dimensional space-time manifold (the metric signature is the same of
Minkowski's metric,  $(-,+,+,+)$) and that space-time is curved, with the curvature locally
determined by the distribution of the sources, the latter described by the energy-momentum
tensor  $T^{(m)}_{\mu\nu}$. An important achievement was to prove that the field equations for a metric tensor $g_{\mu\nu}$ can be obtained by starting from an action linearly depending on
Ricci's scalar $R$ (Hilbert-Einstein action \cite{schroedinger}.)
\begin{equation}\label{HEaction}
  S_{HE}=\frac{1}{2\kappa^2}\int d^4 x \sqrt{-g}\, R + S_{matter}\,, \qquad \kappa^2=8\pi G/c^4\,.
\end{equation}
The choice of Hilbert and Einstein was completely arbitrary, but it was certainly the
simplest one. Some years later the GR formulation, it was clarified by Levi--Civita that
curvature is  not a purely metric notion. It is indeed
related to the linear connection, which plays a central role in the definition of parallel
transport and covariant derivation \cite{levicivita} (this is, in some sense, the precursor idea of what would
be called a "gauge theoretical framework" \cite{gauge}, after the Cartan work \cite{cartan}).
As later clarified, the principles of relativity, equivalence and covariance require
that the space-time structure has to be determined by either one or
both of two fields: a metric $g$ and a linear connection $\Gamma$ (symmetric in order that the theory is torsionless). The metric  $g$  fixes the causal structure of space-time (the
light cones) as well as some metric relations (clocks and rods);
the connection  $\Gamma$  fixes the free-fall, {\it i.e.} the locally
inertial observers.
They have to satisfy a number of compatibility relations, which generally lead to a specific form of the connection in terms of the metric tensor (Levi-Civita connections),
but  they can be also independent, leading to the Palatini approach of GR.
\cite{palatiniorigin}.
It is on this basis that the so-called alternative theories of
gravitation, or {\it Extended Theories of
Gravitation} (ETGs) arise. In fact their starting points is that
gravitation is described by either a metric (purely metric
theories), or by a linear connection (purely affine theories) or by both fields
(metric-affine theories). Here the Lagrangian is a scalar density  of the
curvature invariants, constructed by mean of $\{g, \Gamma\}$.

Attempts to generalize GR along these lines \cite{unification} and investigations about "alternative
theories" continued even after 1960 \cite{Will93}. What arises from these studies is that the search
for a coherent quantum theory of gravitation or the belief that gravity has to be
considered as a sort of low-energy limit of string theories \cite{green},
has renew the idea that there is no reason to follow the simple prescription of
Einstein and Hilbert. Other curvature invariants or non-linear functions of them
should be also considered, especially in view of the fact that
they have to be included in both the semi-classical expansion of a
quantum Lagrangian or in the low-energy limit of a string
Lagrangian.

Not only from a mathematical point of view there is the necessity to generalize  GR, but also
current astrophysical and cosmological observations suggest that, as already pointed out, Einstein's
equations are no longer a good test for gravitation  at Galactic, extra-galactic and cosmic scales, unless one does
not admit that the matter side of field equations contains some
kind of exotic matter-energy which is the {\it dark matter} and {\it dark
energy} side of the Universe.

One can adopt a different point of view, in the sense that instead of changing the matter side of Einstein field equations to fit the missing matter-energy content of the currently observed Universe (by adding any sort of exotic matter and/or energy), one may change the gravitational side of the equations, admitting corrections
coming from non-linearities in the effective Lagrangian. This is a possibility that needs to explored, even if, without a complete theory of gravity, one has to tune up the form of effective theory that is going to study, hence a huge family of allowed Lagrangians can be chosen, trying to fit all possible observational tests, at all scales (Solar, Galactic, extragalactic and cosmic, and so on).

\subsection{Shortcomings in the standard cosmological model}

We shortly review the shortcomings of the cosmological standard model (the cosmology based on GR) or hot Big Bang.
The latter provides a framework for the description the evolution of the Universe. The spacetime evolution
is governed by Einstein's field equation (which contains matter content of the Universe)
 \begin{equation}\label{GRfieldeq}
R_{\mu\nu}-\frac{1}{2}g_{\mu\nu}R + \Lambda g_{\mu\nu}= {\kappa}^2 T_{\mu\nu}\,,
 \end{equation}
This equation follows by varying the action  (\ref{HEaction}) with respect to the metric $g_{\mu\nu}$, while
 \[
T_{\mu\nu}=-\frac{2}{\sqrt{-g}}\frac{\delta S_{matter}}{\delta g_{\mu\nu}}
 \]
is the energy-momentum tensor. $\Lambda$ is the cosmological constant (it enters into (\ref{HEaction}) by replacing $R\to R+\Lambda$).

There are three important epochs to which the early Universe undergoes: radiation, matter and vacuum (energy) domination eras. They are characterized by an appropriate relation between
pressure and energy density of matter, $p=p(\rho)$. Assuming the homogeneity and isotropy  of the (flat) cosmological background Friedman-Robertson-Walker (FRW) Universe, the line element is given by\footnote{The homogeneity, isotropy and expanding nature of the space-time is mathematically described by the more general FRW metric given by
 \be\label{FRWmetrica}
 ds^{2}=-dt^{2}+a^{2}(t)\left[\frac{dr^{2}}{1-\kappa r^{2}} + r^{2}\left(d\theta^{2}+sin^{2}\theta d\phi^{2}\right)\right],
 \ee
where ($r, \theta, \phi$) are the comoving spatial coordinates. $\kappa$ represents the spatial curvature and can take values $+1, 0, -1$ describing open, flat and closed universe, respectively.}
 \begin{equation}\label{FRWmetric}
 ds^2=-dt^2+a^2(t) d{\bf x}^2\,,
 \end{equation}
where $a(t)$ is the scale factor.
For a long time, the success of the Big Bang model was based on three cornerstones: $i)$ Hubble expansion, $ii)$ the distribution of relic
photons (CMBR), and $iii)$ the light element abundance such as ${}^3 He$, ${}^4 He$, $D$,
${}^7 Li$ (Big Bang Nucleosynthesis (BBN)). With the recent developments of modern Cosmology, our view of the Universe evolution has been
completely transformed. For example, the cosmological model present some inconsistencies that can be solved only admitting the existence of the Inflation, i.e. a period of accelerated  expansion of the Universe
which occurred in the very early phase of the Universe evolution.
It is believed that Inflation is responsible for the inhomogeneities in the matter distribution (whose evolution allowed the formation of structures, stars, planets) and the inhomogeneities of the CMB. These perturbations
are generated by quantum fluctuations of the scalar field (the inflaton) which drives the Inflation, and they can be scalar, vectorial or tensorial. The challenge of modern cosmology is to verify all predictions of Inflationary scenario.

The Inflationary paradigm allows to solve some inconsistence of the standard cosmological model:
\begin{itemize}
 \item  Flatness of the Universe -
  $\Omega=\rho/\rho_c=1$ where $\rho=3H^2m_P^2/8\pi$ is the critical density. Without
  Inflation, the adjustment of $\Omega$ should be $\sim 10^{-60}$
  at the Planck era, and $\sim 10^{-15}$ at the primordial
  nucleosynthesis.
  \item The problems of homogeneity, isotropy, and
  horizon (which created headache in the frameworks of FRW cosmology)
  are elegantly solved.
  \item Inflation provides a natural mechanism of generation of
  {\it small density perturbations} with practically
  flat spectrum in agreement with observations.
  \item To solve all problems of standard FRW cosmology, the
  duration of Inflation must be $N=H t \sim 70-100$.
\end{itemize}
We will return on the standard cosmological model and Inflation in next Sections.

\section{FRW Cosmology - Inflation}
\setcounter{equation}{0}

Today the very basic picture of our observable universe is presented quite accurately by the standard Big Bang model of cosmology, also known as $\Lambda CDM$-model or FLRW cosmology. Under the standard Big Bang model, the universe in its early stages is considered to be very hot, uniform in its density and expanding uniformly in all the directions and cooling down at late times. So far, it has passed a large number of increasingly precise tests. It successfully predicts the age, Hubble expansion rate, mass density of the universe and light elemental abundance in the early universe. Also it explains the presence of the Cosmic Microwave Background Radiation (CMBR). The CMBR is a snapshot of the oldest light in our universe leftover after decoupling and imprinted on the Last Scattering Surface (LSS), when the universe was just $3,80,000$ years old. The most remarkable feature of CMBR is its high degree of uniformity everywhere and in all the directions. It has inhomogeneity only at the level of one part in\footnote{Typically inhomogeneity are divided in blue and red/yellow spots. The Blue spots represents
the sky where the temperature is $10^{-5}$ below the mean temperature $T_{0}=2.725$K. This corresponds
to the regions where photons loose their energy while climbing out of the gravitational potential of the overdense
regions in the early universe. The Red/Yellow spots represents the underdense regions where the temperature is
$10^{-5}$ above the mean temperature.} $10^{5}$. These tiny inhomogeneities in the early universe are believed to have grown to
cosmological scales later in the history of the universe which resulted in structure formation: stars, galaxies
and galactic clusters of today. Current precision measurements of these small inhomogeneities in the CMBR has led to
constraining a variety of cosmological parameters and therefore theoretical cosmological models. However, the
standard Big Bang model could poorly explain some of the observed characteristics of the universe, $e.g.$ why the
universe is so uniform and its intrinsic geometry is so flat. These unsolved problems in standard cosmological model are also known as Horizon and Flatness problems. Invocation of a rapid exponential expansion phase, {\it Inflation}, in the very beginning of the universe could solve these problems.
%
%

In the standard model of cosmology, the matter content of the universe is described by a perfect fluid which is characterised
only by the energy density $\rho$ and isotropic pressure $p$. The stress-energy-momentum tensor for a perfect fluid with energy density $\rho$ and
pressure $p$ is given by
\bea
T_{\mu\nu}=(\rho + p) u_{\mu} u_{\nu} +p g_{\mu\nu}, \label{EMtensor}
\eea
where $u_{\mu}$ is the 4-velocity of the fluid in some arbitrary coordinate system $x_{\mu}$ given by
$u_{\mu}= \frac{dx_{\mu}}{d\tau}$.
Here $\tau$ is the proper time of the observer, so that $g^{\mu\nu}u_{\mu}u_{\nu}=-1$. If such a fluid is at rest in the geometry described by metric~(\ref{FRWmetric}) and obeys the equation of state $p_i=\omega_i \rho_i$, then from the covariant conservation of the stress-energy-momentum tensor ($\bigtriangledown_{\mu}T^{\mu\nu}=0$) one finds the equation of motion of energy densities in the FLRW universe
\be\label{continuity}
\dot\rho_i + 3H(1+\omega_i)\rho_i =0,
\ee
where $H=\frac{\dot a}{a}$ is the Hubble parameter describing Hubble expansion rate, $i$ represents the various components
of the cosmological fluid, $e.g.$ matter, radiation and dark energy, and $\omega_i$ represents the respective
equation of state parameter for different
components. For the system as a whole, the total energy density is given by
 \[
\rho=\sum_i \rho_i\,,
 \]
and total pressure by
 \[
p=\sum_i p_i\,.
 \]
The solution of the continuity equation~(\ref{continuity}) is given by
\be\label{solconti}
\rho_i \propto a^{-3(1+\omega_i)}.
\ee
As the universe expands, the matter density, consisting of all non-relativistic matter particles, dilutes as
$\rho_{nr}\propto a^{-3}$; and radiation density, consisting of all relativistic particles, dilutes as $\rho_{r}\propto a^{-4}$, as
for pressureless non-relativistic matter $\omega=0$ and for radiation $\omega=\frac{1}{3}$. For $\omega=-1$ which corresponds to a negative
pressure fluid, a strange behavior occurs, the energy density of the universe remains constant as the universe expands. Such an exotic matter is
known as {\it Dark Energy} or {\it Cosmological Constant} and usually attributed to the present day accelerated expansion of our universe.

In the flat FLRW universe the dynamics of the scale factor $a(t)$ is determined through Friedmann equations
which can be derived by solving Einstein field equation for the FLRW metric (\ref{FRWmetric}) and energy-momentum
tensor (\ref{EMtensor})
\bea
\frac{\dot a^2}{a^2}&=&\frac{8\pi G \rho}{3}+\frac{\Lambda}{3}-\frac{\kappa}{a^2},\label{friedmann}\\
 & & \nonumber \\
\frac{\ddot a}{a} &=& -\frac{4\pi G}{3} (\rho + 3 p) + \frac{\Lambda}{3},\label{acceleration}
\eea
According to acceleration equation if $\Lambda =0$ then the matter and radiation filled universe decelerates
which contradicts the observational data from Type-1a supernovae \cite{accUn}, South
Pole Telescope \cite{Story:2012wx} and from the measurement of high multipole CMB data \cite{Smoot:1998jt,WMAP9,Planck:2015}.
These observations have led to the conclusion that the universe is accelerating in its expansion. In the
$\Lambda CDM$ model the present accelerated expansion is achieved with a small positive cosmological constant
$\Lambda$. Also Eq. (\ref{acceleration}) implies that the present accelerated expansion
can be achieved if the energy density of the universe is dominated by some unknown exotic matter with
negative pressure $p<-\frac{\rho}{3}$ or equation of state parameter $\omega<\frac{-1}{3}$, which
generates repulsive gravity. To consider different matter contributions to the total energy density of the universe,
it is common to define the density parameter as
\be
\Omega_i\equiv\frac{\rho_i}{\rho_c}
\ee
where $\rho_c$ is the critical density for which the universe is spatially flat $i.e.$ from equation~(\ref{friedmann})
$\rho_c=\frac{3 H^{2}}{8\pi G}$ for $\kappa=0$. We define the total density parameter of the universe as
 \[
 \Omega=\sum_i \Omega_i\,.
 \]
If we divide Friedmann equation~(\ref{friedmann}) by $H^{2}$, it can be written as
\be
\sum_i \Omega_i +\Omega_\kappa = \Omega_m +\Omega_r +\Omega_\Lambda +\Omega_\kappa = 1
\ee
where
 \[
 \Omega_\Lambda=\frac{\Lambda}{3H^{2}}=\frac{8\pi G }{3H^{2}}\rho_\Lambda
 \]
 and
 \[
 \Omega_\kappa=-\frac{\kappa}{H^{2}a^{2}}
 \]
are the dark energy and curvature density parameters respectively. The matter density parameter consists of baryonic matter and non-relativistic
cold dark matter (CDM), $i.e.$
 \[
\Omega_m=\Omega_b +\Omega_{CDM}\,.
 \]
The recent Planck observations of CMB~\cite{Planck:2015} combined with
WMAP polarization data~\cite{WMAP9} for low multipoles $l<23$, give the present values of the density parameters at $68\% CL$ as:
\bea
\Omega_{b}h^2 &=& 0.02205\pm 0.00028,\\
 & & \nonumber \\
 \Omega_{CDM}h^2 &=& 0.1199\pm 0.0027,\\
 & & \nonumber \\
 \Omega_\Lambda &=& 0.685^{+0.018}_{-0.016},
\eea
where $h$ is the dimensionless parameter defined through the present value of the Hubble parameter as
 \[
 H_0=100 h~km ~s^{-1} Mpc^{-1}=67.3\pm 1.2~km ~s^{-1} Mpc^{-1}\,.
 \]
Therefore, the present observations suggest that our universe is composed of nearly $4.9\%$ atoms (or baryons), $26.8\%$ (cold) dark matter and
$68.3\%$ of dark energy which adds up to approximately $1$ in the total density parameter. According to Planck combined with BAO data~\cite{Planck:2015}
\be
\Omega_\kappa = 0.000\pm 0.005 ~~(95\%~CL)\,,
\ee
i.e., the observations suggest that the intrinsic geometry of our universe is (very close to) flat $i.e.$ $\kappa\simeq0$ or the universe is at the critical density. Why the universe is so close to flat geometry or at its critical density is known as {\it flatness problem}.

\subsection{Inflation}
%
%
We have seen that the $\Lambda CDM$ model can describe the evolution of the universe in a great detail.
Before we discuss the mathematical description of {\it inflation}, let us briefly discuss the problem of initial
conditions. The conventional model of standard Big Bang
cosmology requires a set of fine-tuned initial conditions so that the universe could evolve to its present state. These initial
conditions are the assumptions of the extreme flatness and homogeneity in the beginning of the universe. The dramatic flatness of the
universe at its beginning can not be predicted or explained by the standard model, instead it must be assumed as an initial condition.
Similarly, the large scale homogeneity of the universe is not predicted or explained by the standard model but it must be assumed.

In the late 1970's, cosmologists realised the problem of initial conditions with the $\Lambda CDM$ model and solution to these
problems could be reached at with the invocation of Inflation, i.e. an accelerated expansion
phase in the early evolution of the universe \cite{Guth:1980zm}
The cosmological inflation is believed to took place in the very early universe around
$10^{-35}$ seconds after the Big Bang. Remarkably, inflation not only explains the large scale
homogeneity and isotropy of the universe but also widely accepted as responsible for the formation and evolution of the
structures in the universe. Inflation can provide the mechanism for producing the tiny density fluctuations which are responsible for
seeding the structures in our universe $e.g.$ stars, galaxies and galactic clusters.

Mathematically, the accelerated expansion of the FLRW universe or {\it condition for inflation} can be given as
\be
\ddot a > 0. \label{acccondi}
\ee
The second time derivative of the scale factor can easily be related to the time variation of the Hubble parameter as
\be
\frac{\ddot a}{a} = H^{2}(1-\epsilon),\label{acccondition}
\ee
where $\epsilon\equiv -\frac{\dot H}{H^{2}}$. Therefore acceleration $\ddot a >0$ corresponds to
\be
\epsilon=-\frac{\dot H}{H^{2}}=-\frac{1}{H}\frac{dH}{dN} < 1,\label{epsilonslow}
\ee
here, we have defined $dN$ as
 \[
 dN = d\ln a = H dt\,,
 \]
which determines the number of e-foldings in an inflationary expansion. More precisely
the number of e-foldings $N$ during inflation in the time interval $t_i < t < t_e$ is given by the integral
\be\label{NefoldGae}
N = \int_{t_i}^{t_e} H dt.
\ee
The equation~(\ref{epsilonslow}) implies that the fractional change in the Hubble parameter per e-folding is small. An inflationary scenario where
this change is too small $i.e.$ $\epsilon<<1$ is termed as {\it slow-roll inflation}. From the acceleration equation~(\ref{acceleration}), we see that, for inflation, the energy density of the universe must be dominated by
a fluid whose equation of state satisfies the condition
\be
\omega\equiv \frac{p}{\rho} < -\frac{1}{3}.  \label{omega_inflation}
\ee

\subsection{Scalar Fields as a Source of Inflation}
%
The existence of scalar fields in the very early universe is
suggested by our best theories of fundamental interactions in Nature, which predict that the universe went through a succession
of phase transitions in its early stages as it expanded and cooled. In general, the phase transition occurs when certain scalar
parameters known as Higgs fields acquire a non-zero value or vacuum expectation value (VEV) via a process called
spontaneous symmetry breaking. The symmetry is manifest as long as the Higgs fields have not acquired vev and it is spontaneously
broken as soon as at least one of the Higgs fields become non-zero. Therefore, the existence of scalar fields in the early universe is
suggested by the occurrence of phase transitions and therefore provides the motivation for considering them as the source of inflation (it can act as a negative pressure source). The simplest inflation models involves a single scalar field $\phi$ which in the inflationary context is termed as {\it inflaton}.
The model is described by the following action
\begin{eqnarray}\label{actioninflation}
S &=& \int d^{4}x \sqrt{-g} \left[-\frac{M_{p}^{2}}{2}R + \frac{1}{2} g^{\mu\nu} \partial_{\mu}\phi \partial_{\nu}\phi + V(\phi)\right]
\\
 &=&
\int d^{4}x \sqrt{-g} \left( \mathcal{L}_{EH} + \mathcal{L}_{\phi}\right)\,.
\end{eqnarray}
In this action the inflaton field has a minimal coupling with the gravity and a canonical kinetic term. $V(\phi)$ is the potential
of the field due to self-interaction and it can be different in different inflation models. Here we will assume an arbitrary $V(\phi)$. We shall set $M_p = 1$, restore it at the end of caluctions.
The energy momentum tensor of the scalar field  is
\begin{equation}\label{emtphi}
T_{\mu\nu}^{\phi} = \partial_{\mu}\phi \partial_{\nu}\phi - g_{\mu\nu}\left[\frac{1}{2}\partial^{\rho}\phi \partial_{\rho}\phi + V(\phi) \right].
\end{equation}
In principle, scalar fields can be dependent on space and time both $i.e.$ $\phi=\phi(t,{\bf x})$, however as we know that the universe is
homogeneous on largest scales, therefore at the background level, homogeneity implies that scalar field can be described by its time dependence only, $i.e.$
$\phi(t,{\bf x})\equiv \phi(t)$. Therefore for the homogeneous background field, the energy momentum tensor for $\phi$ takes the form of a perfect
fluid (\ref{EMtensor}) with energy density and pressure for scalar fields given by
\begin{eqnarray}\label{rhophi}
\rho_{\phi} &=& \frac{\dot \phi^{2}}{2} + V(\phi),  \\
 & & \nonumber \\
 p_{\phi} &=& \frac{\dot \phi^{2}}{2} - V(\phi)\,.  
\end{eqnarray}
The resulting equation of state is
\be
\omega_{\phi}\equiv \frac{p_{\phi}}{\rho_{\phi}}=\frac{\frac{\dot \phi^{2}}{2} - V(\phi)}{\frac{\dot \phi^{2}}{2} + V(\phi)}.\label{omegaphi1}
\ee
If the potential energy of the field dominates over its kinetic energy $i.e.$ $\dot\phi^{2} << V(\phi)$, then the above simple relation (\ref{omegaphi1}) implies that the scalar field can act as a negative pressure source $i.e.$ $\omega_{\phi}<0$ and can provide accelerated expansion $i.e.$ $\omega_{\phi} < -\frac{1}{3}$ (see (\ref{omega_inflation})).
The Friedmann equation and the equation of motion of the scalar field are
\bea\label{Friedmannsingle}
H^{2} &=& \frac{1}{3}\left(\frac{\dot \phi^{2}}{2} + V(\phi)\right)\,, \\
 & & \nonumber \\
\ddot\phi &+& 3H\dot\phi + V'(\phi) = 0\,.\label{eomphi}
\eea
These equations determine the dynamics of the space and scalar field in a FRW universe.
\subsubsection{Slow-roll Inflation}
The equations~(\ref{eomphi}) and (\ref{Friedmannsingle}) can be solved analytically for some specific potentials $V(\phi)$, however
in general, an analytical solution is possible only under slow-roll approximation.
As discussed above, slow-roll inflation occurs when $\dot\phi^{2} \ll V(\phi)$ which implies that the field $\phi$
rolls down the potential slow enough that the potential is nearly constant during inflation. A second order differentiation
of the condition $\dot\phi^{2} \ll V(\phi)$ implies $\ddot \phi \ll V'(\phi)$ which ensures that the accelerated expansion is sustained for
a sufficient period of time. Under slow-roll approximation  equations~(\ref{eomphi}) and (\ref{Friedmannsingle}) become
\bea
3 H \dot\phi &\simeq& -V'(\phi), \label{eomphi1}\\
 & & \nonumber \\
3 H^{2} &\simeq& V(\phi).\label{friedmannsingle1}
\eea
It is worth noting that the time variation of the Hubble parameter and scalar field can be related easily by differentiating equation~(\ref{friedmannsingle1}) w.r.t. time and combining the result with Eq. (\ref{eomphi1}) as
\be\label{Hphidot}
\dot H \simeq -\frac{\dot\phi^{2}}{2}.
\ee
The slow-roll conditions $\dot\phi^{2} \ll V(\phi)$ and $\ddot \phi \ll V'(\phi)$ can be put into useful dimensionless parameters as
\bea
\epsilon &=& -\frac{\dot H}{H^{2}} \simeq \frac{1}{2} \left(\frac{V'(\phi)}{V(\phi)}\right)^{2} \ll 1,\label{epsilon} \\
 & & \nonumber \\
\eta&=& -\frac{\ddot\phi}{H\dot\phi} \simeq \frac{V''(\phi)}{V(\phi)} \ll 1.
\eea
These two conditions ensures that the potential $V(\phi)$ is sufficiently ‘flat’ that the field $\phi$ rolls slowly enough for inflation to
occur. After the end of inflation when field has crossed the flat part of the
potential, it fast rolls, $\dot\phi^{2} \approx V(\phi)$, towards the minimum of the potential and then oscillates and decays into the standard model particles.

It is worth considering the case in which $V(\phi)$ is nearly constant during some part of the period of inflation during which
Hubble expansion is constant. Solving (\ref{friedmannsingle1}), one obtains that during this period scale factor evolves exponentially $a(t) \sim e^{H t}$,  such a spacetime is approximately de-Sitter.
From the acceleration equation (\ref{acccondition}), it is clear that inflation ends when
$\epsilon(\phi_e)=1$,  which represents the violation of slow-roll condition $\epsilon \ll 1$ and as soon as this condition is met $i.e.$ $\epsilon=1$, the kinetic energy of the field $\dot\phi^{2}$ becomes comparable to its
potential energy $V(\phi)$ and the potential becomes steeper and field speeds up towards the minimum of the potential. The number of e-foldings before the inflation ends, as defined in (\ref{epsilon}), is given by (see (\ref{NefoldGae}))
\begin{eqnarray}\label{Nphi}
N(\phi) &=& \int_{t_i}^{t_e} H dt \\
    &=& \int_{\phi_i}^{\phi_e} \frac{H}{\dot\phi} d\phi \simeq \int_{\phi_e}^{\phi_i} \frac{V}{V'} d\phi\,,
    \nonumber
\end{eqnarray}
where we used the slow-roll equations~(\ref{eomphi1}) and (\ref{friedmannsingle1}). To solve the horizon and flatness problems it is required that
the total number of e-foldings during inflation exceeds $60$
\be
N_{tot} \equiv \ln \frac{a_e}{a_i} \gtrsim 60.
\ee
However the precise value of $N_{tot}$ depends on the energy scale of inflation and details of the reheating after inflation.
It is during the slow-roll phase, which lasts nearly $40-60$ e-folds before inflation ends (the precise value again is determined by the
details of reheating and post-inflationary evolution of the universe), when the quantum fluctuations in the field are
imprinted on the CMB and $\phi_i$ corresponds to the field value when these fluctuations in the CMB are created.

\subsection{Inflation models and key inflationary observables}
The models of inflation can be broadly divided into two categories: {\it large field inflation} and {\it small field inflation}.
The class of models in which during inflation $\phi_s > 1 M_p$ are called the large field models. The chaotic potential
$V(\phi)=\lambda_{n}\phi^{n}$ models and exponential potential $V(\phi)=V_{0}e^{\phi}$ models are the large field type models.
In the chaotic inflation scenario, first introduced in
\cite{Linde:1983gd}, as the universe exits the Planck era at $t \sim 10^{-43} sec$ the initial value of the inflaton field
is set chaotically, $i.e.$ it acquires different values in different parts of the universe and the initial displacement of the
field from the minimum of the potential is larger than Planck scale. These models usually satisfy $V''(\phi)>0$.
The small field inflation models, instead are characterized by the fact that the slow-roll trajectory is at the small
field values $\phi_s < 1 M_p$. In these models the field starts close to an unstable maximum of the potential and rolls down to a stable minimum. An example of small field models is 'new inflation' \cite{Linde:1981mu} which arises naturally in the mechanism of spontaneous
symmetry breaking. In general, the form of the potential in these models are $V(\phi)=V_{0} (1-\phi^{n})$ and typically these models satisfy $V''(\phi)<0$.

A basic difference between the large field and small field models is that the large field models predict large amplitude of gravity waves produced during inflation whereas the small field models predict small amplitudes of gravity waves which are too small to
be detected in future observations. In either class of these models the inflation ends as soon as the slow-roll
conditions are violated and the field rolls down to the minimum of the potential, oscillates and decays into
the standard model particles. The decay process of the fields into standard model particles is known as {\it reheating}
and after this universe eventually enters into the radiation domination phase \cite{Bassett:2005xm,Allahverdi:2010xz}.

The models of standard slow-roll inflation are typically
defined through its scalar potential. For any inflation model in order not to be ruled out, it must predict certain
physical quantities in agreement with the observations. These physical quantities/observables are: the amplitude of the power spectrum
of curvature perturbations $\Delta_{\mathcal{R}}^{2}$, spectral index $n_{s}$, running of spectral index $\alpha_{s}$
and tensor-to-scalar ratio $r$. The latest
constraints on the inflationary observables as given by Planck-2015, for the combination {\it Planck TT $+$ lowP},~are \cite{Ade:2015lrj}
\bea
\ln(10^{10} \Delta_{\mathcal R}^{2}) &=& 3.089\pm 0.036, \label{DeltaGae}\\
 & & \nonumber \\
n_{s} &=&  0.9666 \pm 0.0062,\\
 & & \nonumber \\
r_{0.002} &<& 0.1.
\eea
The above values are for 7-parameter $\Lambda CDM$+$r$ model, when there is no scale dependence
of the scalar and tensor spectral indices. The value of amplitude and spectral index are given at $68 \% CL$ at the pivot scale
$k=0.05 Mpc^{-1}$. Whereas the upper bound on tensor-to-scalar ratio is determined at $95 \% CL$ at $k=0.002 Mpc^{-1}$
~\cite{Ade:2015lrj}. There are numerous models of inflation. For a review on variety of models of inflation
we refer the reader to ref.~\cite{Martin:2013tda} and references therein.

Also for 8-parameter $\Lambda CDM$+$r$+$\alpha_{s}$ model, when there is $k-$dependence of the spectral index or there
is a running of the spectral index, the Planck observations give
\bea
n_{s} &=& 0.9667 \pm 0.0132,\\
 & & \nonumber \\
r_{0.05} &<& 0.168.
\eea
The value of amplitude and spectral index are given at $68 \% CL$ at the pivot scale
$k=0.05 Mpc^{-1}$. Whereas the upper bound on tensor-to-scalar ratio is determined
at $95 \% CL$ at $k=0.05 Mpc^{-1}$. However, the amplitude of the power spectrum remains the same.
Notice that with running, the constraint on tensor-to-scalar ratio is relaxed.

Later, the joint BICEP2/Keck Array and Planck analysis put an upper limit on tensor-to-scalar~\cite{BKP:2015}
\be
r_{0.05} < 0.12 ~~~~~~ at~~ 95\% CL.
\ee
Most recently BICEP2/Keck Array Collaboration with its CMB polarization data and
combining it with Planck analysis of CMB polarization and temperature data have
further improved the bound on $r$~\cite{bicep2keck2015,Ade:2015xua}
\be\label{rGae}
r_{0.05} < 0.07 ~~~~~~ at~~ 95\% CL.
\ee
We will use these observed values to constrain parameters of the models studied in this review.

\section{Cosmological Perturbation Theory}\label{cosmoperttheory}
Here we present the calculation of the primordial density fluctuations power spectra generated by quantum fluctuations in
the inflaton field during inflation. Observation of the CMB anisotropies $\frac{\delta\rho}{\rho}\sim 10^{-5}$ proves that
the early universe was not perfectly uniform in its matter distribution. However, as the observed anisotropies are very small,
therefore these can be analyzed in terms of linear quantum fluctuations $\delta\phi(t,{\bf x})$ around the
homogeneous background\footnote{Since the measured CMB perturbations are small a linearized analysis of the KG and Einstein equations suffices, and in particular we do not need a theory of quantum gravity to describe the fluctuations. We quantize the perturbations, but keep the background classical.}. The linear theory of cosmological perturbations is a cornerstone of the modern cosmology. It not only explains the CMB anisotropies but also the formation and evolution of the structures in the universe.
The seed of these anisotropies were stretched to astronomical scales because of the superluminal expansion of the cosmic
space during inflationary quasi de-Sitter expansion. This theory has been extensively studied in literature; the
details can be found in Ref.s \cite{Baumann:2009ds,Riotto:2002yw}.

The linear perturbations of the metric $g_{\mu\nu}$ can be decomposed according to their spin w.r.t. a local rotation
of the spatial coordinates on the hypersurfaces of constant time into three kinds of perturbations: scalar, vector and
tensor. Here we will study only scalar and tensor perturbations in detail. Scalar perturbations explain the CMB temperature
anisotropy (or matter density fluctuations) and the seed for the structure formations in the universe.
One can see from the Einstein field equation (\ref{GRfieldeq}) (with $\Lambda=0$) that the scalar perturbations which give rise to perturbations in
the energy-momentum tensor leads to metric perturbations. On the other hand, metric perturbations
back react through the perturbations in the KG equations of motion (\ref{eomphi}) of the field, giving rise to field (or matter) perturbations. Therefore
\be
\delta g_{\mu\nu}(t,{\bf x}) \Longleftrightarrow \delta\phi(t,{\bf x})\,.
\ee
The tensor perturbations corresponds to primordial gravitational waves which is a generic prediction of
inflationary models. Observational constraints on the amplitude of the primordial gravitational waves can be
used to eliminate various inflation models.

\subsection{Linear Perturbations}\label{linearpert}
Linear order perturbations in the metric and field around the homogeneous background solutions of the
field $\phi(t)$ and the metric $g_{\mu\nu}(t)$ can be given as
\bea
\delta\phi(t,{\bf x}) &=& \phi(t,{\bf x}) - \phi(t), \\
 & & \nonumber \\
\delta g_{\mu\nu}(t,{\bf x}) &=& g_{\mu\nu}(t,{\bf x}) - g_{\mu\nu}(t).
\eea
The most general linearly perturbed spatially flat FRW metric can be written as
\begin{equation}\label{metricpert}
ds^{2} = -(1+2\Phi)dt^2 + 2 a(t) B_{i} dt dx^{i} + a(t)^2 \left[(1-2\Psi)\delta_{ij} + 2 E_{ij}\right] dx^{i} dx^{j},
\end{equation}
where $\Phi, \Psi$ are the scalar perturbations, $B_{i}$ are the vector perturbations and $E_{ij}$ are the tensor perturbations.
According to {\it SVT decomposition} scalar, vector and tensor perturbations are decoupled during inflation and
therefore evolve independently, this is also known as {\it Decomposition Theorem}. This theorem implies that if some physical process
in the early universe sets up tensor perturbations then these do not induce scalar perturbations, on the other hand, evolution of the
scalar perturbations is unaffected by the presence of any possible vector and tensor perturbations \cite{dodelson}. The importance of
SVT decomposition is that the Einstein equations for scalar, vector and tensor perturbations do not mix at linear order and
can therefore be studied separately. In this way the SVT decomposition greatly simplifies the calculations. The vector perturbations
are not sourced by inflation and furthermore they quickly decay with the expansion of the universe \cite{Baumann:2009ds}. Therefore we
will ignore vector perturbations and focus on scalar and tensor perturbations only.

According to SVT decomposition of the metric perturbations in real space, the vector $B_i$ can be decomposed into a gradient of
as scalar, say $B$, and divergence free vector, say $S_i$, as
\be
B_i \equiv \partial_i B - S_i, \quad\quad \text{where} \quad \partial^{i}S_{i}=0,
\ee
and similarly, any second rank tensor $E_{ij}$ can be written in terms of a divergence free vector and a traceless and divergence free tensor as
\be
E_{ij} \equiv 2\partial_{i}\partial_{j}E + 2 \partial_{(i}F_{j)} +h_{ij}\,,
\ee
where
 \[
\partial^{i}F_{i}=0, \quad h_{i}^{i}=\partial^{i}h_{ij}=0\,.
 \]
Since the $g_{\mu\nu}$ is a symmetric tensor therefore in $4$-dimensions it has $10$ independent components or $10$ degrees of freedom (d.o.f.).
The $10$ d.o.f. of metric has been decomposed into $4+4+2$ SVT d.o.f., $i.e.$ $4$ scalar d.o.f. $\Psi, \Phi, B, E$, $2$
vector d.o.f. for $S_{i}, F_{i}$ vectors each and $2$ d.o.f. of tensor $h_{ij}$.
\subsection{Gauge Transformation and Gauge Invariance}\label{GTGI}
In GR the gauge transformations are the general coordinate transformations from one local reference
frame to another. Here we will briefly review the gauge fixing and the behavior of the scalar, vector and tensor
perturbations under general coordinate transformation. We will introduce the gauge invariant quantities
in next Section. Fixing a gauge in GR implies choosing a coordinate
system, a slicing of spacetime into constant time hypersurfaces and threading into lines with fixed spatial coordinate ${\bf x}$.
Now let us consider the infinitesimal coordinate transformations
 \[
 x^{\mu}\rightarrow \tilde x^{\mu} = x^{\mu} + \xi^{\mu}\,,
 \]
where $\xi^{\mu}$ is a spacetime dependent infinitesimal quantity. At a given point on spacetime manifolds, the metric in the
new coordinate system $\tilde x^{\mu}$ can be determined using the invariance of the line-element
 \[
 ds^{2} = g_{\mu\nu}dx^{\mu}dx^{\nu} = \tilde g_{\mu\nu}d\tilde x^{\mu}d\tilde x^{\nu}\,,
 \]
or by applying the usual tensor transformation law
\be\label{gmunutransf}
\tilde g_{\mu\nu}(\tilde x^{\rho}) = \frac{\partial x^{\mu}}{\partial \tilde x^{\alpha}} \frac{\partial x^{\nu}}
{\partial \tilde x^{\beta}} g_{\mu\nu}(x^{\rho}).
\ee
Consider splitting the metric $g_{\mu\nu}(x^{\rho})$ into background and perturbed parts in both
$x^{\mu}$ and $\tilde x^{\mu}$ coordinate systems
\bea\label{tildegmunu}
g_{\mu\nu}(x^{\rho}) &=& g^{(0)}_{\mu\nu}(x^{\rho}) + \delta g_{\mu\nu}(x^{\rho}),\label{gmunu} \\
 & & \nonumber \\
\tilde g_{\mu\nu}(\tilde x^{\rho}) &=& g^{(0)}_{\mu\nu}(\tilde x^{\rho}) + \delta\tilde g_{\mu\nu}(\tilde x^{\rho}).
\eea
Note that we have not put tilde over the background metric because due to homogeneity and isotropy, the background forms of the metric tensor (also vectors and scalars) does not change, so that the background quantities behave the same way in the new coordinate
system $\tilde x^{\mu}$. Partial differentiation of coordinate transformations gives
 \[
\frac{\partial x^{\mu}}{\partial \tilde x^{\alpha}} = \delta^{\mu\alpha} - \frac{\partial \xi^{\mu}}{\partial \tilde x^{\alpha}}\,.
 \]
The background metric can be expanded as
\be
g^{(0)}_{\mu\nu}(x^{\rho}) = g^{(0)}_{\mu\nu}(\tilde x^{\rho} - \xi^{\rho}) \simeq g^{(0)}_{\mu\nu}(\tilde x^{\rho})
- \frac{\partial g^{(0)}_{\alpha\beta}}{\partial \tilde x^{\rho}} \xi^{\rho}. \label{gtaylor}
\ee
Substituting (\ref{gmunu}) into (\ref{gmunutransf}) and comparing with (\ref{tildegmunu}) while
using (\ref{gtaylor}), one infers the transformation law of metric tensor perturbation,
\be
\delta g_{\alpha\beta}(x^{\rho}) \rightarrow \delta \tilde g_{\alpha\beta}(\tilde x^{\rho}) = \delta g_{\alpha\beta} -
\frac{\partial g^{(0)}_{\alpha\beta}(\tilde x^{\rho})}{\partial \tilde x^{\rho}} \xi^{\rho}
- g^{(0)}_{\alpha\nu}(x^{\rho}) \frac{\partial \xi^{\nu} }{\partial \tilde x^{\beta}}
- g^{(0)}_{\mu\beta}(x^{\rho}) \frac{\partial \xi^{\mu} }{\partial \tilde x^{\alpha}}.\label{metrictransf}
\ee
Similarly, for a $4-$vector $u^{\mu}(x^{\rho})$ which transform as
 \[
 \tilde u_{\alpha}(\tilde x^{\rho}) = \frac{\partial x^{\mu}}{\partial \tilde x^{\alpha}} u_{\mu}(x^{\rho})\,,
 \]
one obtains that its perturbation $\delta u_{\alpha}(x^{\rho}) = u_{\alpha}(x^{\rho})- u^{(0)}_{\alpha}(x^{\rho})$ transforms as
\be\label{vectortransf}
\delta u_{\alpha}(x^{\rho}) \rightarrow \delta \tilde u_{\alpha}(\tilde x^{\rho}) = \delta u_{\alpha}(x^{\rho})
- \frac{\partial u^{(0)}_{\alpha}(\tilde x^{\rho})}{\partial \tilde x^{\rho}} \xi^{\rho} - u^{(0)}_{\mu}(x^{\rho})
\frac{\partial \xi^{\mu} }{\partial \tilde x^{\alpha}},
\ee
while, for a scalar $q(x^{\mu})$, which doesn't change under the coordinate transformation
 \[
\tilde q(\tilde x^{\mu}) = q(x^{\mu})\,,
 \]
one gets that its perturbation $\delta q(x^{\mu})= q(x^{\mu}) - q^{(0)}(x^{\mu})$ transforms as
\be\label{scalartransf}
\delta q(x^{\mu}) \rightarrow \delta \tilde q(\tilde x^{\mu}) = \delta q(x^{\mu})
- \frac{\partial q^{(0)}(\tilde x^{\rho})}{\partial \tilde x^{\rho}} \xi^{\rho}.
\ee
Let us now write the temporal and spatial components of the infinitesimal vector $x^{\mu}\rightarrow \tilde x^{\mu} = x^{\mu} + \xi^{\mu}$ as
\be\label{gaugeT}
t \rightarrow t + \alpha, \qquad
x^{i} \rightarrow x^{i} + \delta^{ij}\partial_{j}\beta\,,
\ee
where
 \[
 x^{\mu}\equiv (x^{0},x^{i})=(t,x^{i})\,, \quad \xi^{\mu}\equiv (\xi^{0},\xi^{i})=(\alpha,\partial^{i}\beta)\,.
 \]
$\alpha$ is infinitesimal
temporal shift and $\beta$ is a scalar function.
Using the metric tensor and scalar perturbation transformation laws (\ref{metrictransf}) and (\ref{scalartransf}), we find that
the tensor perturbations $h_{ij}$ are invariant under the gauge transformations (and therefore they already represents gravitational waves
in a gauge invariant manner), whereas the scalar perturbations $\Phi$, $\Psi$, $B$ and $E$ transform as
\bea
\Phi &\rightarrow& \Phi - \dot\alpha,\nonumber \\
\Psi &\rightarrow& \Psi + H \alpha, \nonumber\\
B &\rightarrow& B + a^{-1}\alpha - a \dot\beta, \nonumber \\
E &\rightarrow& E-\beta, \label{metricT}
\eea
Thus we find that only $\alpha$ and $\beta$ contributes to the transformations of the scalar perturbations and we can choose them
appropriately (as we are free to choose them) and can impose two conditions on the scalar functions $\Phi$, $\Psi$, $B$ and $E$
to remove any two of them. This is called the {\it gauge fixing} or {\it gauge choice} which corresponds to choosing a {\it gauge transformation}. It is
possible that the freedom in coordinate choice leads to an appearance of fictitious perturbation modes which do not describe any real
physical inhomogeneities. However, one can construct gauge invariant quantities which do not depend on choice of coordinate system
and represents real inhomogeneities. Two important gauge-invariant quantities were introduced by Bardeen \cite{Bardeen:1980kt}
\bea
\Phi_B &\equiv& \Phi - \frac{d}{dt}[a^{2}(\dot E -B/a)],\\
 & & \nonumber \\
\Psi_B &\equiv& \Psi + a^2 H (\dot E -B/a).
\eea
The gauge invariance of $\Phi_B$ and $\Psi_B$ implies that if they vanish in one particular coordinate system then they will be vanishing
in any coordinate system. Such a construction of gauge invariant quantities allows us to distinguish between physical inhomogeneities
and fictitious perturbations. If there are metric perturbations present even when both $\Phi_B$ and $\Psi_B$ are zero, then they are
fictitious perturbations and can be eliminated using change of coordinates.

Using (\ref{scalartransf}), we find that the perturbations of the scalar field $\phi$ transform as
\be
\delta \tilde \phi(\tilde x^{\mu}) = \delta \phi(x^{\mu}) - \frac{\partial \phi^{(0)}(x^{\rho})}{\partial x^{\rho}} \xi^{\rho}.
\ee
Since the background field $\phi^{(0)}=\phi(t)$ is time-dependent only, therefore
\be
\delta \tilde \phi(\tilde x^{\mu}) = \delta \phi(x^{\mu}) - \dot \phi(t) \alpha,
\ee
where $\alpha=\xi^0$. Also the matter perturbations or the perturbations to the total stress energy tensor
$T_{\mu\nu}$ are given in terms of the perturbations of the energy density $\delta\rho$, perturbations of
pressure $\delta p$ and perturbations of momentum density $\delta q$. Under gauge transformation these
perturbations transform as
\bea
\delta\rho &\rightarrow& \delta\rho - \dot \rho \alpha,\nonumber \\
 & & \nonumber \\
\delta p &\rightarrow& \delta p -\dot p \alpha,\nonumber \\
 &  & \nonumber \\
\delta q &\rightarrow& \delta q + (\rho + p)\alpha.\label{matterT}
\eea
\subsection{Gauge Invariant Variables}\label{GIV}
We discussed and explained the fictitious and real perturbations in previous Section.
In order to avoid the fictitious gauge modes, it is preferable to use the
gauge-invariant combinations of the matter and metric perturbations \cite{Bardeen:1980kt}. An important gauge-invariant
quantity is the {\it comoving curvature perturbation} $\mathcal{R}$ \cite{Bardeen:1983qw} defined as
\be
\mathcal{R} \equiv \Psi - \frac{H}{\rho+p}\delta q, \label{comcurvpert}
\ee
this can also be given in terms of the metric perturbations in the longitudinal gauge as \cite{Mukhanov:1990me}
\be
\mathcal{R} = \Psi - \frac{H}{\dot H}(\dot\Psi + H\Phi ),
\ee
and for perfect fluid, this can further be simplified to give (we use $\Psi=\Phi$, see Eq. (\ref{ijGT1}))
\be
\mathcal{R} = \Phi - \frac{H}{\dot H}(\dot\Phi + H\Phi ).\label{curvpertphi11}
\ee
The condition (\ref{comcurvpert}) can be constructed by considering the slicing of the spacetime into
constant $\delta q$ (or constant-$\phi$) hyperserfaces which provide the constraint
\be
\delta q \rightarrow \delta q + (\rho + p)\alpha=0 \quad \Longrightarrow \quad \alpha =-\frac{\delta q}{\rho+p}
\ee
substituting this $\alpha$ into the metric transformation relation $\Psi\rightarrow \Psi+H\alpha$ gives the relation (\ref{comcurvpert}) for $\mathcal{R}$.
Since $\delta q$ is the scalar $0i-$component of the perturbed energy momentum tensor $T_{i}^{0}=\partial_{i}\delta q$ and during inflation
$T^{0}_{i}=-\dot\phi \partial_{i}\delta\phi$, therefore both of these relations implies $\delta q= -\dot\phi \delta\phi$. Also from (\ref{rhophi}) we have $\rho +p = \dot\phi^{2}$. Then the comoving curvature perturbations (\ref{comcurvpert}) during inflation becomes
\be
\mathcal{R} \simeq \Psi + \frac{H}{\dot\phi}\delta\phi\,. \label{comcurvpert1}
\ee
Geometrical interpretation of $\mathcal{R}$ is that it measures the spatial curvature of the comoving hypersurface where $\delta\phi=0$, $i.e.$
\be
\mathcal{R} = \Psi|_{\delta\phi=0}.
\ee
Another important gauge-invariant quantity is {\it curvature perturbations on constant energy density hypersurfaces} $\mathcal{\zeta}$ defined as
\be
-\mathcal{\zeta} \equiv \Psi + \frac{H}{\dot\rho}\delta\rho\,, \label{zetacurvpert}
\ee
Similar to $\mathcal{R}$, the quantity $\mathcal{\zeta}$ can be constructed by considering the slicing of the spacetime into constant
energy density hyperserfaces which provide the constraint
\be
\delta\rho \rightarrow \delta\rho - \dot \rho \alpha=0 \quad \Longrightarrow \quad \alpha =\frac{\delta \rho}{\dot\rho}
\ee
substituting this $\alpha$ into the metric transformation relation $\Psi\rightarrow \Psi+H\alpha$ gives the relation (\ref{zetacurvpert})
for $\mathcal{\zeta}$. Since during slow-roll (from equation~(\ref{rhophi})) one has
 \[
 \delta\rho=\dot\phi \delta\dot\phi+V' \delta\phi \simeq V'\delta\phi
 \]
and
 \[
 \dot\rho = \dot\phi \ddot\phi + V' \dot\phi \simeq V' \dot\phi\,.
 \]
These equations implies
 \[
 \frac{\delta\rho}{\dot\rho}\simeq \frac{\delta\phi}{\dot\phi}\,.
 \]
As a consequence, $-\zeta $ becomes
\be
-\mathcal{\zeta} \simeq \Psi + \frac{H}{\dot\phi}\delta\phi. \label{zetacurvpert1}
\ee
Geometrical interpretation of $\mathcal{\zeta}$ is that it measures the spatial curvature of the uniform density hypersurface, $i.e.$
\be
-\mathcal{\zeta} = \Psi|_{\delta\rho=0}.
\ee
We see that the curvature perturbations $\mathcal{R}$ and $\mathcal{\zeta}$ by construction are invariant under the gauge transformations (\ref{gaugeT}),
which can be verified using (\ref{metricT}) and (\ref{matterT}) into their expressions (\ref{comcurvpert}) and (\ref{zetacurvpert}).

Also using the linearized Einstein field equations it can be shown that the gauge invariant curvature
perturbations $\mathcal{\zeta}$ and $\mathcal{R}$ are related as \cite{Baumann:2009ds}
\be
-\zeta= \mathcal{R} + \left(\frac{k}{aH}\right)^{2} \frac{2\rho}{3(\rho+p)} \Psi_B,\label{zetaR}
\ee
which implies that {\it on superhorizon scale $k\ll aH$, $\mathcal{\zeta}$ and $\mathcal{R}$ are equal}. Also we saw that under slow-roll
they are equal, $cf.$ equations~(\ref{comcurvpert1}) and (\ref{zetacurvpert1}). We conclude that:

\vspace{0.1in}

{\it The curvature perturbations $\mathcal{\zeta}$ and $\mathcal{R}$ also share an important property that on superhorizon scales they are conserved for adiabatic matter perturbations.}

\vspace{0.1in}

In general, it is possible that the pressure perturbations (in any gauge) can be split into adiabatic and non-adiabatic
(entropic) parts as
\begin{eqnarray}
\delta p &\equiv &  \delta p_{ad} +\delta p_{nad} \\
 & & \nonumber \\
 & = & c_{s}^{2} \delta\rho + \delta p_{nad}\,,
\end{eqnarray}
where $c_{s}^{2} \equiv \frac{\dot p}{\dot \rho}$ and the adiabatic pressure perturbations are defined as
\be
\delta p_{ad}\equiv \frac{\dot p}{\dot \rho}\delta \rho\,.
\ee
They satisfy the condition
\be
\frac{\delta p}{\dot p} =\frac{\delta\rho}{\dot\rho}
\ee
which implies that a given time displacement $\delta t$ causes the same relative fractional change $\frac{\delta X}{\dot X}$
in all scalar quantities $X\equiv (\rho, p,...)$. The non-adiabatic part of the pressure perturbations $\delta p_{nad}$ are defined as
\be
\delta p_{nad} \equiv \dot p \Gamma \equiv \delta p - \frac{\dot p}{\dot \rho}\delta \rho\,,
\ee
where
\be
\Gamma \equiv \frac{\delta p}{\dot p} - \frac{\delta\rho}{\dot\rho},
\ee
is the {\it entropy perturbation}, also known as {\it isocurvature perturbation}. $\Gamma$, defined in this way, is
gauge-invariant and represents the displacement between hypersurfaces of uniform pressure and uniform density.

Using the perturbed Einstein field equations (\ref{pertEFE}), as discussed in detail in the next Section,
it can be shown that the evolution of the gauge invariant curvature perturbations in the longitudinal gauge
is given by~\cite{Kaiser:2010yu,Baumann:2009ds}
\be
\mathcal{\dot R} = -\frac{H}{\rho+p} \delta p_{nad} + \left(\frac{k}{aH}\right)^{2} \left[\frac{H^{2}}{3(\rho+p)}\delta q \right],\label{Rdot}
\ee
therefore if there are no non-adiabatic matter perturbations $\delta p_{nad} = 0$ or no isocurvature perturbations $\Gamma=0$,
the curvature perturbations $\mathcal{R}$ (also $\mathcal{\zeta}$, $cf.$ equation~(\ref{zetaR}))
are conserved on superhorizon scales $k \ll aH$.

\vspace{0.2in}

{\bf Physical Interpretation of Adiabatic (Curvature) and Isocurvature perturbations:}
If the curvature perturbations are such that they can not give rise to variations in the
relative density between different components of the cosmological fluid (photons, baryons, neutrinos and CDM particles)
after inflation, then the curvature perturbations are adiabatic :
\be
\delta \left(\frac{n_m}{n_r}\right) =0 \quad\Longrightarrow \quad \frac{\delta n_m}{n_m} = \frac{\delta n_r}{n_r},\label{nmnr}
\ee
where
 \[
\delta n = n(t,{\bf x})-n^{(0)}(t)
 \]
and the index $m$ collectively stands for non-relativistic matter components
$e.g.$ baryons and CDM and index $r$ for relativistic matter components $e.g.$ photons and neutrinos. Since
 \[
  n_{(m,r)}\propto a^{-3}\,, \quad
 \rho_m \propto a^{-3}\,, \quad \rho_r \propto a^{-4}\,,
 \]
the condition (\ref{nmnr}) gives
\be
\frac{\delta\rho_m}{\rho_m} = \frac{3}{4} \frac{\delta\rho_r}{\rho_r}. \label{rhomr}
\ee
In single-field slow inflationary scenario, the condition (\ref{rhomr}) holds and therefore the perturbations produced by
single-field inflation are purely adiabatic.
However, in inflationary models with more than one field, the perturbations are not necessarily adiabatic. If during inflation there
are more than one field and all are evolving in time, the fluctuations orthogonal to background trajectory can affect the relative density
between different components of the cosmological fluid even if the total density (and therefore spatial curvature) is unperturbed
\cite{Gordon:2000hv}. For example, the relative density perturbations (isocurvature/entropy perturbations) between photon and CDM can be defined as
\be
\Gamma_{m\gamma} \equiv \frac{\delta\rho_{cdm}}{\rho_{cdm}} - \frac{3}{4} \frac{\delta\rho_\gamma}{\rho_\gamma}.
\ee
Since adiabatic and isocurvature perturbations give different peak structure in the CMB power spectrum, therefore
different type of perturbations can be distinguished from the CMB measurements. In fact, CMB observations suggest
that even if the isocurvature perturbations are present, their amplitude is vanishingly small compared to amplitude
of the adiabatic (curvature) perturbations \cite{Ade:2015lrj}. The theoretical predictions of the isocurvature
perturbations are extremely model dependent. Not only the presence of more than one scalar field may give rise to entropic
perturbations but these may also be generated in non-minimally coupled inflation
models~\cite{Kaiser:2010yu}. Also the post-inflationary evolution may generate them.

%
\subsection{Curvature Perturbation and Scalar Power Spectrum}\label{CPSPS}
For a metric with small perturbations, the Einstein tensor $G_{\nu\mu}$ can be written as
 \[
 G_{\nu\mu}=G^{(0)}_{\nu\mu} + \delta G_{\nu\mu} + ....\,,
 \]
where $\delta G_{\nu\mu}$ represents the terms
with linear metric perturbations $\delta g_{\mu\nu}$. The stress energy tensor $T^{\nu}_{\mu}$ can be split
in a similar fashion and we get the linearized Einstein field equations
\be
\delta G^{\nu}_{\mu}= 8\pi G ~\delta T^{\nu}_{\mu}. \label{pertEFE}
\ee
The gauge freedom allows to choose the two functions $\alpha$ and $\beta$ which provides two conditions on the
scalar functions $\Phi$, $\Psi$, $B$, $E$ and therefore allows to remove any two of them. The gauge freedom greatly simplifies the
calculations and knowing the solutions of the gauge-invariant variables, one can calculate the density and metric perturbations
in any coordinate system in a simple way \cite{mukhanov}. One of many useful gauges is the {\it conformal Newtonian gauge}
or {\it longitudinal gauge} which is defined by the conditions
 \[
 B=0\,, \quad E=0\,.
 \]
In this gauge the FRW line element assumes the simple form
\be
ds^{2} = -(1+2\Phi)dt^2 + a(t)^2 (1-2\Psi)\delta_{ij} dx^{i} dx^{j}. \label{NGFRWmetric}
\ee
Now we calculate the perturbed Einstein field equations (\ref{pertEFE}). For the metric (\ref{NGFRWmetric}), the components of the
perturbed Einstein tensor can be obtained as
\bea
\delta G^{0}_{0} &=& -2\nabla^{2}\Psi + 6 H^{2}\Phi+ 6 H \dot\Psi, \label{00G} \\
\delta G^{0}_{i} &=& -2\partial_{i}(H\Phi+\dot\Psi), \label{0iG} \\
\delta G^{i}_{j} &=& \partial^{i}\partial_{j}(\Psi-\Phi) + [\nabla^{2}(\Phi-\Psi)+2\ddot\Psi+(4\dot H + 6H^{2})\Phi
 + H(2\dot\Phi+2\dot\Psi)]\delta^{i}_{j}. \label{ijG}
\eea
Using the perfect fluid description as defined in equation~(\ref{EMtensor}) and the stress-energy-momentum tensor for the
scalar field $\phi$ as defined in equation~(\ref{emtphi}), the components of the perturbed $T_{\mu\nu}$ are given by
\bea
\delta T^{0}_{0}&=& -\delta\rho=\dot\phi^{2} \Phi - \dot\phi \delta\dot\phi - V' \delta\phi, \label{00T} \\
\delta T^{0}_{i}&=& \delta q = -\dot\phi \partial_{i} \delta\phi, \label{0iT} \\
\delta T^{i}_{j}&=& \delta p = [-\dot\phi^{2}\Phi + \dot\phi \delta\dot\phi - V' \delta\phi]\delta^{i}_{j}. \label{ijT}
\eea
where the relation
 \[
 \delta T^{\nu}_{\mu} = \delta (g^{\nu\tau}T_{\mu\tau})=\delta g^{\nu\tau}T_{\mu\tau}+g^{\nu\tau}\delta T_{\mu\tau}
 \]
has been used.
To compute the curvature perturbation $\mathcal{R}$, one first considers the $ij$-component of the perturbed
Einstein field equation (\ref{pertEFE}). From equation~(\ref{ijT}), we see that the stress energy tensor has no off-diagonal components, therefore taking the
off-diagonal components, $i.e.$ $i\neq j$, of the equations~(\ref{ijG}) and (\ref{ijT}), we have
\be
\partial^{i}\partial_{j}(\Psi-\Phi)=0 \quad \Longrightarrow \quad \Phi=\Psi,\label{ijGT1}
\ee
therefore we can work with any of the variable $\Phi$ or $\Psi$, let's work with $\Psi$.
We note that if the spatial part of the stress energy tensor is diagonal, $i.e.$ $\delta T^{i}_{j}\propto \delta^{i}_{j}$,
the variable $\Phi$ or $\Psi$ can be seen as a generalisation of the Newtonian potential
which therefore explains the name {\it Newtonian gauge} for this choice of coordinate system.
Considering the diagonal components, $i.e.$ $i=j$, of equations~(\ref{ijG}) and (\ref{ijT}), one gets
\be
\ddot\Psi + 4 H\dot\Psi + (2\dot H + 3 H^2)\Psi = -\dot\phi^{2}\Phi + \dot\phi \delta\dot\phi - V' \delta\phi. \label{ijGT}
\ee
Since $\phi$ is background quantity which is only time dependent, therefore equations~(\ref{0iG}) and (\ref{0iT}) for
$0i$-components give
\be
\dot\Psi + H\Psi = 4 \pi G ~\dot\phi \delta\phi = \epsilon H^{2} \frac{\delta\phi}{\dot\phi},\label{0iGT}
\ee
where  $\epsilon = 4\pi G \frac{\dot\phi^2}{H^2}$ (slow-roll parameter).
Similarly the equations~(\ref{00G}) and (\ref{00T}) for $00$-component gives
\be
\nabla^{2}\Psi -3H\dot\Psi -3H^2 \Psi = 4\pi G (\dot\phi \delta\dot\phi - \dot\phi^2 \Psi +V' \delta\phi).\label{00GT}
\ee
For the purpose of analysis, it is convenient to work in terms of
the Fourier decomposition of the metric and the field perturbations, and see what happens to a
perturbation corresponding to a given comoving spatial scale $k$ with corresponding comoving wavelength $\lambda=\frac{2\pi}{k}$.
Using Fourier transformation, we can decompose the perturbations $\Psi$ and $\delta\phi$ into a superposition of plane-wave states with
comoving wavevector ${\bf k}$ :
\be\label{FTpert}
\Psi(t,{\bf x}) = \int \frac{d^{3}{\bf k}}{(2\pi)^{3/2}} \Psi_{\bf k}(t)  e^{i {\bf k.x}}.
\ee
A similar expression holds for $\delta\phi$. The evolution of a mode amplitude $\Psi_{\bf k}$
or $\delta\phi_{\bf k}$ depends only on the comoving wavenumber $k = |{\bf k}|$ whereas the corresponding actual physical wavenumber
is $\frac{k}{a(t)}$ as $\lambda \propto k^{-1} \propto a $. Using (\ref{FTpert}) it is easy to show that both the perturbations
satisfy the Poisson equation:
\be
\nabla^{2}\Psi_{k} = -k^{2} \Psi_{k}~, \quad \quad\quad \nabla^{2}\delta\phi_{k} = -k^{2} \delta\phi_{k}.
\ee
Using these Poisson's equations for perturbations we can simply work in terms of $\Psi_{\bf k}$ and $\delta\phi_{\bf k}$.
We now add equations~(\ref{0iGT}) and (\ref{00GT}) to arrive at the equation of motion of gravitational potential $\Psi$ as
\be
\ddot\Psi_k +\left(H-2\frac{\ddot\phi}{\dot\phi}\right)\dot\Psi_k +2\left(\dot H - H\frac{\ddot\phi}{\dot\phi}\right)
\Psi_k +\frac{k^{2}}{a}\Psi_k = 0,\label{0i00GT1}
\ee
where we have used the background equation for scalar field $V' \simeq -3 H \dot\phi$ and the relation
$\dot H \simeq -4\pi G~ \dot\phi^{2}$. Using the slow-roll parameter relation
 \[
 \delta=\eta-\epsilon=\frac{-\ddot\phi}{H\dot\phi}\,,
 \]
the above equation~(\ref{0i00GT1}) can also be given as
\be
\ddot\Psi_k +H\left(1-2\epsilon + 2\eta\right)\dot\Psi_k +2 H^{2} \left(\eta-2\epsilon \right) \Psi_k +\frac{k^{2}}{a}\Psi_k = 0,\label{0i00GT}
\ee
Since the slow-roll parameters satisfy $\epsilon\ll1$ and $\eta\ll1$, it is easy to infer from the above equation
(\ref{0i00GT}) that on superhorizon scales $k \ll (aH)$,
\be
\dot\Psi_k \simeq 2(2\epsilon-\eta) H \Psi_k \quad \Longrightarrow \quad \dot\Psi_k \ll H\Psi_k
\ee
which implies that on superhorizon scales the time variations of the perturbations $\Psi_k$ can be safely neglected compared to $H\Psi_k$. This relation
holds true for field perturbations as well, $i.e.$ $\delta\dot\phi_k \ll H \delta\phi_k$. Therefore on superhorizon scales,
from equation~(\ref{0iGT}), we can relate the gravitational potential and field perturbations as
\be
\Psi_k \simeq \epsilon H \frac{\delta\phi}{\dot\phi},\label{0iGTapprox}
\ee
This can be used to compute the comoving curvature perturbation $\mathcal{R}_k$ on superhorizon scale (\ref{comcurvpert1}) as
\begin{eqnarray}\label{comcurvpertSH}
\mathcal{R}_k &\simeq & \Psi_k + \frac{H}{\dot\phi}\delta\phi_k \\
 &\simeq & (1+\epsilon)\frac{H}{\dot\phi}\delta\phi_k \nonumber \\
 &\approx & \frac{H}{\dot\phi}\delta\phi_k\,.
\end{eqnarray}
Before we go any further, it is convenient to define the Power Spectrum. It characterizes the properties of
perturbations. Any generic quantity $f(t,{\bf x})$ in the Fourier space can be expanded as
\be\label{FT}
f(t,{\bf x}) = \int \frac{d^{3}{\bf k}}{(2\pi)^{3/2}} f_{\bf k}(t)  e^{i {\bf k.x}},
\ee
and the power spectrum $\mathcal{P}_{f}(k)$ of the quantity $f_{\bf k}(t)$ is defined through
\be\label{PS1}
\langle|f^{\ast}_{\bf k} f_{\bf k'}|\rangle \equiv \delta^{(3)} ({\bf k}-{\bf k'}) \frac{2\pi^{2}}{k^{3}} \mathcal{P}_{f}(k),
\ee
where $\langle|f^{\ast}_{\bf k} f_{\bf k'}|\rangle$ implies the vacuum expectation value of the quantity $f_{k}(t)$ in the vacuum quantum state $|0\rangle$ of the system and $\delta^{(3)} ({\bf k}-{\bf k'})$ is the three dimensional Kronecker delta function. The definition (\ref{PS1}) lead to the power spectrum:
\be\label{PSgeneric}
\mathcal{P}_{f}(k) = \frac{k^{3}}{2\pi^{2}}\langle|f_k|^{2}\rangle.
\ee
Therefore, using (\ref{PSgeneric}), we may write the {\it power spectrum of comoving curvature perturbation} $\mathcal{R}$ as
\be\label{PScurvpert}
\mathcal{P}_{\mathcal{R}}(k) = \frac{k^{3}}{2\pi^{2}}\langle|\mathcal{R}_k|^{2}\rangle\,.
\ee
Hence, using (\ref{comcurvpertSH}), the power spectrum of comoving curvature perturbation on superhorizon scale $k\ll (aH)$ becomes
\begin{eqnarray}\label{PScurvpertSH}
\mathcal{P}_{\mathcal{R}}(k) &\simeq & \frac{k^{3}}{2\pi^{2}}\frac{H^{2}}{\dot\phi^{2}}\langle|\delta\phi_k|^{2}\rangle \\
&\simeq & \frac{k^{3}}{4 \pi^{2} \epsilon M_{p}^{2}}\langle|\delta\phi_k|^{2}\rangle\,. \nonumber
\end{eqnarray}
Now we are left to calculate the time evolution of the field perturbation mode amplitudes $\delta\phi_k$. Consider perturbing the
KG equation of motion (\ref{eomphi}) for scalar field $\phi$, $i.e.$ taking the variation of KG equation, we get
\be
\delta\ddot\phi_k + 3 H \delta\dot\phi_k +\frac{k^2}{a^2}\delta\phi_k + V'' \delta\phi_k= -2 V' \Psi_k + 4\dot\phi \dot\Psi_k
\ee
where we have used the background equation~(\ref{eomphi1}). Since on superhorizon scales $|2 V' \Psi_k|\gg |4\dot\phi \dot\Psi_k|$
(which follows from the condition $\dot\Psi_k \ll H\Psi_k$ upon using the relation $V'\simeq -3H\dot\phi$), using
equation~(\ref{0iGTapprox}) and (\ref{eomphi1}), the perturbed KG equation on superhorizon scale can be written as
\be
\delta\ddot\phi_k + 3 H \delta\dot\phi_k + (V'' +6\epsilon H^{2} )\delta\phi_k = 0.\label{deltaphik}
\ee
In the above equation (\ref{deltaphik}), replacing  the variable $\delta\phi_k$ with $\frac{\delta\sigma_k}{a}$
and introducing the conformal time $d\tau=\frac{dt}{a}$ one infers
\be\label{deltasigmak}
\delta\sigma_{k}''- \frac{1}{\tau^{2}}\left(\nu^{2}-\frac{1}{4}\right) \delta\sigma_k = 0,
\ee
where prime denotes the derivatives w.r.t. conformal time $\tau$ and
\be\label{nuscalar}
\nu^{2} = \left(\frac{9}{4}-\frac{m_{\phi}^{2}}{H^{2}}\right) \simeq \frac{9}{4}+9\epsilon-3\eta.
\ee
In deriving the above relation (\ref{deltasigmak}),
we have used the relations
 \begin{eqnarray}
 \eta &=& \frac{V''}{V}\simeq \frac{m_{\phi}^{2}}{3H^{2}}\,, \nonumber \\
 & & \nonumber \\
  \frac{a''}{a} &=&\frac{1}{\tau^{2}}\left(\nu^{2}-\frac{1}{4}\right) \simeq
\frac{1}{\tau^{2}}(2+3\epsilon) \,, \nonumber
 \end{eqnarray}
which can be obtained using the definition of the conformal time during quasi de-Sitter expansion. Under quasi de-Sitter expansion during which the Hubble rate is not exactly constant and follow the relation $\dot H = -\epsilon H^{2}$, the definition of conformal time establishes the relation for the scale factor as
 \[
 a(\tau)=-\frac{1}{H \tau}\frac{1}{1-\epsilon}\,.
 \]
We see that the perturbed KG equation~(\ref{deltasigmak}) is
Bessel equation and its solution can be given in terms of Hankel functions
\be
\delta\sigma_k= \sqrt{-\tau}[c_1(k)H_{\nu}^{(1)}(-k\tau) + c_2(k)H_{\nu}^{(2)}(-k\tau)], \label{soldeltasigma}
\ee
where $H_{\nu}^{(1)}$ and $H_{\nu}^{(2)}$ are the Hankel's functions of the first and second kind, respectively.
We assume that in the ultraviolet regime, $i.e.$ on subhorizon scales, $k\gg aH$ ($-k\tau\gg1$) the solutions matches
the plane wave solutions $e^{-ik\tau}/\sqrt{2k}$. The assumption that in the ultraviolet regime when the mode
wavelengths are of sub horizon size the modes should behave like plane waves as we expect in the flat Minkowski spacetime is called the {\it Bunch-Davies boundary condition}. In the limit $-k\tau\gg1$ Hankel's functions are given by
\begin{eqnarray}
H_{\nu}^{(1)}(-k\tau \gg 1) &\sim & \sqrt{\frac{2}{-k\tau\pi}} e^{i\left(-k\tau-\frac{\pi}{2}\nu-\frac{\pi}{4}\right)}\,, \\
 & & \nonumber \\
H_{\nu}^{(2)}(-k\tau \gg 1) &\sim & \sqrt{\frac{2}{-k\tau\pi}} e^{i\left(-k\tau-\frac{\pi}{2}\nu-\frac{\pi}{4}\right)}\,.
\end{eqnarray}
Setting
 \[
 c_1(k) = \frac{\sqrt{\pi}}{2} e^{i\left(\nu+\frac{1}{2}\right)\frac{\pi}{2}}\,, \qquad c_2(k)=0\,,
 \]
from equation~(\ref{soldeltasigma}) one dervies the exact solution for $\delta\sigma_k$
\be
\delta\sigma_k = \frac{\sqrt{\pi}}{2} e^{i\left(\nu+\frac{1}{2}\right)
\frac{\pi}{2}}\sqrt{-\tau}H_{\nu}^{(1)}(-k\tau). \label{soldeltasigmak1}
\ee
As we are interested in the modes which have become superhorizon $k\ll aH$ ($-k\tau\ll1$) during inflation,
knowing that in the limit $-k\tau\ll1$ Hankel's function have solution
\be
H_{\nu}^{(1)}(-k\tau \ll 1) \sim \sqrt{\frac{2}{\pi}}  \frac{\Gamma(\nu)}{\Gamma(3/2)} 2^{\nu-\frac{3}{2}}
e^{-i\frac{\pi}{2}} (-k\tau)^{-\nu},
\ee
the solution (\ref{soldeltasigmak1}) on superhorizon scales becomes
\be
\delta\sigma_k \simeq \frac{\Gamma(\nu)}{\Gamma(3/2)} 2^{\nu-\frac{3}{2}} e^{i\left(\nu+\frac{1}{2}\right)\frac{\pi}{2}}
\frac{1}{\sqrt{2k}} (-k\tau)^{\frac{1}{2}-\nu}. \label{soldeltasigmak}
\ee
Since $\epsilon\ll1$ and $\eta\ll1$, we can set $\nu\sim\frac{3}{2}$ in the factors but will not do the same in the exponent because
exponent term $(-k\tau)^{\frac{1}{2}-\nu}$ gives the small scale dependence of the power spectrum of perturbations.
Going back to original variable $\delta\phi_k$, we find the the fluctuations on superhorizon scales in cosmic time
\be
|\delta\phi_k(t)| \simeq \frac{H}{\sqrt{2k^{3}}} \left(\frac{k}{aH}\right)^{\frac{3}{2}-\nu}\,.\label{deltaphiSH}
\ee
Therefore the power spectrum of fluctuations, from (\ref{PScurvpertSH}), becomes
\begin{eqnarray}\label{scalaramp}
 \mathcal{P}_{\mathcal{R}}(k) &\simeq & \frac{1}{8\pi^{2} \epsilon} \frac{H^{2}}{M_{p}^{2}} \left(\frac{k}{aH}\right)^{n_{s}-1} \\
  & & \nonumber \\
 &\equiv &\Delta_{\mathcal{R}}^{2}\left(\frac{k}{aH}\right)^{n_{s}-1}\,, \nonumber
\end{eqnarray}
where we have defined the spectral index $n_{s}$ of the comoving curvature perturbations, which determines the
tilt of the power spectrum or the small deviation of the power spectrum from scale invariance, as
\begin{eqnarray}\label{SIndex}
n_{s}-1 &\equiv & \frac{d\ln\mathcal{P}_{\mathcal{R}}}{d\ln k} \\
 & & \nonumber \\
&=& 3-2\nu = 2\eta-6\epsilon \nonumber \,.
\end{eqnarray}
Since the slow-roll parameters $\epsilon$ and $\eta$ are much smaller than unity, therefore $n_{s}-1\simeq0$
which implies inflation is responsible for producing the curvature perturbations with an almost scale invariant spectrum. For comparison with the observations, the power spectrum (\ref{scalaramp}) can be given as
\be\label{PS}
\mathcal{P}_{\mathcal{R}}(k) = \Delta_{\mathcal{R}}^{2}(k_{0})\left(\frac{k}{k_{0}}\right)^{n_{s}-1},
\ee
where
 \[
 k_{0}=a_{0}H_{0}
 \]
is the pivot scale. The pivot scale corresponds to a wavelength $\lambda_{0}\propto k_{0}^{-1}$ at
which the instrument measuring the CMB radiation has the maximum sensitivity. $\Delta_{\mathcal{R}}^{2}(k_{0})$ is the amplitude of
the power spectrum at the pivot scale $k_{0}$.

It is possible that spectral index may depend on scales $k$. The running (variation) of the spectral index $\alpha_s$ with modes $k$ is defined as
\be
\alpha_s \equiv \frac{dn_s}{d\ln k},
\ee
the running of the spectral index with scale arises only at the second order in slow-roll parameters and is therefore
expected to be very small $\alpha_s=\mathcal{O}(\epsilon^{2})$. Using the fact that $\frac{dN}{dt}=H$,
at horizon crossing $k=a(t_k)H(t_k)$ we find
 \[
 \frac{d\ln k}{dt} = H\left(1+\frac{\dot H}{H^2}\right)\approx H\,.
 \]
The $\alpha_s$ can be written as
\begin{eqnarray}
\alpha_s &=& \frac{dn_s}{d\phi}\frac{d\phi}{dt}\frac{dt}{d\ln k} \\
  &=&\frac{\dot\phi}{H}\frac{dn_s}{d\phi} \nonumber \\
  &=& -\frac{V'}{V}\frac{dn_s}{d\phi}\,,
\end{eqnarray}
where in the last equality we have used the background slow-roll equations~(\ref{eomphi1}) and (\ref{friedmannsingle1}). Now using spectral index relation (\ref{SIndex}) and the
definition of slow-roll parameters in terms of scalar potential, the $\alpha_s$ in terms of slow-roll
parameters can be given as
\be
\alpha_s=16 \epsilon\eta -24\epsilon^{2} - 2\xi\,,
\ee
where the slow-roll parameter $\xi$ is defined as
 \[
 \xi=\frac{V'V'''}{V^{2}}\,.
 \]
\subsection{Tensor Power Spectrum}
Along with density fluctuations (or scalar perturbations), inflation also predicts the existence of gravitational waves which are identified with
the tensor perturbations in the metric. According to SVT decomposition theorem all perturbations (scalar,vector and tensor)
evolve independently. The line element for tensor perturbations around the flat background is given by
\be
ds^{2} = -dt^{2} + a^{2} (\delta_{ij}+ h_{ij}) dx^{i} dx^{j},\label{tensormetric}
\ee
where $h_{ij}\ll 1$. The tensor perturbations $h_{ij}$ has $6$ d.o.f., but as we have seen the
tensor perturbations are traceless and divergence free, that is
 \[
 \delta^{ij} h_{ij}=0\,, \qquad  \partial^{i}h_{ij}=0\,.
  \]
These 4 conditions reduces the tensor d.o.f. to 2 physical d.o.f. which corresponds to 2 polarisations of the gravitational waves, indicated by
$s=+,\times$. The components of the perturbed Ricci scalar and Ricci curvature are
\be
\delta R_{00}=0,\quad \delta R_{i0}=0, \quad \delta R =0,
 \ee
 \[
\delta R_{ij} = -\frac{1}{2 a^2(t)}\triangledown^2 h_{ij} +\frac{1}{2}\ddot{h}_{ij} -\frac{\dot{a}}{2a}\dot{h}_{ij}
+ 2 \left(\frac{\dot a}{a}\right)^2 h_{ij}\,.
 \]
For a diagonal stress-energy tensor, as provided by inflaton, Eq. (\ref{emtphi}), the tensor modes do not have any source term in their equation of motion. This statement can be verified
very easily by calculating the perturbed Einstein field equations for the tensor perturbations metric (\ref{tensormetric})
where we find $\delta R=0$ and $\delta R_{\mu\nu}=0$ for all components except $i \neq j$ and $\delta T^{i}_{j}=0$
for $i\neq j$. Therefore we have a glimmer of decomposition theorem and we can state that the e.o.m. for tensor metric
perturbations have no scalar (inflaton) source in it and they evolve independent of scalar perturbations. Using the above mentioned perturbed
components of Ricci scalar and Ricci curvature into the perturbed field equations (\ref{pertEFE}) the e.o.m. (\ref{tensoreom}) for tensor metric perturbations $h_{ij}$ can be obtained.

In a more simpler manner the equation of motion for $h_{ij}$ can be obtained
from second order expansion of the Einstein-Hilbert action \cite{Baumann:2009ds,Riotto:2002yw}
\begin{eqnarray}\label{tensoraction}
S^{(2)} &=& \frac{M_{p}^{2}}{2} \int dx^{4} \sqrt{-g} \frac{1}{2} \partial_{\rho}h_{ij} \partial^{\rho} h_{ij} \\
 &=& \frac{M_{p}^{2}}{4} \int d\eta dx^{3} \frac{a^{2}}{2} [(h_{ij}')^{2}- (\partial_l h_{ij})^2]\,. \nonumber
\end{eqnarray}
This is the same actions as for the free massless scalar field in FRW universe. We define the following
Fourier expansion
\be
h_{ij} = \int \frac{d^{3}k}{(2\pi)^{3}} \sum_{s=+,\times} h_{{\bf k}}^{s}(\tau) e^{s}_{ij}(k) e^{i{\bf k.x}},\label{tensorFE}
\ee
where $e^{s}_{ij}$ are the polarization tensors which satisfy the following properties
\be\label{poltensor}
e_{ij}(k) =  e_{ji}(k), \quad e_{ii}(k)=0, \quad k^{i}e_{ij}(k)=0, \quad e^{s}_{ij}(k)e^{s'}_{ij}(k) = 2\delta_{ss'}.
\ee
Using (\ref{tensorFE}) and (\ref{poltensor}), the action (\ref{tensoraction}) leads to the e.o.m. for the quantity $h_{\bf k}$
\be
h^{s\prime\prime}_{\bf k} + 2\frac{a'}{a} h^{s\prime}_{\bf k} + k^{2} h^{s}_{\bf k} =0.\label{tensoreom}
\ee
Defining the canonically normalized field $\nu_{\bf k}^{s} \equiv \frac{1}{2}a h_{\bf k}^{s} M_{p}$, the e.o.m. (\ref{tensoreom}) becomes
\be
\nu_{\bf k}^{s\prime\prime} + \left(k^{2}-\frac{a''}{a}\right)\nu_{\bf k}^{s} = 0,\label{MSeqnu}
\ee
where
 \[
\frac{a''}{a} = \frac{2}{\eta^{2}}(2+3\epsilon)
 \]
during quasi de-Sitter epoch when $\dot H= -\epsilon H$. On super horizon scale $k\ll aH$, $k^{2}$ term in the  equation~(\ref{MSeqnu}) can be neglected and then it exactly matches with
the Mukhanov-Sasaki equation (\ref{deltasigmak}) for the massless scalar field in FRW universe during quasi de-Sitter epoch
whose solution on superhorizon scales can be given in analogy with the solution (\ref{deltaphiSH}) for $\delta\sigma_k$ as
\be
|\nu_{\bf k}^{s}| = \frac{1}{M_{p}} \frac{a H}{\sqrt{2k^{3}}}\left(\frac{k}{aH}\right)^{\frac{3}{2}-\nu_{T}}.\label{solnuk}
\ee
Here the quantity $\nu_{T}$, given by
 \[
 \nu_{T}\simeq \frac{3}{2}+\epsilon\,,
 \]
has been obtained using the relation
 \[
 \frac{a''}{a}= \frac{1}{\eta^{2}}\left(\nu_{T}^{2}-\frac{1}{4}\right)\simeq \frac{1}{\eta^{2}}(2+3\epsilon) \,.
 \]
Also, since the equation (\ref{MSeqnu}) or the action (\ref{tensoraction}) matches with the equations for
massless scalar field, therefore there will be no appearance of slow-roll parameter $\eta$ in $\nu_{T}$ through $m_{\phi}$
in contrast to relation (\ref{nuscalar}).

To characterize the tensor perturbations, we define the power spectrum of tensor perturbations as
\be
\mathcal{P}_T \equiv \frac{k^{3}}{2\pi^{2}} \sum_{s=+,\times} |h_{\bf k}^{s}|^{2} =
2\times \frac{k^{3}}{2\pi^{3}} \frac{4 |\nu_{\bf k}^{s}|^{2}}{a^{2}},
\ee
where the factor of 2 is due to the sum over the two polarization states of the gravitational wave.
Substituting for the solution (\ref{solnuk}), we get the amplitude of the tensor power spectrum on superhorizon scales as
\begin{eqnarray}\label{tensoramp}
\mathcal{P}_T &=& \frac{2}{\pi^{2}}\frac{H^{2}}{M_{p}^{2}} \left(\frac{k}{aH}\right)^{n_{T}} \\
 & & \nonumber \\
 &\equiv& \Delta_{T}^{2}\left(\frac{k}{aH}\right)^{n_{T}}\,. \nonumber
\end{eqnarray}
Similar to scalar spectral index $n_s$, we can define the {\it tensor spectral index} $n_{T}$ as
\begin{eqnarray}
n_{T}& \equiv & \frac{d\ln \mathcal{P}_{T}}{d\ln k} \\
 & & \nonumber \\
 &=& 3-2\nu_{T}=-2\epsilon\,. \nonumber
\end{eqnarray}


\vspace{0.2in}

{\bf Tensor-to-scalar ratio and energy scale of inflation:} Amplitude of the tensor perturbations are often normalized relative to the measured amplitude of the scalar
perturbations $\Delta_{\mathcal{R}}^{2}\simeq 1.95\times 10^{-9}$. The {\it tensor-to-scalar ratio} $r$ is defined as
the ratio of the two amplitudes
\be
r\equiv \frac{\Delta_{T}^{2}}{\Delta_{\mathcal{R}}^{2}}=16\epsilon\,,\label{tsr}
\ee
which determines the relative contribution of the tensor modes to mean squared low multipole CMB anisotropy.
In the last equality in above equation~(\ref{tsr}), we have used the amplitude relations (\ref{scalaramp}) and
(\ref{tensoramp}) for scalar and tensor perturbations.
Since scalar amplitude is fixed from the observations $\Delta_{\mathcal{R}}^{2}\simeq 1.95\times 10^{-9}$ and,
from (\ref{tensoramp}), amplitude of the tensor perturbations $\Delta_{T}^{2} \propto H^{2}\approx V(\phi)$,
therefore the value of tensor-to-scalar ratio is a direct measure of {\it energy scale of inflation}:
\be
V(\phi)^{1/4} \sim \left(\frac{r}{0.01}\right)^{1/4} 10^{16} ~GeV\,.\label{energyscale}
\ee
The value of tensor-to-scalar ratio $r>0.01$ implies inflation occurring at the GUT energy scale $10^{16} ~ GeV$.

\vspace{0.2in}

{\bf The Lyth bound and large-field inflation:} Inflation models which can predict large amplitude of the gravity waves (or large $r$) are extremely sensitive
to super Planckian physics. Here we will derive the Lyth bound which relates the tensor-to-scalar ratio with
super Planckian displacement of the inflaton value $\Delta\phi$ during inflation.
During slow-roll inflation, using (\ref{Hphidot}) and (\ref{epsilon}), the slow-roll parameter $\epsilon$ can be given as
\begin{eqnarray}
\epsilon &\simeq & \frac{1}{2M_{p}^{2}}\frac{\dot \phi^{2}}{H^{2}} \\
 &=& \frac{1}{2M_{p}^{2}}\left(\frac{d\phi}{dN}\right)^{2}\,, \nonumber
\end{eqnarray}
where we have used the relation
 \[
N=H dt = \frac{H}{\dot \phi}d\phi\,.
 \]
Therefore the tensor-to-scalar ratio can be directly related to the evolution of the inflaton as a function of e-foldings $N$
\bea
r=16\epsilon \simeq \frac{8}{M_{p}^{2}} \left(\frac{d\phi}{dN}\right)^{2},
\eea
which implies that the total change in the field during inflation between the times when observable
CMB modes leaves the horizon at $N_{cmb}=N_{s}$ and the end of inflation at $N_{e}$ can be given by
the following integral
\be
\frac{\Delta \phi}{M_{p}}= \int_{N_{s}}^{N_{e}} \sqrt{\frac{r}{8}} ~dN.
\ee
Since during slow-roll inflation $r$ doesn't evolve much with change in $N$, therefore the above integral,
for $\Delta N = N_{s}-N_{e}\approx 60$, gives
\be
\frac{\Delta \phi}{M_{p}}\simeq \mathcal{O}(1)\left(\frac{r}{0.01}\right)^{1/4}\,,\label{deltaphir}
\ee
so the large value of tensor-to-scalar ratio, $r>0.01$, implies large field inflation $\Delta \phi> M_{p}$.
Or $\Delta \phi>M_{p} \Rightarrow (\phi_{s}-\phi_{e})>M_{p} \Rightarrow \phi_{s}> M_{p}$, since $\phi_{e}>0$, implies
inflaton field values are super Planckian during the time observable CMB modes leave the horizon.

\vspace{0.1in}

We will use the formalism and expressions for power spectrum, spectral index and its running, and tensor-to-scalar ratio derived here extensively in Section [\ref{ETGmodels}] where we disccuss various Extended Theories of Gravity models.

\section{INFLATION IN MODIFIED GRAVITY AND SUPERGRAVITY}
In the remaining part of this review, we discuss different extended theories of gravity and supergravity theory. We then study the single and double field models of inflation in the context of modified gravity theories. In order to motivate these models, as they are not generic in the particle physics models, we derive them from supergravity.

\subsection{Extended Theories of Gravity}
\setcounter{equation}{0}

Due to the problems of Standard Cosmological Model, and, the absence of a definitive quantum theory of gravity,
{\it Extended Theories of Gravity} (ETGs), which are based on corrections and
generalizations of Einstein's theory, seems to be a very fruitful approach. They rely in constructing the (effective) action of gravitational interactions
by taking into account higher-order curvature invariants, as well as scalar fields that are
minimally or non-minimally coupled to gravity \cite{CF1}.

The modifications of GR are mainly motivated for including in a theory the Mach Principle.
According to it, the average motion of distant astronomical objects affects the determination of the local inertial frame \cite{bondi}. As a consequence, the gravitational coupling can be scale-dependent (varying gravitational coupling), presumably related to
some scalar field, which leads to a revision of  the concept of "inertia''
and the Equivalence Principle. Besides, it is a consolidate fact that the effective actions of every unification scheme (such as Superstrings, SUGRA, Kaluza-Klein theories and GUT) include non-minimal couplings to the geometry or higher-order terms in the
curvature invariants. The origin of these additional terms are due to
one-loop or higher-loop corrections in the high-curvature regimes (i.e. the interactions among quantum fields and background geometry or the gravitational self-interactions yield corrective terms in the Hilbert-Einstein Lagrangian)
\cite{CF1,birrell}. Specifically, in constructing an effective gravitational action by taking into account quantum corrections, one finds that higher-order terms in curvature invariants, such as
$R^{2}$, $R_{\mu\nu} R^{\mu\nu}$, $R^{\mu\nu\alpha\beta}R_{\mu\nu\alpha\beta}$, $R \,\Box R$, or $R
\,\Box^{k}R$, or non-minimally coupled terms between scalar fields
and geometry, such as $\phi^{2}R$, have to be necessarily added in the action.
A relevant aspect of these models is that, by mean of conformal
transformations, the higher-order terms and non-minimally coupled terms
always correspond to Einstein's gravity plus one or more than
one minimally coupled scalar fields \cite{TeyssandierTourrenc83}.
It must be mentioned that the debate on the physical meaning of conformal
transformations is far to be solved \cite{MagnanoSokolowski94}.
Besides the fundamental physics motivations, all these extended theories of gravity have
acquired a huge interest in cosmology due to the fact that they
in a natural way exhibit inflationary behaviours able to overcome the
shortcomings of  Cosmological Standard Model, and
match with the Cosmic Microwave Background (CMB) Radiation observations \cite{starobinsky}.

One of the simplest modifications to GR is the so called $f(R)-$gravity in which the Lagrangian density $f(R)$ can be a generic function of Ricci scalar $R$ \cite{Bergmann:1968ve}. A very well
known model with $f(R)=R+\frac{1}{M^{2}}R^{2}$, ($M>0$), is the Starobinsky model of inflation which can lead to an accelerated expansion of the universe due to the presence of the term $\frac{1}{M^{2}}R^2$ \cite{starobinsky2}. This model is well consistent with observations of the CMB anisotropies and therefore can be a viable alternative to the scalar field model of inflation. It is known that the $f(R)$ gravity theories in the metric formalism (in which the field equations are obtained by varying the action w.r.t. the metric $g_{\mu\nu}$) are equivalent to scalar-tensor theory, the Brans-Dicke theory, with the Brans-Dicke parameter $\omega_{BD}$ equals to zero \cite{DeFelice:2010aj}.
Another class of models with a coupling between field and curvature scalar are the {\it non-minimally coupled} inflation models, whose Lagrangian density is $f(\phi)R$. A simplest model of inflation with non-minimal coupling is the Higgs inflation model where the Higgs scalar $\phi$ can give rise to a viable inflationary phase as a result of coupling with the curvature scalar of the form $f(\phi)R=R+\xi\phi^{2}R$, where $\xi$ is the non-minimal coupling parameter \cite{Bezrukov1}. Interestingly, the Starobinsky model of inflation is shown to be equivalent to Higgs inflation model in the conformal Einstein frame \cite{DeFelice:2010aj,Bezrukov1}.  Both of these models lead to the same scalar potential in Einstein frame: it is possible to transform indeed  $f(R)$ and $f(\phi)R$ gravity actions into an Einstein gravity action via a conformal transformation of the metric $g_{\mu\nu}$ and redefinition of the field $\phi$.

We recall the main properties of conformal transformations:
\bea\label{conf_trafo}
\tilde{g}_{\mu\nu}(x) &=& \Omega^2(x) g_{\mu\nu}(x)\,, \\
\tilde g^{\mu\nu} &=& \Omega^{-2} g^{\mu\nu}, \nonumber \\ 
ds^{2}&=&\Omega^{-2} d\tilde s^{2}\,, \nonumber \\
 \sqrt{-g}&=&\Omega^{-4}\sqrt{-\tilde g}\,,\nonumber \\
R &=& \Omega^{2}\left[\tilde R + 6 \frac{\tilde\Box\Omega}{\Omega} - 12 \frac{
\tilde g^{\mu\nu}\partial_{\mu}\Omega\partial_{\nu}\Omega}{\Omega^{2}}\right]\,. \nonumber
\eea
Hereafter tilde represents quantities in the Einstein frame (EF).
To better understand the mechanism of conformal transformations, we take $f(R)$ and $f(\phi)R$ gravity action as examples and show that it can be recast into Einstein frame action. Exhaustive studies of $f(R)$, $f(\phi)R$ and more generally  $f(\phi,R)$ theories can be found in Refs. ~\cite{DeFelice:2010aj}


\subsubsection{Example I: $f(R)$ gravity}

Among the different approaches proposed to generalize Einstein's General Relativity, the $f(R)$-theories of gravity have received a growing attention (see for example \cite{salvbook,altri-f(R)}). The reason relies on the fact that they allow to explain, via gravitational dynamics, the observed accelerating phase of the Universe \cite{accUn}, without invoking exotic matter as sources of dark matter or extradimensions.
In these models, the Hilbert-Einstein (\ref{HEaction}) is generalized as
 \begin{equation}\label{Jordan-frame}
 S=\frac{1}{8\pi G}\int d^4 x \sqrt{-g} \, f(R)+S_{(m)}\,,
 \end{equation}
This model of $f(R)$ gravity is equivalent to scalar-tensor theories \cite{chiba}.
The variation of (\ref{Jordan-frame}) with respect to the metric yields the fourth
order field equations
\begin{eqnarray}\label{HOEQ}
  {\cal G}_{\mu\nu} &=& \kappa^2 T^{(m)}_{\mu\nu}\,, \\
  & & \nonumber \\
  {\cal G}_{\mu\nu} &\equiv & f' R_{\mu\nu}-\frac{f}{2}\, g_{\mu\nu}-\nabla_\mu \nabla_\nu f' +g_{\mu\nu}\Box f'\,,\nonumber
\end{eqnarray}
where the prime indicates the derivative with respect to the scalar curvature $R$.
Notice that the Bianchi's identities are fulfilled
 \[
 \nabla^\alpha {\cal G}_{\alpha\beta}=0=\nabla^\alpha T_{\alpha\beta}^{(m)}\,.
 \]
The trace equation is given by
\begin{equation}\label{TrHOEQ}
   3\Box f'(R)+f'(R)R-2f(R)\,=\kappa^2 \, T^{(m)}\,.
\end{equation}
Field equations can be cast in the form\footnote{In the right-hand side of Eq. (\ref{eq:field})  two effective  fluids  appear: a {\it curvature fluid} and a standard {\it matter fluid}. This representation allows to treat fourth order gravity as standard Einstein gravity in presence of two effective sources \cite{santuzzo}.
This means that such fluids can admit features that could be unphysical for standard matter. Consequently all the thermodynamical quantities associated with curvature should be considered {\it effective} and not bounded by the standard constraints related to matter fields. Moreover, this description does not compromise any of the thermodynamical features of standard matter since Bianchi's identities are separately fulfilled for both fluids.}
\begin{eqnarray}
G_{\mu \nu} & \equiv  & R_{\mu\nu}-\frac{1}{2}\, g_{\mu\nu} R = \kappa^2\left(T^{curv}_{\mu \nu} + T^{m}_{\mu \nu}\right)\,, \label{eq:field}\\
T^{curv}_{\mu\nu} & = &  \frac{1}{\kappa^2 f'(R)} \left\{ g_{\mu\nu} \left[ f(R) - R f'(R)\right] + f'(R)^{; \rho \sigma}
        \left( g_{\mu\rho} g_{\nu\sigma} - g_{\rho\sigma} g_{\mu\nu}\right)\right\} \label{eq:curvstress} \\
T^{m}_{\mu\nu} &=& \frac{T^{(m)}_{\mu\nu}}{f'(R)}\,. \label{eq:mattstress}
\end{eqnarray}
For a Universe described by the FRW metric, the cosmic acceleration is achieved when the right handed side of the acceleration equation
remains positive
 \[
 \frac{\ddot a}{a}\propto -(\rho_{tot}+p_{ tot})
 \]
where
 \[
 \rho_{tot}=\rho^{m}+\rho^{curv}\,,
 \]
 \[
  p_{ tot}=p^{m}+p^{curv}\,.
  \]
In particular, if the Universe is filled by dust ($p^{m}=0$), one has
 \[
 \rho^{m}+\rho^{curv}+3p^{curv}<0 \quad  \to \quad w^{curv} < -\frac{\rho^{m}+\rho^{curv}}{3\rho^{curv}}\,,
  \]
where
 \begin{eqnarray}
 \rho^{curv} &=& \frac{1}{f'}\left[\frac{f-Rf'}{2}-3H {\dot R} f''\right]\,, \label{rhocurbS} \\
 p^{curv} &=& {\dot R}^2 f'''+2 H {\dot R} f'' + {\ddot R} f'' +\frac{1}{2}(f-Rf')\,, \label{pcurvS} \\
 w^{curv}&=&\frac{p^{curv}}{\rho^{curv}}  = -1+\frac{{\ddot R}f''+{\dot R}({\dot R}f'''-Hf'')}{\frac{1}{2}(f-Rf')-3H{\dot R}f''}\,.
 \label{wcurvatureSalv}
 \end{eqnarray}
Owing to the freedom to choose the form of the function $f(R)$, many models can be investigated. All these models have to fulfill the conditions $f' > 0$, in order that the effective gravitational coupling is positive, and $ f'' > 0$ to avoid the Dolgov-Kawasaki instability \cite{dolgov-kawasaki}.

\vspace{0.2in}

{\it Effective potentials in the Einstein frame} -  Let us now discuss as the further gravitational degrees of freedom coming from $f(R)$ gravity can be figure out as an additional scalar field.
Taking
 \begin{equation}\label{conftransf}
  e^\chi\equiv \Omega^2\,, \qquad   {\tilde g}_{\mu\nu} = e^{2\chi}g_{\mu\nu}\,, \quad
  \chi=\frac{1}{2}\ln |f'(R)|\,,
 \end{equation}
and setting
 \[
 k \varphi = \chi \quad (k=\frac{1}{\sqrt{6}})\,,
 \]
it can be shown that the Lagrangian density of $f(R)$ in (\ref{Jordan-frame}) can be recast in the (conformally) equivalent form \cite{salvbook}
\begin{equation}\label{euivform}
    \sqrt{-g}f(R)=\sqrt{-{\tilde g}}\left(-\frac{1}{2}{\tilde R}+\frac{1}{2}\nabla_\mu \varphi \nabla^\mu \varphi - V\right)\,.
\end{equation}
The field equations are
 \begin{equation}\label{einsteintransf}
    {\tilde G}_{\mu\nu}=\kappa^2 \left[\nabla_\mu \varphi \nabla_\nu \varphi-\frac{1}{2}{\tilde g}_\mu\nu \nabla_\rho \varphi \nabla^\rho  \varphi+{\tilde g}_{\mu\nu}
    V(\varphi)\right]\,,
 \end{equation}
where $ {\tilde G}_{\mu\nu}$ is the Einstein tensor written in terms of the metric ${\tilde g}_{\mu\nu}$.
The potential $V$ is given by
\begin{equation}\label{potV}
    V= \frac{f-R f'}{2f^{'\, 2}}\,.
\end{equation}
As an application, consider the model $f=R+\alpha R^n$. Inverting $\chi$ in (\ref{conftransf}) ($f' = e^{2k \varphi}$), one obtains
 \[
 R=\left[\frac{1}{\alpha n}(e^{2k\varphi}-1)\right]^{\frac{1}{n-1}}\,,
 \]
so that the potential (\ref{potV}) reads
\begin{eqnarray}\label{V1gae}
  V &=& \frac{\alpha(1-n)}{2}e^{-4k\varphi}\left[\frac{1}{\alpha n}(e^{2k\varphi}-1)\right]^\frac{n}{n-1} \\
  & & \nonumber \\
   &=& \frac{2^{\frac{1}{n-1}}\alpha (1-n)}{(\alpha n)^{\frac{n}{n-1}}} e^{k \frac{4-3n}{n-1} \varphi}\left[\sinh k\varphi\right]^{\frac{n}{n-1}}\,. \nonumber
\end{eqnarray}
For $k\varphi\ll 1$ the potential assumes a power law behavior
\begin{equation}\label{potV2}
    V\simeq V_0 \varphi ^\delta\,,
\end{equation}
where
 \[
   V_0 \equiv \frac{2^{\frac{1}{n-1}}\,\alpha (1-n)}{(\alpha n)^{\frac{n}{n-1}}}\,,
 \quad \delta \equiv \frac{n}{n-1}\,.
 \]
Such a form of potential has been widely studied in literature in the framework of alternative theories
of gravity \cite{salvbook}.
For $n=2$ the potential (\ref{V1gae}) tends to a constant value for large $k\varphi$
 \begin{equation}\label{Vconstant}
    V \to \frac{V_0}{4} \qquad\quad \text{as}\,\,\,\, k\varphi \gg 1\,.
 \end{equation}
Therefore, in this regime the potential plays the role of a cosmological constant.

\subsubsection{Example II: $f(\phi)R$ gravity}

The action of a single-field non-minimal coupled is given by (in the Jordan frame (JF))
\be
S_{J} = \int d^4 x \sqrt{- g} \left[\frac{M_{p}^{2}}{2} f(\phi) R- \frac{1}{2} g^{\mu \nu}\partial_\mu \phi
\partial_\nu \phi- V(\phi)\right].\label{actionJ}
\ee
Using the conformal transformations with $\Omega^{2}=f(\phi)$, the action (\ref{actionJ}) can be
cast in the form
\be
S_{E} = \int d^4 x \sqrt{- \tilde g} \left[\frac{M_{p}^{2}}{2}  \tilde R- \frac{3 M_{p}^{2}}{4}
\frac{\tilde g^{\mu \nu} \partial_\mu f \partial_\nu f}{f^{2}} - \frac{1}{2}
\frac{\tilde g^{\mu \nu} \partial_\mu \phi \partial_\nu \phi}{f}- \frac{V(\phi)}{f^{2}}\right].\label{actionNCKT}
\ee
To get a canonical kinetic term we redefine field $\phi$ to $\tilde\phi$ through
\be
\frac{1}{2} \tilde g^{\mu \nu} \partial_\mu \tilde\phi \partial_\nu \tilde\phi = \frac{3 M_{p}^{2}}{4}
\frac{\tilde g^{\mu \nu} \partial_\mu f \partial_\nu f}{f^{2}} + \frac{1}{2}
\frac{\tilde g^{\mu \nu} \partial_\mu \phi \partial_\nu \phi}{f},\label{fieldrelation}
\ee
therefore, with the redefined kinetic term (\ref{fieldrelation}), we get the EF action (\ref{actionNCKT}) as
\be
S_{E} = \int d^4 x \sqrt{- \tilde g} \left[\frac{M_{p}^{2}}{2}  \tilde R- \frac{1}{2} \tilde g^{\mu \nu}
\partial_\mu \tilde\phi \partial_\nu \tilde\phi - \tilde V(\tilde\phi)\right],
\ee
where
 \[
 \tilde V(\tilde\phi) = \frac{V(\phi(\tilde\phi))}{f(\phi(\tilde\phi))^2}\,.
\]
Using the fact $\partial_{\mu}f(\phi)=\frac{\partial f}{\partial\phi}\partial_{\mu}\phi$, the equation (\ref{fieldrelation})
can be solved to give
\be
\frac{\partial\tilde\phi}{\partial\phi} = \sqrt{\frac{3 M_{p}^{2}}{2 f^{2}} \left(\frac{\partial f}{\partial\phi}\right)^{2} + \frac{1}{f}}.
\ee
For a given form of $f(\phi)$ the above equation can be integrated to give the EF field  $\tilde\phi$ in terms of Jordan frame (JF) field $\phi$.

\vspace{0.2in}

{\it Multi-scalar field} - In the context of {\it multi-scalar field inflation with non-minimal coupling} the action is written as
\be
S = \int d^4 x \sqrt{- g} \left[\frac{1}{2} M_{p}^{2} f(\phi^K) R- \frac{1}{2} G_{IJ} g^{\mu \nu}\partial_\mu \phi^I \partial_\nu \phi^J- V(\phi^K)\right] .
\ee
where $I,J,K=1,2,3,....,N$ for a model with $N$ scalar fields. Under conformal transformations, the above action
transforms to an Einstein frame action
\be
\tilde S = \int d^4 x \sqrt{- \tilde g} \left[\frac{1}{2} M_{p}^{2} \tilde R- \frac{1}{2}
\tilde G_{IJ}(\phi^K) \tilde g^{\mu \nu} \partial_\mu \phi^I \partial_\nu \phi^J - \tilde V\right],\label{multiaction}
\ee
where
 \[
{\tilde V} = \frac{V}{f^2}
\]
is the EF potential and
\be
\tilde G_{IJ} = \frac{1}{f} G_{IJ}+ \frac{3 M_{p}^{2}}{2} \frac{f_{,I} f_{,J}}{f^2}.
\ee
where $f_{,I}=\partial f/\partial\phi^{I}$. These multifield models with action (\ref{multiaction}), where there is no coupling between scalar fields and curvature scalar $R$ and kinetic terms in the fields are non-canonical,
arise naturally in Higher dimensional theories such as supergravity and string theories. We will study such
a model in next Sections with two fields where there is no cross term in the kinetic
energy of the fields and illustrate how such a model can be derived from supergravity with an appropriate choice of
K\"ahler potential and superpotential. Multifield dynamics in the context of Higgs inflationary scenario
have been studied~\cite{Greenwood:2012aj} and it is shown that for $N-$fields model which obey an $SO(N)$ gauge
symmetry in field space, the multifield effects damp out very quickly at the onset of inflation.

\subsection{Inflation from Supergravity Theory}\label{IfST}
Supergravity (SUGRA) is a local version of the $\mathcal{N}=1$ Supersymmetry (SUSY) in four dimension
\cite{Wess:1984jr,Nilles:1983ge,Lahanas:1986uc,freedman}. Supersymmetry is a symmetry which relates
fermionic and bosonic degrees of freedom. $\mathcal{N}$ represents the number of independent SUSY
transformations and therefore independent SUSY transformation parameters. Global SUSY
extension of standard model (SM) of particle physics can not only solve the hierarchy problem but also
account for the large amount of dark matter in our universe. SUSY in the context of cosmology is also a
welcome tool. If nature is found to be supersymmetric then gravitational sector should be supersymmetric too.
The local version of SUSY automatically engages the theory of gravity as spin$-3/2$ gauge field $\psi_{\mu}^{\alpha}$, termed as {\it gravitino},
of the SUGRA transformations has superpartner spin$-2$ tensor field $g_{\mu\nu}$, termed as {\it graviton} which can be identified with
the metric tensor. Therefore, local SUSY is a perfect landscape for establishing connections between high energy particle
physics and cosmology. The basic difference between a local and a global SUSY is that the symmetry transformation parameter in local SUSY is explicitly spacetime dependent.

The presence of many scalar fields in the supersymmetry allows to realize inflation within its framework. Since in
models of inflation, the inflationary energy scale is very high and close to the fundamental scale of
gravity, the Planck scale, where all the fundamental forces are expected to unify, the effects of an
unknown theory of quantum gravity can not be neglected. The $\mathcal{N}=1$ SUGRA in four dimensions may offer
an effective description of quantum gravity. Also it is worth noting that SUSY plays a crucial role in the
structure of string theory and the low-energy limit of the string theory compactifications include supergravity .

Realizing inflation in SUGRA is not so trivial because of the presence of an exponential factor in the
scalar potential of the supergravity. For canonical K\"ahler potential $\delta^{ij}\phi_{i}\phi_{j}^{\ast}$,
any scalar field or inflaton acquires mass of the order of Hubble parameter and it violates one of the
slow-roll condition. Therefore, it is not possible to have nearly flat potential for successful inflation in these models.
This problem in realizing inflation in SUGRA is known as {\it $\eta$ problem}~\cite{Copeland:1994vg,Yamaguchi:2011kg}.
This difficulty can not be resolved without invoking some symmetry or fine tuning of the scalar potential.
To resolve this problem people tune the K\"ahler potential and superpotential in SUGRA models to obtain a suitable
scalar potential which can provide slow-roll inflation. The K\"ahler potential must be fixed by the model builder
and they are not fixed by the symmetries of the theory. There are no legitimate reasons to justify the choices of the
K\"ahler potentials and superpotentials.

During the development of Starobinsky model of inflation in early 1980, the no-scale SUGRA was also discovered and
developed and applied to particle physics problems \cite{Ellis:1983sf,Ellis:1983ei,Ellis:1984bm}.
As required for successful inflation, in SUGRA models of inflation the effective potential should vary slowly enough
for a sufficient period over a large range of inflaton field values during inflation. This occurs naturally in no-scale supergravity models \cite{Ellis:1984bm,Lahanas:1986uc}. Also in these models the energy scale of the
effective potential can be naturally much smaller than $M_{p}\sim 10^{19} GeV$ as required by CMB observations.
These models are {\it called} no-scale because the scale at which the SUSY breaks is undetermined at the tree level
and could be anywhere between the experimental lower limit $1~TeV$ from the LHC \cite{Buchmueller:2012hv} and $10~TeV$ from the measurement of tensor-to-scalar ratio \cite{Ellis:2013xoa}. These no-scale SUGRA
models have an attractive feature that they arise naturally in generic four dimensional reductions of
string theory \cite{Witten:1985xb} and therefore they were proposed as a framework for constructing models
of inflation \cite{Ellis:1984bf}. There are several inflationary models in the context of no-scale
supergravity~\cite{Lahanas:1986uc,Olive:1989nu,Ellis:1984bm}.

\subsubsection{Supergravity Framework for Inflation}

In this section, we briefly summarize the SUGRA results which are relevant to motivate certain inflation
models from SUGRA theory. In order to derive the Lagrangian for modified gravity and multi field inflation models,
the most relevant part of the SUGRA Lagrangian is its scalar part which gives the kinetic and the potential terms
for the inflaton. The chiral multiplet for SUGRA algebra has the field content $(\phi^{i}, \chi^{i}, F^{i})$, where $\phi^{i}$
are the complex scalar fields, $\chi^{i}$ are the Weyl fermions and $F^{i}$ are complex scalar auxiliary fields.

The scalar part of the SUGRA Lagrangian is determined by three functions, K\"ahler potential $ K(\Phi_i, \Phi_i^{\ast})$,
superpotential $W(\Phi_i)$ and gauge kinetic function $ f (\Phi_i)$. The superpotential $W$ and gauge kinetic function $f$
are the holomorphic functions of complex scalar fields $\phi_i$ , while the K\"ahler potential is not holomorphic and a
real function of $\phi_i$ and their conjugates $\phi_i^{\ast}$ \footnote{A holomorphic function, say $h(z)$, is a complex valued function
of one or more complex fields that is complex differentiable at each point $z_0$ in its domain. They satisfy the
Cauchy-Riemann equations of complex analysis or equivalently $\frac{\partial h}{\partial z^{\ast}}=
\frac{\partial h^{\ast}}{\partial z}=0$ and $\frac{\partial h}{\partial z}|_{z\rightarrow z_{0}}=h'(z_0)$,
$\frac{\partial h^{\ast}}{\partial z^{\ast}}|_{z^{\ast}\rightarrow z^{\ast}_{0}}=h'(z^{\ast}_{0})$.}.

The interactions or the coupling of all the chiral superfields $\phi^{i}$ are determined by a real function
called {\it K\"ahler function}
\be
G(\phi_{i},\phi^{*}_{i}) \equiv K(\phi_{i},\phi^{*}_{i}) + \ln W(\phi_{i}) +\ln W^{\ast}(\phi^{*}_{i}),\label{KF}
\ee
The K\"ahler functions has a property that they are invariant under the so called K\"ahler transformations
\bea
W(\phi_{i}) &\rightarrow& e^{-U(\phi_{i})} W(\phi_{i}), \\
 & & \nonumber \\
K(\phi_{i},\phi^{*}_{i}) &\rightarrow& K(\phi_{i},\phi^{*}_{i}) + U(\phi_{i}) + U^{\ast}(\phi^{\ast}_{i}),\label{KTransf}
\eea
where $U(\phi_{i})$ are arbitrary holomorphic function of the scalar fields $\phi_{i}$. The invariance
of $G(\phi_{i},\phi^{*}_{i})$ is manifest if $U(\phi_{i})=\ln W$. The K\"ahler transformation sends
$W\rightarrow e^{-U} W \equiv 1$.

The canonical SUGRA Lagrangian for the complex scalar fields in curved spacetime is given by
\be
e^{-1}\mathcal{L} = -\frac{1}{2}R + \mathcal{L}_{kin} - V(\phi_i,\phi_i^{\ast}). \label{sugraL}
\ee
where $e=\sqrt{-g}$ is the determinant of the tetrad $e_{\mu}^{a}$ \footnote{The quantity $e_{\mu}^{a}$ are called
{\it tetrad} or {\it vierbeins}, where $a$ is the local Lorentz index and $\mu$ is the
gauge (curved) index. In order to deal with the spinors in curved spacetime it is necessary to formulate the theory
in terms of tetrads which are related to curved spacetime metric as
\be
g_{\mu\nu}(x) = e_{\mu}^{a}(x) e_{\nu}^{b}(x) \eta_{ab},\nonumber
\ee
where $\eta_{ab}$ is the Minkowski spacetime metric and the tetrads $e_{\mu}^{a}$ are defined as the transformation
from a local Lorentz inertial frame $\xi^{a}(x_{0};x)$ at the point $x_{0}$ to a general non-inertial frame $x^{\mu}$,
$i.e.$ $ \xi^{a}\rightarrow x^{\mu}$, as
\bea
e_{\mu}^{a}(x_{0}) \equiv \frac{\partial \xi^{a}(x_{0};x)}{\partial x^{\mu}} |_{x=x_{0}}.\nonumber
\eea}.
The first term in eq~(\ref{sugraL}) is the familiar vacuum Einstein-Hilbert action and, second and third terms are the kinetic and potential terms, respectively.
The Kinetic terms $\mathcal{L}_{kin}$ of the scalar fields are determined in terms of the K\"ahler potential and given by\footnote{In general in the kinetic term (\ref{sugraLK}) the partial derivative $\partial_{\mu}$ is actually a Lorentz covariant derivative
$D_{\mu} \equiv \partial_{\mu} + \frac{1}{2}\omega_{\mu}^{ab} \Sigma_{ab}$, where $\omega_{\mu}^{ab}$ are a set of gauge fields known
as {\it spin connections} and $\Sigma_{ab}$ are the generators of the Lorentz $SO(1,3)$ group which signifies the {\it spin} of
the associated gauge fields. Since for scalars $\Sigma_{ab}=0$, therefore covariant derivative $D_{\mu}$ in the scalar
Lagrangian equals a partial derivative $\partial_{\mu}$.}
\be
\mathcal{L}_{kin}= -K_{ij^{\ast}} g^{\mu\nu} \partial_{\mu}\phi^{i}\partial_{\nu}\phi^{\ast j}\,, \label{sugraLK}
\ee
where $K_{ij^{\ast}}$ is the K\"ahler metric given by
\be
K_{ij^{\ast}} = \frac{\partial^{2}K}{\partial \phi^{i} \partial \phi^{\ast j}}.\label{KMetric}
\ee
And the scalar potential $V(\phi_i,\phi_i^{\ast})$ can be split into two different contributions
\be
V(\phi_i,\phi_i^{\ast}) = V_{F}(\phi_i,\phi_i^{\ast}) + V_{D}(\phi_i,\phi_i^{\ast}),
\ee
referred to as the F-term and D-term potentials. The F-term potential is determined in terms of superpotential $W$ and K\"ahler potential $K$ as
\be
V_{F} = e^{K} \left[ D_{\phi_i}W K^{ij^{\ast}} D_{\phi_j^{*}}W^{*} - 3 |W|^{2} \right],\label{sugraVF}
\ee
where $K^{ij^{\ast}} \equiv K_{ij^{\ast}}^{-1}$ is the inverse of the K\"ahler metric $K_{ij^{\ast}}$ and
\be
D_{\phi_i}W = \frac{\partial W}{\partial \phi_i} + \frac{\partial K}{\partial \phi_i}W.
\ee
The D-term potential is related to gauge symmetry and given in terms of K\"ahler potential $K$ and
gauge kinetic function $f$,
\be
V_{D} = \frac{1}{2} \sum_{a} \left[Re f_{a}(\phi_i)\right]^{-1} g_{a}^{2} D_{a}^{2},\label{sugraVD}
\ee
where the subscript $a$ represents a gauge symmetry, $g_{a}$ is a gauge coupling constant and $T_{a}$ is an associated generator. $\xi_{a}$ is a Fayet-Iliopoulos term which is non-zero only when the gauge symmetry is Abelian, $i.e.$ a $U(1)$-symmetry. It can be shown that the potentials (\ref{sugraVF}),(\ref{sugraVD}) and the kinetic term (\ref{sugraLK}) are invariant
under the K\"ahler transformations (\ref{KTransf}). The F-term scalar potential (\ref{sugraVF}) in terms of a physically relevant
quantity, the K\"ahler function $G$~(\ref{KF}), can also be written as
\be
V_{F}=e^{G}\left[\frac{\partial G}{\partial \phi^{i}} K^{i}_{j*} \frac{\partial G}{\partial \phi^{*}_{j}} - 3 \right]\,. \label{sugraVF1}
\ee
We now briefly consider the {\it $\eta-$problem}, and the ways to solve it \cite{Copeland:1994vg,Stewart:1994ts,Yamaguchi:2011kg}.
Consider the canonical K\"ahler potential
\be
K_{ij^{\ast}} = \delta^{ij}\phi_{i}\phi_{j}^{\ast},\label{CKP}
\ee
for which the kinetic term (\ref{sugraLK}) of the scalar fields $\phi_{i}$ becomes canonical. The F-term potential
$V_{F}$ (\ref{sugraVF}) can be written as
\bea
V_F &=& e^{\delta^{ij}\phi_{i}\phi_{j}^{\ast}} \times \left \{ \left[ \frac{\partial W}{\partial \phi_i} + \phi_i^{\ast} W \right]
        \left[ \frac{\partial W^{\ast}}{\partial \phi_j^{\ast}} + \phi_j W^{\ast} \right] \delta_{ij} - 3 |W|^2 \right \} \nonumber \\
        & & \nonumber \\
      &=& V_{g} + V_{g} \sum_{i} |\phi_i|^2  + {other~terms},\label{VFexpand}
\eea
where $V_{g}$ is the global SUSY F-term potential given by
\be
 V_{g} = \sum_i \left| \frac{\partial W}{\partial \phi_i} \right|^2.
\ee
Since at the background level from Friedmann equation we have $V\simeq V_{g} \simeq 3H^{2}$, with little
algebraic simplification of equation~(\ref{VFexpand}), it can be shown that
\be
\eta =\frac{V_{F}''}{V_{F}} = 1 + \eta_{0} + {other~terms},
\ee
where $\eta_{0}= V_{g}'' / V_{g}$ and we find that one of the slow-roll approximation, $i.e.$ $\eta \ll 1$, is violated which is required
for successful inflation. $\eta \sim 1$ implies that any scalar field including the one which acts as the inflaton receives the
effective mass $V_{F}''\simeq 3\eta H^{2} \sim H^{2}$ of the order of Hubble parameter. This is the main problem that makes it difficult
to incorporate inflation in SUGRA.

Though there are several ways to get around this problem, in the most widely used method, one uses the K\"ahler potential
other than the canonical one (\ref{CKP}). If one choose a K\"ahler potential which is not canonical, the kinetic term (\ref{sugraLK}) of scalar
field also becomes non-canonical, which however can be made canonical by redefining the scalar field. The canonical normalization of
kinetic terms changes the scalar potential to an effective form which could be nearly flat even if it was originally very steep.
On the other hand it is also possible to impose some symmetry on K\"ahler and/or superpotential which can ensure
the slow-roll potential necessary for successful inflation \cite{Stewart:1994ts,Gaillard:1995az,Antusch:2009ty}. The $\eta$ problem is specific
to F-term potential. Successful inflation can also be achieved using the D-term potential, if it can produce positive
energy \cite{Stewart:1994ts}.

In the F-term inflation models, the difficulty is not limited to $\eta$ problem.
In the case of canonical K\"ahler potential the exponential factor is $e^{K}=e^{\delta^{ij}\Phi_{i}\Phi_{j}^{\ast}}$, therefore in the
large field limit $\phi_{i}>1$ (in $M_{p}=1$ unit) the potential becomes too steep to give a nearly flat potential suitable for inflation.
Thus, it is very difficult to incorporate chaotic inflation models in supergravity, which require inflaton field values
larger than unity during inflation. There have been proposed several models of chaotic inflation in SUGRA where the inflaton
field can have values larger than unity while producing the nearly flat inflaton potential. In these models either
K\"ahler potentials are fine-tuned without any symmetry reasons \cite{Murayama:1993xu,Goncharov:1983mw} or there are models in which the K\"ahler
potential $K(\phi,\phi^{\ast})$ follow Nambu-Goldstone type of shift symmetry $\phi \rightarrow \phi + i c$,
where $c$ is some real parameter~\cite{Kawasaki:2000yn,Kawasaki:2000ws}.

\subsubsection{No-scale SUGRA Models}
During the early Universe evolution, the supergravity theory should play a relevant role, 
since it allows to infer an effective inflationary potential that varies slowly over a
large range of inflaton field values during $60$ $e-$fold expansion of the Universe. This occurs in no-scale supergravity models. 
They are termed as no-scale because the scale at which supersymmetry
is broken is undetermined in the first approximation. Moreover,
the energy scale of the effective potential is much smaller than the Planck scale $\sim 10^{19}$ GeV, 
as required by the CMB data. No-scale models arise in generic four-dimensional
reductions of string theory \cite{Witten:1985xb}. The peculiar features
of the no-scale SUGRA models, in the context of inflation, is that they do not depend sensitively on the supersymmetry-breaking scale, $i.e.$ it could be anywhere between the experimental lower limit $\sim 1$ TeV, provided by LHC collider \cite{[11]}, and $\sim 10^{10}$ TeV,
provided by tensor-to-scalar ratio. The main idea of no-scale models is that they are constructed in such a way that the F-term potential $V_{F}$
vanishes for all values of the scalar fields. Therefore, from (\ref{sugraVF}), the condition for a model to be
no-scale can be given in terms of the K\"ahler function as~\cite{freedman}
\be
\frac{\partial G}{\partial \phi^{i}} K^{i}_{j*} \frac{\partial G}{\partial \phi^{*}_{j}} = 3.\label{NScondi}
\ee
For constant superpotential models, $\partial_{\phi_{i}}W = 0$, the no-scale condition (\ref{NScondi}) can be given
in terms of K\"ahler potential as
\be
\frac{\partial K}{\partial \phi^{i}} K^{i}_{j*} \frac{\partial K}{\partial \phi^{*}_{j}} = 3.\label{NScondiK}
\ee
Examples of no-scale K\"ahler potentials with single complex scalar field $\phi$ and double complex scalar
fields $\phi_{i}=(\phi_1,\phi_2)=(T,\phi)$ which satisfy the no-scale condition (\ref{NScondiK}) are
\be
K= -3 \ln (T+T^{\ast}),
\ee
and
\be
K= -3 \ln \left[T+T^{\ast}- \frac{\phi \phi^{\ast}}{3}\right],\label{NSKP2}
\ee
respectively. It is shown that in the case of no-scale two-field K\"ahler potential (\ref{NSKP2}), if we give vev to
$T$ field such that $2\langle Re T \rangle = C$ and $\langle Im T \rangle=0$, with the following Wess-Zumino
choice of superpotential
\be
W= \frac{\hat{\mu}}{2} \Phi^{2} - \frac{\lambda}{3}\Phi^{3},\label{WZSP}
\ee
where the scalar component of the chiral superfield $\Phi$ is $\phi$, the Starobinsky inflationary potential
for field $\chi$ (where field $\chi$ arises from redefinition of $\phi$ due to canonical normalization of the kinetic
term for $\phi$) can be obtained
\be
V_{F} = \frac{\mu^{2}}{4}\left(1-e^{-\sqrt{\frac{2}{3}}\chi}\right)^{2},\label{staropot22}
\ee
with the choice $\lambda=\mu/3$, where $\mu=\hat{\mu}/\sqrt{C/3}$, the vev of the $T$ field is absorbed in the mass scale
$\mu$ \cite{Ellis:2013xoa}. Since here in deriving Starobinsky model from two-field K\"ahler potential, the superpotential (\ref{WZSP})
is not constant, such a model defines an {\it almost no-scale} model. Such features will be discussed in details 
in next Section.

Let us briefly discuss the importance of no-scale supergravity
The smallness of some physical quantities
$e.g.$ {\it cosmological constant $\Lambda$} can either be explained by some
symmetry argument or by fine tuning. One naturally prefers symmetry reasons for the smallness of the physical quantities.
From the form of the F-term scalar potential (\ref{sugraVF}), it is clear that for the potential to be vanishing
either one can choose some parameter inside $G$, through K\"ahler potential and/or superpotential, to be fine-tuned
or alternatively impose the no-scale condition (\ref{NScondi}). If one imposes the no-scale condition on the model
then the fine tuning is no-longer required for obtaining the vanishing vacuum energy. Also the potential becomes flat as
required for successful inflation, thereby solving the $\eta$ problem naturally~\cite{Lahanas:1986uc}. To explain
vanishingly small vacuum energy and nearly flat inflaton potential we require $V_{F}>0$, which from equation~(\ref{sugraVF1})
implies the condition
\be
e^{G} \frac{\partial G}{\partial \phi^{i}} \neq 0,
\ee
thus breaking the supergravity and generating the mass of gravitino $m_{3/2}=e^{G/2}\neq0$
which represents the scale of supergravity breaking.
Although the mass of gravitino $m_{3/2}$ is non-vanishing but undetermined at the tree level
despite the fact that supergravity is broken and the classical potential $V_{F}$ is vanishing for all values of the
scalar fields and therefore it is said to have flat directions \cite{Gaillard:1995az}. In no-scale SUGRA all
the mass scales below the Planck scale are determined with quantum corrections
\cite{Ellis:1983sf,Ellis:1983ei,Ellis:1984bm,Lahanas:1986uc,Breit:1985ns,Binetruy:1985ap,Binetruy:1987cn}.

The importance of no-scale models lie in the fact, that fine tuning is not needed to achieve positive vanishing
cosmological constant and the gravitino mass $m_{3/2}$ can be determined dynamically. Also there are flat
directions with gentle slope without fine tuning which are essential for successful inflation.

\section{Analysis of different ETG models}\label{ETGmodels}
\setcounter{equation}{0}

In this Section we shall analyze in details some specific models of extended theories of gravity.
We compute the key inflationary observables e.g. tensor-to-scalar ratio, amplitude of curvature purturbation, spectral index and its running; and for all models we show that they can be obtained from SUGRA.

\subsection{The $f(R)=R+R^2/M^2$ model}

We discuss in this Section an inflationary model that is a no-scale
supergravity realization of the $R + R^2$ Starobinsky model. This is
a seminal result obtained in Ref. \cite{Ellis:2013xoa}.

\begin{figure}
  \centering
  \includegraphics[width=10.0cm]{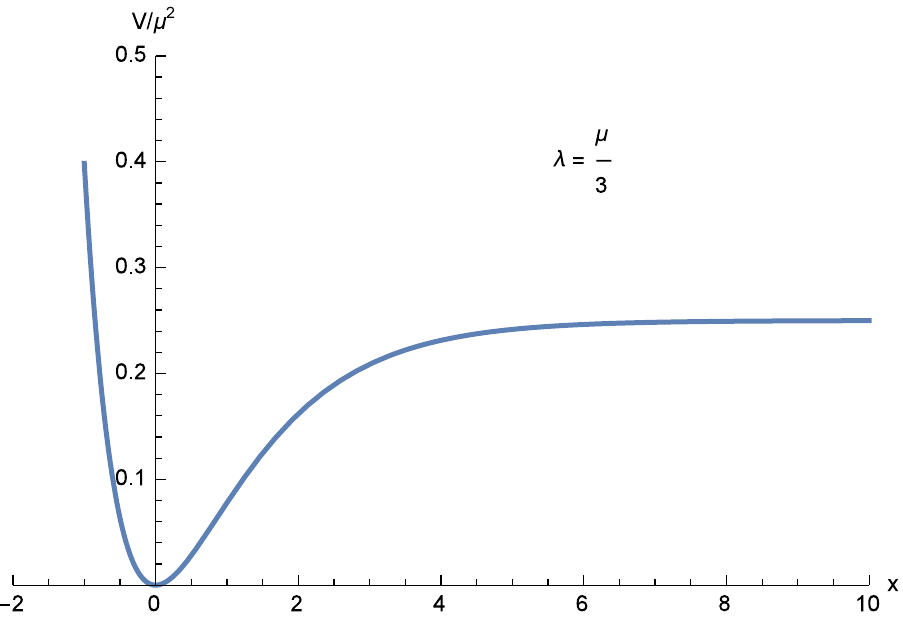}\\
  \caption{$V/\mu^2$ vs $x$.}\label{FigVEllis}
\end{figure}

In the minimal no-scale $SU(2,1)/SU(2) \times U(1)$ model, there are two complex scalar fields: $T$, the modulus field, and $\phi$, the inflaton field. The Kahler function is given by $K = -3 \ln(T +T^* -|\phi|^2/3)$.
The kinetic term for the scalar fields, $T$ and $\phi$, is (see the previous Section for details)
 \be\label{eq(4)}
 {\cal L}_{KE}= X^\dagger_\mu M X^\mu\,,
 \ee
where
 \begin{eqnarray} \label{XMX}
 X^\mu &=& \left( \begin{array}{c} \partial^\mu \phi \\ \partial^\mu T \end{array} \right)\,, \quad
 X_\mu^\dagger = \left( \partial_\mu \phi^* \quad \partial_\mu T^* \right)\,, \\
  & & \nonumber \\
  M& \equiv & {\cal Y} \left( \begin{array}{cc} T+T^* & \,\, -\phi \\
                    -\phi^* & \,\, 3 \end{array} \right)\,, \\
  & & \nonumber \\
  {\cal Y} &\equiv & \frac{1}{(T+T^*-|\phi|^2/3)^2}\,. \nonumber
  \end{eqnarray}
The effective potential assumes the form
 \begin{equation}\label{eq(5)}
  V={\cal Y}\, {\hat V}\,, \qquad {\hat V}=\left|\frac{\partial W}{\partial \phi}\right|^2\,.
 \end{equation}
Assuming that the $T$ field has a vacuum expectation value (vev) given by \cite{[28],[29]}
 \[
2\langle Re T\rangle  = c\,, \quad \langle Im T \rangle = 0\,,
\]
Neglecting the kinetic mixing term between the $T$ and $\phi$ fields in (\ref{eq(4)}), one gets the effective Lagrangian for $\phi$
\begin{equation}\label{eq(6)}
  {\cal L}_{eff} = \frac{1}{(c-|\phi|^2/3)^2}\left[c|\partial_\mu \phi|^2-{\hat V}\right]\,.
\end{equation}
Following \cite{[13]}, one assumes the minimal Wess-Zumino superpotential (\ref{WZSP}) for the inflaton field.
The potential for the inflaton is better studied by redefining the field $\phi$ in terms of the new field $\chi$, i.e.
 \begin{equation}\label{phivschi}
   \phi=c\sqrt{3} \tanh \frac{\chi}{\sqrt{3}}\,,
 \end{equation}
so that the effective Lagrangian density (\ref{eq(6)}) and the potential (\ref{eq(5)}) assume the form
 \begin{eqnarray}\label{Leffchi}
   {\cal L}_{eff} &=& \mbox{sech}^2 \frac{\chi-\chi^*}{\sqrt{3}}\left[|\partial_\mu \chi|^2-\frac{3}{c}\left|
   \sinh\frac{\chi}{\sqrt{3}}\left({\hat \mu}\cosh\frac{\chi}{\sqrt{3}}-\lambda c \sqrt{3}\sinh \frac{\chi}{{\sqrt{}3}}\right)\right|^2\right]\,, \\
   V &=& \mu^2\left|\sinh\frac{\chi}{\sqrt{3}}\left(\cosh\frac{\chi}{\sqrt{3}}-\frac{3\lambda}{\mu}\sinh \frac{\chi}{{\sqrt{}3}}\right)\right|^2\,, \label{VEllis}
 \end{eqnarray}
where we have defined $\mu\equiv \sqrt{3/c}\, {\hat \mu}$. Next one writes $\chi$ in terms of the real and imaginary parts
 \[
\chi=\frac{1}{\sqrt{2}}(x+i y)\,.
 \]
The Lagrangian density (\ref{Leffchi}) reads
\begin{equation}\label{Leffxy}
  {\cal L}_{eff}=\frac{1}{2}\mbox{sec}^2\sqrt{\frac{2}{3}}y\left[(\partial_\mu x)^2+(\partial_\mu y)^2-
  \frac{\mu^2}{2}e^{-\sqrt{2/3}x}\left(\cosh\sqrt{\frac{2}{3}}x-\cos\sqrt{\frac{2}{3}}y\right)\right]\,,
\end{equation}
where we used the specific value $\lambda=\mu/3$. Setting to zero the imaginary part, $y=0$, one gets
 \begin{equation}\label{VxEllis}
   V=\mu^2 e^{-\sqrt{2/3}x} \sinh^2\frac{x}{\sqrt{6}}\,.
 \end{equation}
In Fig. \ref{FigVEllis} is plotted the effective potential $V/\mu^2$ as a function of the real field $x$.

One can now compute the spectral index $n_s$ and for tensor-to-scalar ratio $r$ in terms of the slow-roll
inflation parameters $\epsilon$ and $\eta$. They are given by \cite{Ellis:2013xoa}
 \begin{eqnarray}\label{espilonetaEllis}
   \epsilon &=& \frac{1}{3}\mbox{csch}^2\frac{x}{\sqrt{6}}\, e^{-\sqrt{2/3}x}\,, \\
   & & \nonumber \\
   \eta &=&  \frac{1}{3}\mbox{csch}^2\frac{x}{\sqrt{6}}\left(2 e^{-\sqrt{2/3}x} - 1\right)\,,
 \end{eqnarray}
and the parameters $\{n_s, r\}$ assume the values (in Planck units)
 \[
 n_s = 0.965\,, \quad r=0.0035\,, \quad x=5.35\,, \qquad \mbox{for}\quad N=55\,.
 \]
Let us now see how this inflationary model can be related to the Starobinsky model. We have already
discussed the Starobinsky model as an example of extended theories of gravity (see Section 4). Here is interesting
to show how the free parameter $M$ of the original Starobinsky model (i.e. $R+R^2/M^2$) can be related to the fundamental parameters coming from SUGRA. To this end, consider the conformal transformation
 \[
 g_{\mu\nu} \quad \to \quad {\tilde g}_{\mu\nu}=\left(1+\frac{\varphi}{3M^2}\right)g_{\mu\nu}
 \]
and the redefinition of the field $\varphi$
 \[
 \varphi \quad \to \quad \varphi^\prime=\sqrt{\frac{3}{2}}\, \ln\left(1+\frac{\varphi}{3M^2}\right)\,.
 \]
The Starobinsky action
 \begin{equation}\label{Starobinsyaction}
   S=\frac{1}{2}\int d^4 x \sqrt{-g}\left(R+\frac{1}{6M^2}\, R^2\right)
 \end{equation}
assumes the form
 \begin{equation}\label{Starobiskyaction2}
   S=\frac{1}{2}\int d^4 x \sqrt{-{\tilde g}}\left[{\tilde R}+(\partial_\mu \varphi^\prime)^2-V(\varphi^\prime)\right]\,,
 \end{equation}
where the potential $V(\varphi^\prime)$ is given by
 \begin{equation}\label{VStarobSclar}
   V(\varphi^\prime)=\frac{3}{4}M^2\left(1-e^{-\sqrt{2/3}\varphi'}\right)^2\,.
 \end{equation}
The form of this potential is identical to the potential (\ref{VxEllis}) obtained along the {\it real direction}
of the no-scale Wess-Zumino model. Moreover, one can identify $M^2$ with $\mu$ according to the relation
 \[
 M^2=\frac{\mu^2}{3}={\hat \mu}^2\,,
 \]
where the last equality follows for $c=\langle (T+T^*)\rangle=1$. Thus, the Starobinsky parameter $M$ turns out to be directly related to the no-scale Wess-Zumino parameter ${\hat \mu}$ in the superpotential (\ref{WZSP}).

The Starobinsky model predicts that the tensor-to-scalar ratio is $r \simeq 10^{-3}$, which is in agreement with present experiment (BICEP2 reported a large value of the tensor-to-scalar ratio, $r= 0.2^{+0.07}_{-0.05}$ \cite{BICEP2}, whereas the recent joint analysis by Planck + BICEP2 + Keck Array give the upper bound $r_{0.05}<0.12 (95\% CL)$ \cite{BKP:2015,Ade:2015lrj,Planck:2015}).

\subsection{The model $\xi \phi^a R^b$}
We start with a general Higgs inflationary scenario $\xi \phi^{a} R^{b}$ in the framework of
$f(\phi,R)$ gravity theory, which is a generalization of the Higgs inflation model $\xi \phi^2 R$~\cite{Chakravarty:2013eqa}.
In the light of discoveries by CMS \cite{cms} and ATLAS \cite{Atlas} it is of interest to consider the
Standard Model Higgs boson as the candidate for inflaton. However in the standard single-field slow-roll inflation
with inflaton quartic potential, the idea of considering standard model Higgs as inflaton does not work as
the inflaton quartic coupling should be of the order $\lambda \sim 10^{-12}$ to explain the amplitude of
CMB perturbations measured by WMAP/Planck \cite{WMAP9,Planck:2015} while the 125 GeV Higgs has a quartic coupling $\lambda \sim 0.13$ at
the electroweak scale which can however go down to smaller values at the Planck scale due to renormalization
\cite{Sher}. However just from the standard model renormalization
one cannot have the Higgs coupling  $\lambda \sim 10^{-12}$ over the entire range of the rolling field $\sim(10-1 )M_P$ during
inflation and the standard slow-roll inflation with a Higgs field does not give the observed amplitude and spectrum
of density perturbations \cite{Isidori:2007vm}.
A way out of fine tuning the scalar self coupling to unnaturally small values was found out and it was shown that if one couples the
scalar field to the Ricci scalar $\xi \phi^2 R$ then the effective potential in the Einstein frame becomes a slow
roll one with the effective scalar coupling being $\lambda/\xi^{2}$ and the amplitude of the density perturbations
constrain this ratio \cite{Spokoiny:1984bd}. Therefore for large $\xi$ small self-coupling $\lambda$ can be obtained as required. The equivalence of the density perturbation in Jordan and Einstein frame
and density perturbations from inflation in the curvature coupled theories $\xi \phi^2 R$ are studied
in \cite{Kaiser:1994vs,Komatsu:1997hv,Komatsu:1999mt}.
Bezrukov and Shaposhnikov \cite{Bezrukov} revived the large curvature coupling model to motivate the idea
that the standard model Higgs field could serve as the inflaton in the early universe \cite{Bezrukov,Barvinsky:2008ia}.
However the large value of non minimal coupling which fix the amplitude poses problem of unitarity violation
at Planck scale \cite{Hertzberg:2010dc}. Various ways to solve the unitarity violation have been explored in
\cite{Barvinsky:2008ia,giudice}.

Here we assume that the dominant interaction between Higgs field and gravity is through operators of the form
\be
{\cal L}=\frac{\xi ({\cal H}^\dagger {\cal H})^{a/2} R^b}{M_p^{a+2b-4}}. \label{nonminimal}
\ee
This form (\ref{nonminimal}) of Higgs Curvature interaction has been mentioned in \cite{Atkins:2010yg}.
The complete dynamics of the Higgs field involves the role of the Goldstone modes as has been studied in detail in
\cite{Mooij:2011fi,Kaiser:2010yu,Greenwood:2012aj}. The multifield dynamics of the Goldstone modes gives rise to sizable non-gaussianity.
We will study the dynamics of the Higgs mode and impose a charge conservation and CP symmetry such that the Goldstone modes of the Higgs field do not acquire vevs. We will take the background Higgs field to be
\be
{\cal H} =\begin{pmatrix}
  0  \\
  \phi
 \end{pmatrix}
\ee
where $\phi$ is the Higgs mode with  mass 126 GeV. This inflation model falls in the class of inflation in $f(\phi,R)$ theories studied in Ref. \cite{DeFelice:2010aj}. The motivation is that we use the  Higgs quartic coupling $\lambda ({\cal H}^\dagger {\cal H})^2$ where the standard model value of $\lambda(\mu\sim M_P)$ can  lie in the range $\lambda = (10^{-5}-0.1)$ depending on the value of top quark mass \cite{Sher} or  on new physics \cite{Chakrabortty:2012np}. We take curvature coupling
$\xi$ to be unity and check the possibility of generating the observed density perturbations from Higgs inflation by varying parameters $a$, $b$ and $\lambda$. The non minimal coupling $\xi$ has been taken unity in order to improve the unitarity behavior which increases the natural cutoff scale $\Lambda$ from $\Lambda \simeq \frac{M_{p}}{\xi} \simeq 10^{15}$ to $\Lambda \simeq M_{p} \simeq 10^{19}$.
%

In non-minimal $\xi \phi^2 R$ theory we can always make a conformal transformation to the Einstein frame so one can compute the density perturbations either in Einstein frame or Jordan frame and the gauge invariant curvature perturbations should be same in both the frames \cite{Kaiser:1994vs}. With the $\xi \phi^a R^b$ model we find that no conformal transformation exists which can in general remove this term to go to an Einstein frame. We find that in $\xi \phi^a R^b$ theory such a conformal transformation is only possible if the metric is quasi-de Sitter. The accurate comparison with the experimental data should be made however with the Jordan frame results. The curvature perturbation in both Einstein and Jordan frame for the $\xi \phi^2 R$ theory has been derived in \cite{Kaiser:1994vs,Gong:2011qe,prokopec}.

\subsubsection{Model in the Jordan Frame}\label{s_jordan}

Consider the action for a
scalar field interacting with gravity of the form
\bea
S_J=\int d^4x\sqrt{-g}\left[-\frac{f(\phi,R)}{2\kappa^2} + \frac{1}{2}g^{\mu\nu}\partial_{\mu}\phi\partial_{\nu}\phi
+V(\phi)\right] \label{action_J} \ ,
\eea
where
\bea\label{eqn_f}
\frac{1}{\kappa^2}f(\phi,R) &=& \frac{1}{\kappa^2}R + \frac{\xi \phi^a R^b}{M_p^{a+2b-4}} \\
 & & \nonumber \\
V(\phi) &=& \frac{\lambda \phi^4}{4} \,,
\eea
and $\xi$ is a dimensionless coupling constant. Setting
 \[
 F\equiv\frac{\partial{f}}{\partial{R}}=1+\frac{\xi b \phi^a R^{b-1}}{M_p^{a+2b-2}}
 \]
the field equations read
\bea\label{eqn_J1}
 F R_{\mu\nu}& - & \frac{1}{2}fg_{\mu\nu}- \triangledown_{\mu}\triangledown_{\nu}F+g_{\mu\nu}\square F = \kappa^2 \left(\triangledown_{\mu}\phi \triangledown_{\nu}\phi - \frac{1}{2}
g_{\mu\nu} \triangledown^{\rho}\phi \triangledown_{\rho}\phi - V g_{\mu\nu}\right) , \\
\square{\phi}&=& V_{,\phi} - \frac{f_{,\phi}}{2\kappa^2},\label{eqn_J2}
\eea
For the unperturbed background FRW metric (\ref{FRWmetric}) the above Eqs. (\ref{eqn_J1}) and (\ref{eqn_J2}) give
\bea
3 F H^2 + \frac{1}{2}\left(f-RF\right)+ 3H\dot{F}&=&\kappa^2\left(\frac{1}{2}\dot{\phi}^2 + V(\phi)\right)\,, \label{leading_J1}\\
 & & \nonumber \\
- 2 F\dot{H} -\ddot{F} + H \dot{F}&=& \kappa^2 \dot{\phi}^2\,, \label{leading_J2}\\
 & & \nonumber \\
\ddot{\phi} + 3H\dot{\phi} + V_{,\phi} - \frac{f_{,\phi}}{2\kappa^2} &=&0 \label{leading_J3}\,.
\eea
We assume that the second term of $F$ $i.e.$ $\frac{\xi b \phi^a R^{b-1}}{M_p^{a+2b-2}}$ is dominant for some values of
$a$ and $b$. We find this assumption to be valid while solving numerically for the values of $a$ and $b$ in our model which give rise
to the experimentally observed density perturbations.
From Eq. (\ref{leading_J1}), under this assumption and considering the slow-roll parameters as small (see below their definition, Eq. (\ref{slow_rolls})), the Hubble parameter in the Jordan frame reads
\bea
H = \frac{\lambda^{\frac{1}{2b}}}{\sqrt{12}\big[\xi(2-b)\big]^{\frac{1}{2b}}}
\left(\frac{\phi}{M_p}\right)^{\frac{4-a}{2b}} M_p \label{Hubble_J} \ .
\eea
From equation (\ref{leading_J3}) under the slow-roll assumption we get
\be\label{phidotgae}
\dot \phi=-\frac{\lambda \phi^3}{3 H}\Big[1- \frac{a}{2(2-b)}\Big].
\ee
Now we perturb Eqs. (\ref{eqn_J1}) and (\ref{eqn_J2}) by perturbing the scalar field $\phi=\phi(t)+\delta \phi(x,t)$
and the metric (\ref{metricpert})~\cite{DeFelice:2010aj}.
One can derive the Einstein field equations for
the $f(R, \phi)$ theory keeping the linear terms in the metric and scalar field perturbations \cite{hwang,hwang1}
and the equation of motion of scalar perturbation. The relevant quantity is the curvature perturbation
 \[
 \mathcal{R}=\Psi-H\delta \phi/\dot{\phi}\,,
 \]
where we work in the gauge $\delta \phi=0$
and $\delta R=0$. This  sets $\mathcal{R} =\Psi$ (notice that $\delta F=0$ since
$\delta F = \left(\partial F/\partial \phi\right) \delta \phi + \left(\partial F/\partial R\right) \delta R$). Owing to this gauge the equation for $\Phi$ reads
\bea\label{eqR1}
\Phi = \frac{\dot{\mathcal{R}}}{H + \dot{F}/(2F)}
\eea
while the equation for $\Phi$, $\Psi$, $E$ and $B$, is given in terms of the combination
\bea
A\equiv 3(H\Phi-{\dot \Psi})-\frac{\Delta {\tilde B}}{a^2} = - \frac{1}{H + \dot{F}/(2F)} \left(\frac{\triangle}{ a^2(t)}\mathcal{R}
+ \frac{\left(3 H \dot{F}-\kappa^2\dot{\phi}^2\right)\mathcal{\dot{R}}}{2F\left(H + \dot{F}/(2F)\right)} \right) \label{eqR2}\,.
\eea
where
 \[
 {\tilde B}\equiv a(t) (E+a(t) {\dot B})\,.
 \]
The differential equation for curvature perturbation is
\be \label{eqR}
\ddot{\mathcal{R}} + \frac{(a^3(t) Q_s)\dot{}}{ a^3(t) Q_s} \dot{\mathcal{R}} + \frac{k^2}{a^2(t)}\mathcal{R}=0,
\ee
where
 \[
Q_s \equiv \frac{\dot{\phi}^2 + 3 \dot{F}^2/(2\kappa^2 F)}{\left(H+\dot{F}/(2F)\right)^2}\,.
  \]
Introducing the variables
 \[
 \omega = a(t) \sqrt{Q_s}, \quad \sigma_k = \omega \mathcal{R}\,,
 \]
one may re-write the equation (\ref{eqR}) as
\bea
\sigma_k'' + \left(k^2 - \frac{\omega''}{\omega}\right)\sigma_k = 0 , \label{eqR_final}
\eea
where prime denotes the derivative with respect to the conformal time defined as
$d\eta = dt/a(t)$ and
\bea
\frac{\omega''}{\omega} &=& \frac{a''(t)}{a(t)} + \frac{a'(t)}{a(t)} \frac{Q_s'}{Q_s} + \frac{1}{2}\frac{Q_s''}{Q_s}
-\frac{1}{4} \left(\frac{Q_s'}{Q_s}\right)^2 \\
 & & \nonumber \\
 &=& \frac{1}{\eta^2} \Big[\nu_{\mathcal{R}}^2 - \frac{1}{4}\Big]\,, \nonumber
 \eea
where the last term follows from under quasi de-Sitter expansion $a(\eta)=\frac{-1}{H \eta (1-\epsilon_1)}$, so that
 \[
 \frac{a''(t)}{a(t)} =\frac{1}{\eta^2}\big[2+3\epsilon_1 \big]\,, \quad \frac{a'(t)}{a(t)}=a(t) H\,,
 \]
and
\bea
\nu_{\mathcal{R}}^2 = \frac{9}{4}\Big[1+\frac{4}{3}\left(2\epsilon_{1} + \epsilon_{2} - \epsilon_{3} + \epsilon_{4}\right)\Big].
\eea
\bea
\epsilon_1=-\frac{\dot{H}}{H^2} \ , \ \epsilon_2 =\frac{\ddot{\phi}}{H \dot\phi} \ , \ \epsilon_3=
\frac{\dot F}{2 H F} \ , \ \epsilon_4=\frac{\dot E}{2 H E} \ \label{slow_rolls};
\eea
\bea
 E = F+ \frac{3\dot{F}^2}{2
\kappa^2\dot{\phi}^2} =\frac{Q_s (1+\epsilon_3)^2 }{\dot{\phi}^2/(F H^2)} \ .
\eea
Here $\epsilon_i$ are slow-roll parameters and $\dot{\epsilon_{i}}$ terms have been neglected.
Equation (\ref{eqR_final}) then has solutions in the Hankel functions of order $\nu_R$
\bea
\sigma=\frac{\sqrt{\pi |\eta|}}{2} e^{i (1+2 \nu_{\mathcal R})\pi/4} \left[ c_1\, H_{\nu_{\mathcal R}}^{(1)}(k |\eta|)+
c_2\, H_{\nu_{\mathcal R}}^{(2)}(k |\eta|) \right]
\eea
Applying the Bunch-Davies boundary condition
 \[
 \sigma(k \eta \rightarrow -\infty)=e^{i k \eta}/\sqrt{2 k}\,,
 \]
we fix the integration constants $c_1=1$ and $c_2=0$. Using the relation
 \[
 H_\nu(k |\eta|)= \frac{-i}{\pi} \Gamma(\nu) \left(\frac{k |\eta|}{2} \right)^{-\nu}\,,
 \]
for the super-horizon modes $k \eta \rightarrow 0$, we obtain the
expression for the power spectrum of curvature perturbations which is defined as (\ref{PScurvpert})
\bea
{\mathcal P}_{\mathcal{R}} &=& \frac{k^3}{2 \pi^{2}} \langle|\mathcal{R}|^{2}\rangle \\
 &\equiv&  {\Delta }_{\mathcal R}^2
\left(\frac{k}{a(t)H} \right)^{n_{\mathcal R}-1}, \nonumber
\eea
where the amplitude of the curvature power spectrum turns out to be
\bea
{\Delta}^{2}_{\mathcal{R}}=\frac{1}{Q_s}\left(\frac{H^{2}}{4\pi^{2}}\right)
\eea
and the spectral index is
\bea\label{spectral}
n_{\mathcal R}-1 &=& 3-2\nu_{\mathcal{R}} \\
 & & \nonumber \\
 &\simeq & -4\epsilon_1-2\epsilon_2+2\epsilon_3-2\epsilon_4 \nonumber \\
 & & \nonumber \\
 &\simeq &- 6 \epsilon_1 \nonumber \,.
\eea
In this $f(\phi, R)$ model one gets
 \[
 \epsilon_1\approx -\epsilon_3\,, \quad \epsilon_2\approx-\epsilon_4\,,
 \]
where
\bea
\epsilon_1 &=&  b^{-1} (a-4) (2-b)^{(1-b)/b} (a+2 b-4)\lambda^{(b-1)/b}
\xi ^{1/b} \left(\frac{\phi}{M_p}\right)^{\frac{a+2 b-4}{b}} \label{sl1} \\
\epsilon_2 &=& b^{-1}(a+6b-4) \left(2-b\right)^{(1-b)/b}(a+2b-4) \lambda^{(b-1)/b}\xi^{1/b}
\left(\frac{\phi}{M_p}\right)^{\frac{a+2 b-4}{b}} \label{sl2}  .
\eea
The expressions for the amplitude of power spectrum  and the number
of e-folding are
\bea
\Delta_{\mathcal{R}}^2 = \frac{b [(2-b)/\lambda]^{3-\frac{4}{b}}  M_{p}^{8+\frac{4 (a-4)}{b}}
 \xi ^{-\frac{4}{b}} \phi^{-\frac{4 (a+2 b-4)}{b}}}{288 (a-4)^2 (a+2 b-4)^2 \pi ^2}
\label{DeltaJ}
\eea
and
\bea
N_J= \int_{\phi_J}^{\phi_f} \frac{H}{\dot \phi} d \phi =  \frac{b [(2-b)/\lambda]^{\frac{b-1}{b}}
\xi^{-\frac{1}{b}}}{2(a+2b-4)^2}\left(\frac{\phi}{M_p}\right)^{\frac{4-a-2b}{b}}
\Bigg|^{\phi_J}_{\phi_f} \label{Nj}
\eea
respectively.  Here  $\phi_J$ and $\phi_f$ are the values of scalar field $\phi$
at the beginning and the end of inflation respectively.

Let us compute the tensor perturbation of the field equations (\ref{eqn_J1}).
Setting $D_{ij}= h_{ij}/a^2$ and using the polarization tensors $e_{ij}^{1}$ and $e_{ij}^2$, such that the tensor $D_{ij}$ can be written as\footnote{For gravity wave propagating in $\hat z$ direction, the components of polarization tensor are given by
\bea
e_{xx}^1=-e_{yy}^1=1, \ \ e_{xy}^2=e_{yx}^2=1, \ \  e_{iz}^{1,2}=e_{zi}^{1,2}=0.
\eea
}
 \[
 D_{ij}= D_1 e_{ij}^1 + D_2 e_{ij}^2\,,
\]
the equation for the tensor perturbation reads
\bea
\ddot{D}_{\lambda} +\Xi\, \dot{D}_{\lambda} + \frac{\kappa^2}{a^2}D_{\lambda}=0, \label{tensor_eqn1}
\eea
where $\lambda \equiv1,2$ corresponds to two polarizations of gravity wave, and
 \[
 \Xi=  \frac{1}{a^3 F}\frac{d}{dt}\, a^3 F\,.
 \]
Substituting
 \[
 z=a \sqrt{F}\,, \qquad v_k=\frac{1}{\sqrt{2}}\, z D_{\lambda} M_P\,,
 \]
we get
\bea
v_{\lambda}''+ \left(k^2-\frac{z''}{z}\right)v_{\lambda}=0, \label{tensor_eqn2}
\eea
where $^{\prime}$ stands for the derivative with respect to conformal time.
To obtain the amplitude of power spectrum  of $D_{\lambda}$, one has to
sum over all polarization states. Equation (\ref{tensor_eqn2}) yields
\bea
\Delta^{2}_{T} &=& 4 \times \left(\frac{2}{M_p^2}\right) \frac{k^3}{2\pi^2}\frac{1}{a^2 F} v_{\lambda}^2 \\
 & & \nonumber \\
 &\simeq & \frac{2}{\pi^2} \left(\frac{H}{M_P}\right)^2\frac{1}{F}\,. \nonumber
\eea
The ratio of the amplitude of tensor perturbations to scalar perturbations $r$ is given by
\be
r \equiv  \frac{\Delta^{2}_{T}}{\Delta^{2}_{\mathcal{R}}}= \frac{8 Q_s}{M_{p}^{2}F} \simeq  48 \epsilon_3^2\,.
\ee

\subsubsection{Model in the Einstein Frame} \label{s_einstein}

Our aim is to compute the action (\ref{action_J}), with $V=\lambda \phi^4/4$, in the JF
\be
S_J = \int d^4 x \sqrt{-g}\left[-\frac{{M_p}^2}{2} R\left(1 + \frac{\xi \phi^a R^{b-1}}{M_p^{a+2b-2}}\right)
+ \frac{1}{2}\partial_\mu \phi \partial^\mu \phi + \frac{\lambda \phi^4}{4}\right] \label{action1}
\ee
We use the a conformal transformation of the metric and Ricci scalar, see Eqs. (\ref{conf_trafo}), with the conformal factor given by
\bea
\Omega^2  = 1 + \frac{\xi \phi^a R^{b-1}}{{M_p}^{a+2b-2}} \label{Omega_eqn}\ .
\eea
For quasi de-Sitter space we may ignore the second and third terms in the bracket in Eq. (\ref{conf_trafo}) (we have $\{{\dot \Omega}/\Omega, {\ddot \Omega}/\Omega\}\ll 1$, as will be showed in (\ref{check})). We therefore write (\ref{Omega_eqn}) in EF as
\be\label{Omega_eqn1}
\Omega^2= 1 + \frac{\xi^{1+q} \phi^{p}{\tilde{R}}^{q}}{M_p^{p+2q}} \,,
\ee
where
 \begin{eqnarray}
 p = \frac{a}{2-b}\,,~~~~~~~ q = \frac{b-1}{2-b}\,,~~~~~~\tilde R = 12 \tilde H^{2}\simeq Constant\,.
 \end{eqnarray}
Via conformal transformation we write the action (\ref{action1}) in term of new field $\chi$
\bea\label{Einstein}
S_E  = \int d^4 x \left(- \frac{M_p^2}{2} \tilde R + \frac{1}{2} \partial^{\mu}{\chi}\partial_{\mu}\chi + U(\chi)\right),
\eea
where
\bea\label{diff}
\frac{d \chi}{d\phi} &=& \frac{1}{\Omega^2}\left(\Omega^2 +  \frac {3 p^2 \xi'^{2}}{2} \left(\frac{\phi}{M_p}\right)^{2 p -2}\right)^{1/2}\,,~~~~~~~ \xi' \equiv  \xi^{1+q}\left(\frac{12 \tilde H^{2}}{M_p^2}\right)^{q} \nonumber \\
 & & \nonumber \\
 U(\chi) &=& \frac{1}{\Omega^4} \frac{\lambda}{4}\phi(\chi)^4 \label{Pot1}.
\eea
For $\phi\gg M_P/\xi'^{1/p}$, Eq. (\ref{diff}) can be integrated to give
\bea\label{eqnh}
\phi(\chi)= \frac{M_p}{\xi^{\prime\, 1/p}} \exp\left(\sqrt{\frac{2}{3}}\frac{\chi}{M_p p}-\frac{1}{2}\right).
\eea
Considering
 \[
\tilde g_{\mu\nu} = diag(-M^2(t),\tilde a^2(t),\tilde a^2(t),\tilde a^2(t))
 \]
and varying the action (\ref{Einstein}) with
respect to $M(t)$ or $a(t)$ and setting $M=1$  in the final equation which corresponds FRW metric, we get the Friedmann equation
\bea\label{eq1}
12 \tilde H^2  - \zeta^{-1} M_p^2\lambda \left(1+\frac{2q}{p}\right) =0  \ ,
\eea
 where
\bea
\zeta = 12^{4q/p} \left(\frac{\tilde H^2}{M_p^2}\right)^{4q/p} \xi^{\frac{4(1+q)}{p}}
\exp\left(2 \sqrt{\frac{2}{3}}\frac{(p-2)\chi}{p M_p}\right).
\eea
All the derivative terms of  Hubble parameter $\tilde H$ have been neglected (this is corresponds to slow
roll condition, $i.e.$, $\dot{\chi}^2$ is much smaller than potential term). From (\ref{eq1})  the Hubble parameter assumes the form
\bea
\tilde{ H}=M_p \frac{\left[\left(1+2 q/p\right) \lambda \right]^{\frac{p}{2(p + 4q)}}}
{\sqrt{12} \ \xi^{\frac{2(1+q)}{p+ 4 q}}}
\exp\Bigg[\sqrt{\frac{2}{3}}\left(\frac{2-p}{p+4q}\right) \frac{\chi}{M_p}\Bigg] \label{Hubble}\ .
\eea
Using Eqs. (\ref{Omega_eqn1}) and (\ref{Hubble}) into (\ref{Pot1}) one infers the scalar potential
\bea
U(\chi)=\frac{1}{4} M_p^4 \lambda^{\frac{p}{p+4q}}\xi^{-\frac{4(1+q)}{p+4q}}
\left(1+\frac{2q}{p}\right)^{- \frac{4q}{p+4q}} \exp\Big[2 \sqrt{\frac{2}{3}}
\left(\frac{2-p}{p+4q}\right) \frac{\chi}{M_p}\Big] \label{Pot2}.
\eea
In deriving (\ref{Pot2}) the large field approximation $\exp(\sqrt{\frac{2}{3}} \frac{\chi}{M_p}) \gg 1$ for $\chi \gg M_p$ has been used. Also the parameters $a$ and $b$, which appear through $p$ and $q$ as defined above, in the potential (\ref{Pot2}) are the parameters of the JF action (\ref{action1}) which appear in the EF potential via conformal transformation of the metric (\ref{conf_trafo}).
The spectral index and curvature perturbation can be computed by using above potential (\ref{Pot2}).
The slow-roll parameters for large $\chi \gg M_p$ comes out to be
\begin{eqnarray}
 \epsilon &=& \frac{M_p^2}{2}\left(\frac{U'}{U}\right)^2=\frac{4}{3}\left(\frac{a+2b-4}{a+4b-4}\right)^2\,, \\
 & & \nonumber \\
  \eta &=& M_p^2 \left(\frac{U''}{U}\right)=\frac{8}{3}\left(\frac{a+2b-4}{a+4b-4}\right)^2\,, \nonumber
\end{eqnarray}
and the amplitude of curvature perturbations (\ref{scalaramp})
\begin{eqnarray}\label{curvp_E}
\Delta^{2}_{\mathcal R} &=& \frac{1}{8\pi^{2} \epsilon} \frac{\tilde H^{2}}{M_{p}^{2}} \\
  &=& \frac{1}{128 \pi^{2} }\left(\frac{y+2}{2y-x+4}\right)^{\frac{x+2y+4}{x}}
 \lambda^{\frac{2y-x+4}{x}}\xi^{-\frac{4}{x}} \left(\frac{x}{y}\right)^{2}
 e^{-2\sqrt{\frac{2}{3}}\frac{y}{x} \frac{\chi}{M_p}} \,, \nonumber
\end{eqnarray}
where
 \[
 x= a+ 4b-4, \qquad y=a+2b-4\,.
 \]
The spectral index in the term of slow-roll parameters is (\ref{SIndex})
\be
n_s=1-6\epsilon + 2\eta.\label{sindexGHI}
\ee
The number of e-folding is calculated as
\be\label{N_E}
N_{E} =\int^{{\chi}_0}_{{\chi}_e} \frac{U(\chi)}{U'(\chi)} d\chi = -\frac{1}{2}\sqrt{\frac{3}{2}}\left(\frac{x}{y}\right)\left(\frac{{\chi}_0 -{\chi}_e}{M_p}\right)
\ee
For  ${\chi}_0\sim 13 M_p$ and ${\chi}_e\sim 1 M_p$, the number
of e-folding is found to be around $60$.
From equation~(\ref{Omega_eqn}) and (\ref{eqnh}), one
can calculate the order of terms like $\ddot{\Omega}/\Omega$ and $(\dot\Omega/\Omega)^2$
for $\phi \gg \frac{M_p}{\xi^{1/p}}$. For $\lambda=10^{-3}$ and $\xi=1$, one gets
\begin{eqnarray}\label{check}
\frac{\ddot{\Omega}}{\Omega} &\sim& \frac{U}{9 M_p^2} (\epsilon +\sqrt{3\epsilon} (\eta-\epsilon)) = 4.1 \times 10^{-11} M_p^2\,, \\
 & & \nonumber \\
\left(\frac{\dot{\Omega}}{\Omega}\right)^2  &\sim& \frac{U}{9 M_p^2} \epsilon = 3.3 \times 10^{-11} M_p^2 \,, \nonumber
\end{eqnarray}
whereas the value of curvature scalar $\tilde R =12 \tilde{H}^2$ at the same values of parameter is
$4.1 \times 10^{-8} M_p^2$. Thus the approximation we have made is consistent. We now use the measured values of these CMB anisotropy parameters to get the numerical values for the parameters  $(a,b,\xi,\lambda)$.

\subsection*{Results and Discussion} 

Planck$+$WP measurements \cite{Planckgroup} provides that the curvature
perturbation is $\Delta_{\mathcal R}^2 = 2.195^{+0.533}_{-0.585} \times 10^{-9}$,
spectral index is $n_{\mathcal R}= 0.9603 \pm 0.0073$ and the tensor to scalar ratio $r < 0.11 (95\% CL)$ (see (\ref{DeltaGae})-(\ref{rGae})). Equation (\ref{N_E}) implies that to get $60$ e-foldings, the scalar field $\chi$ should
roll from $13 M_p$ to $1 M_p$. In Table \ref{values1_E} are reported the values of parameters $\{\lambda, a, b\}$
giving the measured values of $\Delta_{\mathcal R}^2$ and $n_s$. Here $\xi=1$ and $\lambda$ $\xi=1$ assumes different values.

The slow-roll parameters are $\epsilon \sim 0.02$  and $\eta \sim 0.04$.
In particular, for $\epsilon\simeq 0.02$ the tensor-to-scalar ratio in EF is predicted to be large $r\simeq0.3$.
\begin{table}
\center
\caption[The values of the model parameters $a$ and $b$ in the Einstein frame for different
    values of parameter $\lambda$ and $\xi=1$.]{The values of parameters $a$ and $b$ in the Einstein frame at $\chi_0=13 M_p$ with $\xi=1$ for different
    values of $\lambda$. The parameters $a$ and $b$ of the Jordan frame action appears in the Einstein frame potential
    via conformal transformation of the metric.}
    \begin{tabular}{|l|l|l|l|l|l|}
        \hline
        $ \ \lambda$ & $ \ 0.1  $  &  $ \ 10^{-2}  $  &   $ \ 10^{-3} $ & $ \ 10^{-4} $ & $\ 10^{-5}$ \\ \hline
       \ a &$ \ 3.385  $ & $\ 3.026$  & $ \  2.735$ & $\ 2.494$&$\ 2.292$\\ \hline
       \ b & $ \ 0.277$ & $\ 0.439$  & $\ 0.571 $& $\ 0.679 $&$\ 0.770$ \\ \hline
       \ a+2b & $ \ 3.939 $ & $ \ 3.904$ & $ \ 3.877 $ & $ \ 3.852 $ & $ \ 3.832 $ \\ \hline
     \end{tabular}
    \label{values1_E}
\end{table}
We evaluate now the observables in JF using (\ref{DeltaJ}) and (\ref{spectral}) for $\xi=1$. In Table \ref{values3_J} are displayed the values of  the scalar field $\phi$ in the JF corresponding to $\chi_{e}=1 M_p$ and $\chi_0=13 M_p$ for different values of $\lambda$. Using these values of the  range of the roll in $\phi$ we see that the number
of e-foldings $N_J$ in the JF, corresponding to $N_E=60$ is $N_J\sim 830$. The values of
$\{\lambda, a, b\}$, required to get the curvature perturbation and spectral index, are
shown in the Table \ref{values3_J}. The slow-roll parameters are found to be $\epsilon_1 \simeq -
\epsilon_3 \simeq 0.007$ and $\epsilon_2 \simeq - \epsilon_4 \simeq -0.013$ for chosen
range of $\lambda$. For these values, the tensor-to-scalar ratio is small $r\simeq0.002 $ in JF.
Two comments are in order: 1) The assumptions $F=1+\frac{\xi b \phi^a R^{b-1}}{M_p^{a+2b-2}} \gg 1$ and $\frac{\kappa^2 \dot{\phi}^2}{F H^2} \ll 6 \epsilon_3^2$ are verified for the best fit values of the above parameters. 2) In the limit $a\simeq2$ and $b\simeq1$, the correct value of $\Delta_{\mathcal R}^2$ and $n_{\mathcal R}$ are obtained for $\lambda \sim 0.1$ only for large value of $\xi \sim 10^4$, in agreement with the prediction of Higgs inflation models $\xi\phi^{2}R$ ~\cite{Bezrukov,Bezrukov1}.
\begin{table}
\center
\caption[The values of the model parameters $a$ and $b$ in the Jordan frame for different
    values of parameter $\lambda$ and $\xi=1$.]{The values of parameters $a$ and $b$ are evaluated in the Jordan frame
    at  $\xi=1$ and $\phi_J|_{_{\chi_0=13 M_p}}$ for different values of $\lambda$.}
    \begin{tabular}{|l|l|l|l|l|l|}
        \hline
       $ \ \lambda$ & $ \ 0.1  $  &  $ \ 10^{-2}  $  &   $ \ 10^{-3} $ & $ \ 10^{-4} $ & $\ 10^{-5}$ \\ \hline
        \ $\phi_{f} |_{_{(\chi_e =1 M_p)}}$  &$ \ 0.0146 M_p $ & $\ 0.0253 M_p$  & $ \ 0.044 M_p $ & $\ 0.077 M_p$&$\ 0.134 M_p$\\ \hline
         \ $\phi_{J}|_{_{(\chi_0=13 M_p)}}$ &$ \ 3.566 M_p $ & $\ 6.187 M_p $  & $ \ 10.77 M_p $ & $\ 18.8 M_p$&$\ 32.77 M_p$\\ \hline
       \ a &$ \ 3.56398962 $ & $\ 3.2751299$  & $ \ 3.0257694 $ & $\ 2.809561$&$\ 2.620851$\\ \hline
       \ b & $ \ 0.21800513 $ & $\ 0.3624348$  & $\ 0.4871146$& $\ 0.595217$&$\ 0.689566$ \\ \hline
       \ a+2b & $ \ 3.999999 $ & $ \ 3.999999$ & $ \ 3.999998 $ & $ \ 3.999995 $ & $ \ 3.99998 $ \\ \hline
    \end{tabular}
    \label{values3_J}
\end{table}

To summarize, this model ${\xi \phi^a R^b}$ is a 
generalization of the Higgs inflation model ${\xi \phi^2 R}$ with $\lambda\phi^{4}$ potential. We find that 
if the Higgs self coupling $\lambda$ is in the range $(10^{-5}-0.1)$, parameter $a$ in the range $(2.3 -3.6)$ and $b$ in the range
$(0.77-0.22)$ at the Planck scale, one can have a viable inflation model even for $\xi \simeq 1$. $\lambda\sim0.1$ in this 
model solve the fine tuning problem of Higgs self-coupling in the standard slow-roll inflation which predict $\lambda\sim10^{-12}$.
The tensor-to-scalar ratio $r$ in this model in EF is large $r\simeq0.3$, therefore model with generalized scalar-curvature
couplings is ruled out by observational limits on $r$ like the pure  $\frac{\lambda}{4} \phi^4$ theory. However, with
independent calculations in JF gives small $r\simeq0.002$ which is allowed from the observations. Therefore, JF result 
contradicts the EF result. The observations should be compared with EF results or JF results is still matter of debate.
However in this model, by requiring the curvature coupling
parameter to be of order unity, we have evaded the problem of unitarity violation in scalar-graviton scatterings which
plague the $\xi \phi^2 R$ Higgs inflation models. Therefore, the Higgs field may still be a good candidate for being the
inflaton in the early universe if one considers higher dimensional curvature coupling. Also we find that, upto 
slow-roll approximation, for the same set of parameter values ($\xi,\lambda$) the set of ($a$,$b$) values is nearly
the same in Jordan and Einstein frames. Therefore, the Einstein and Jordan frames are equivalent. In this model, we found
a symmetry $a+2b\approx4$ which holds true in both the frame and it implies that the curvature coupling $\xi$
is nearly scale invariant.

\subsection{The power law model $R+R^\beta$}

We now study a model $R+\frac{1}{M^{2}}R^{\beta}$ of inflation which is a generalization
of Starobinsky model $R+\frac{1}{M^{2}}R^{2}$ and so we call it as power law Starobinsky model~\cite{Chakravarty:2014yda}.
$M$ and $\beta$ are the two dimensionless parameters. In general scalar-curvature theories the scalar plays the role of the inflaton after transforming to Einstein frame whereas in pure curvature theories like $R + \frac{1}{M^{2}}R^{\beta}$ model the longitudinal part of the graviton is the equivalent scalar in the Einstein frame plays the role of inflaton.
Along with the independent observable predictions of this model, we will
show that generalized Higgs inflation model $\xi\phi^{a}R^{b}$ is equivalent to power law Starobinsky model.

The Starobinsky model of inflation \cite{starobinsky2} with an $\frac{1}{M^{2}}R^{2}$ interaction term is of interest
as it requires no extra scalar fields but relies on the scalar degree of the metric tensor to generate the 'inflaton' potential.
The $R^{2}$ correction to Einstein gravity is, as we have before seen, equivalent to scalar-tensor theory with a scalar potential which is an exponentially corrected plateau potential \cite{Bezrukov}. This model is favored by the Planck constraint on the tensor to scalar ratio which ruled out potentials like $m^{2}\phi^{2}$ and $\lambda \phi^{4}$ in the context of standard slow-roll inflation.
The characteristic feature of the Starobinsky equivalent models was the prediction that the tensor-to-scalar ratio was $r \simeq 10^{-3}$. BICEP2 reported a large value of $r= 0.2^{+0.07}_{-0.05}$ \cite{BICEP2} but the recent joint analysis
by Planck + BICEP2 + Keck Array give only an upper bound of $r_{0.05}<0.07 (95\% CL)$ \cite{BKP:2015,Ade:2015xua}.
In an analysis of the genus structure of the B-mode polarisation of Planck + BICEP2 data by Colley et al. put the tensor-to-scalar ratio
at $r=0.11\pm0.04 (68\% CL)$ \cite{Colley:2014nna}. In the light of the possibility that $r$ can be larger than the
Starobinsky model prediction of $r\sim0.003$, generalisations of the Starobinsky model are of interest.
The quantum correction on $\phi^{4}$-potential in Jordan frame was studied in \cite{Kehagias:2013mya,Joergensen:2014rya,Codello:2014sua,Gao:2014fha} where they have shown the equivalence of the $\xi \phi^{2} R + \lambda \phi^{4(1+\gamma)}$ model with $\frac{1}{M^{2}}R^{\beta}$ model. The generalized Starobinsky model with $R^{p}$ correction has been studied in the ref. \cite{stelle}.

The action is given by \cite{DeFelice:2010aj,Nojiri:2010wj}
\begin{equation}\label{eq1gae}
S_{J} = \int d^4x\sqrt{-g} f(R) \,, \qquad f(R)= -\frac{M_p^{2}}{2} \left(R+ \frac{1}{6M^{2}} \frac{R^{\beta}}{M_p^{2\beta-2}} \right)
\end{equation}
Consider the conformal transformation
 \[
 \Omega = F = \frac{\partial f(R)}{\partial R}
 \]
and the new scalar field $\chi$ defined by
 \[
 \Omega \equiv \exp\left(\frac{2\chi}{\sqrt{6} M_p}\right)\,.
 \]
The action~(\ref{eq1gae}) gets transformed in the EF (see Section 4) as
\begin{figure}[t!]
 \centering
\includegraphics[width= 10.0cm]{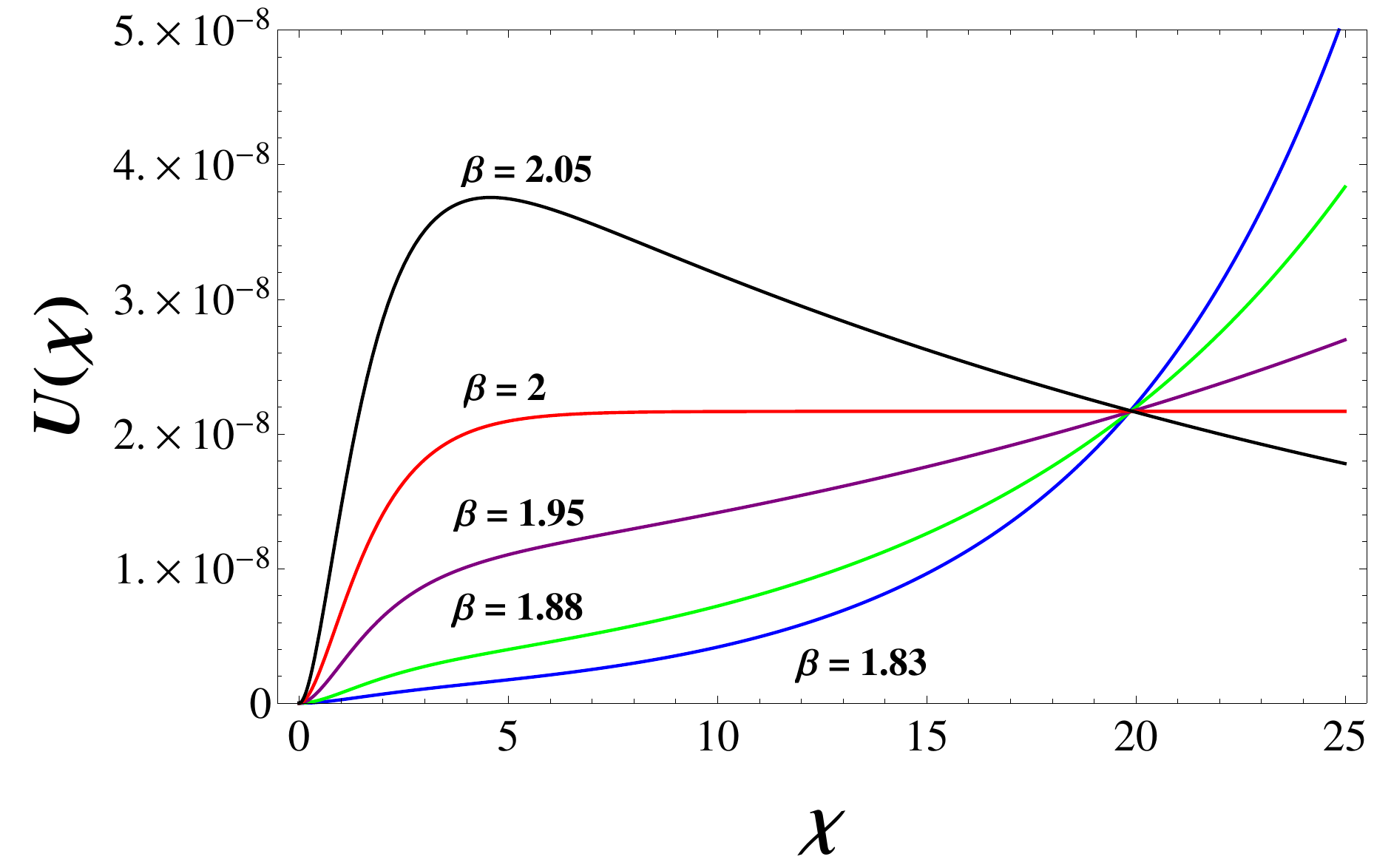}\hspace{0.5cm}
\caption[The nature of the power law potential of inflaton for different values of parameters $\beta$ and $M$.]
   {The nature of the potential (\ref{U1}) for different $\beta$ values (with $M=1.7\times10^{-4}$).
   The potential and the field values are in $M_p=1$ units.}
\label{fig1a}
\end{figure}
\begin{equation}
S_{E}=\int d^{4}x\sqrt{-\tilde{g}}\left[\frac{-M_p^{2}}{2}\tilde{R}
+\frac{1}{2}\tilde{g}^{\mu\nu}\partial_{\mu}\chi\partial_{\nu}\chi+U(\chi)\right]\,,
\label{Ein}
\end{equation}
where $U(\chi)$ is can be written in the EF as 
\begin{eqnarray}
U(\chi) &=& \frac{\left(R F(R)-f(R)\right)M_p^{2}}{2 F(R)^2}  \nonumber\\
  &=&
 \frac{(\beta-1)}{2} \left(\frac{6M^{2}}{\beta^{\beta}}\right)^{\frac{1}{\beta-1}}
\exp\left[ \frac{2\chi}{\sqrt6}\bigg(\frac{2-\beta}{\beta-1}\bigg)\right]
\left[1 - \exp\bigg(\frac{-2\chi}{\sqrt6}\bigg)\right]^{\frac{\beta}{\beta-1}}\label{U1}\,
\end{eqnarray}
with $f(R)$ given in (\ref{eq1gae}) ($M_p=1$).
Also we see that in the limit $\beta\rightarrow 2$ potential (\ref{U1}) reduces to exponentially
corrected flat plateau potential of the Starobinsky model.
Assuming large field limit $\chi \gg \frac{\sqrt6}{2}$ and $ 1 < \beta < 2$, the potential (\ref{U1}) reduces to
\bea
U(\chi) \simeq \frac{(\beta-1)}{2} \left(\frac{6M^{2}}{\beta^{\beta}}\right)^{\frac{1}{\beta-1}} \exp\Bigg[\frac{2\chi}{\sqrt6}
\bigg(\frac{2-\beta}{\beta-1}\bigg)\Bigg]\label{U2}
\eea
We shall use Eq. (\ref{U2}) to compare with SUGRA version of the power law potential in the large field limit.


In Fig.\ref{fig1a} it is plotted the potential for small deviations from the Starobinsky model value $\beta=2$.
We see that the potential is very flattest  for $\beta=2$ but becomes very steep even with small deviation
from Starobinsky model value $\beta=2$. The scalar curvature perturbation $\Delta^{2}_{\mathcal{R}}
\propto \frac{U(\chi)}{\epsilon}$ is fixed from observations which implies that the magnitude of the
potential $U(\chi)$ would have to be larger as $\epsilon$ increases for steep potential to maintain the
level of observed amplitude $\Delta^{2}_{\mathcal{R}}$. The tensor perturbation which depends on the magnitude of
$U(\chi)$ therefore increases rapidly as $\beta$ varies from $2$. The variation of $r$
with $\beta$ is shown in the Fig. \ref{fig3gae}.

From Eq. (\ref{U1}), in the large field approximation, the slow-roll parameters in Einstein frame
can be obtained as
\bea
\epsilon &=& \frac{1}{2}\left(\frac{U'}{U}\right)^2 \simeq \frac{1}{3}\left[\frac{\beta(3-2\beta)}
{(\beta-1)^{2}} \exp\left(\frac{-2\chi}{\sqrt{6}}\right) + \frac{\beta-2}{\beta-1}\right]^{2},\label{slow1}\\
 & & \nonumber \\
\eta &=& \frac{U''}{U} \simeq \frac{-2}{3} \left[\frac{\beta(3-2\beta)^{2}}{(\beta-1)^{3}}
\exp\left(\frac{-2\chi}{\sqrt{6}}\right) - \frac{(\beta-2)^{2}}{(\beta-1)^{2}}\right],\\
 & & \nonumber \\
\xi &=& \frac{U'U'''}{U^{2}} \simeq \frac{4\sqrt{\epsilon}}{3\sqrt{3}} \left[\frac{\beta(3-2\beta)^{3}}
{(\beta-1)^{4}} \exp\left(\frac{-2\chi}{\sqrt{6}}\right) + \frac{(\beta-2)^{3}}{(\beta-1)^{3}} \right].
\eea
The field value $\chi_{e}$ at the end of inflation can be fixed from Eq. (\ref{slow1}) by using the end
of inflation condition $\epsilon \simeq 1$. And the initial scalar field value $\chi_{s}$ corresponding to
$N = 60$ e-folds before the end of inflation, when observable CMB modes leave the horizon,
can be fixed by using the e-folding expression
 \[
N=\int_{\chi_e}^{\chi_s} \frac{U(\chi)}{U'(\chi)} d\chi\,.
 \]
Under slow-roll approximation, one can use the standard Einstein frame relations for the amplitude of the curvature perturbation $\Delta_{\mathcal R}^{2}$, the spectral index $n_{s}$ and its running $\alpha_{s}$, and the tensor-to-scalar ratio $r$ to fix the parameters of the model.


 \begin{figure}[t!]
 \centering
 \includegraphics[width= 7.0cm]{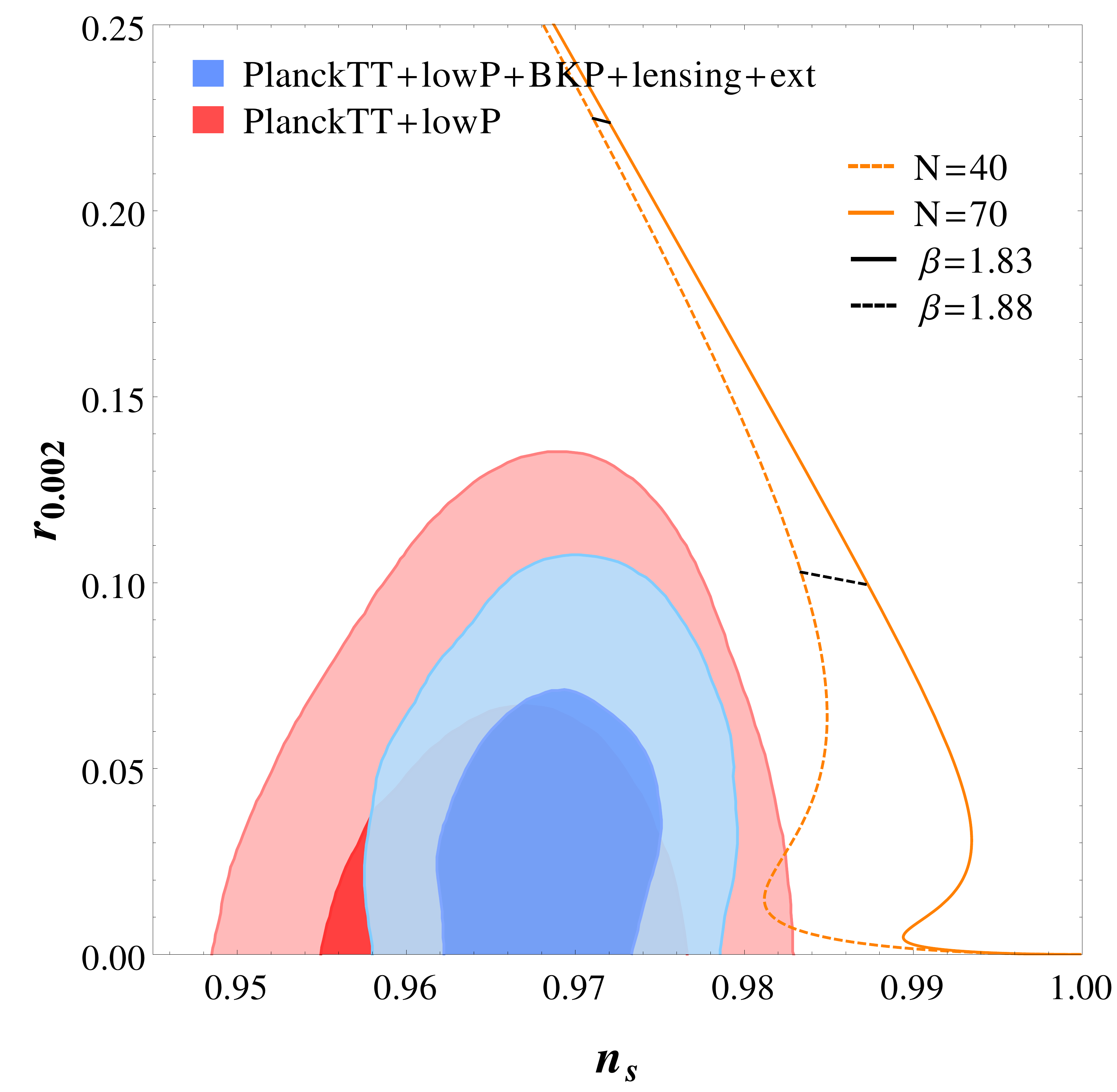}
\caption[The power law model predictions of $n_{s}$ and $r$ are compared with Planck-2015 and joint BKP analysis.]
   {The regions of $(n_{s},r)$ allowed by Planck-2015 and joint BKP analysis at $68 \% CL$ and $95 \% CL$ are shown \cite{Ade:2015xua}.
   The colored contour lines are the predictions for our model for two sets of $\beta$ and $N$ values corresponding
   to $M\approx 10^{-4}$ which satisfies the observed amplitude of the CMB power spectrum.}
 \label{fig2}
\end{figure}

From CMB observations, for 7-parameter $\Lambda CDM$+$r$ model, when there is no scale dependence
of the scalar and tensor spectral indices the bound on $r$ is  $r_{0.002} < 0.07$ ($95 \% CL$, PlanckTT+lowP) and
the amplitude, the spectral index and the running of spectral index are $10^{10}\ln(\Delta_{\mathcal R}^{2}) = 3.089\pm 0.036$, $n_{s}= 0.9666 \pm 0.0062$ and $\alpha_s = - 0.0084 \pm 0.0082$, respectively at ($68 \% CL$, PlanckTT+lowP) \cite{Ade:2015lrj,BKP:2015,Ade:2015xua}.
Since the scalar potential $U(\chi)$ depends on both the parameters $M$ and $\beta$ whereas the slow-roll
parameters depend only on $\beta$, therefore parameter $M$ affects only the scalar amplitude
$\Delta_{\cal R}^{2} \propto \frac{U(\chi)}{\epsilon}$ whereas $r$, $n_{s}$ and $\alpha_{s}$ which depend
only on slow-roll parameters remain unaffected by $M$. Therefore taking amplitude from the observation and
fixing the number of e-foldings $N$ fixes the value of $M$ and $\beta$.
We find that the values of $M \simeq 1.7\times10^{-4}$ and $\beta \simeq 1.83$ which satisfy
the amplitude, the spectral index and its running for $N\approx 60$ gives large $r\approx 0.22$. Also we see that for $\beta \simeq 1.88$, tensor-to-scalar ratio can be reduced to $r\simeq 0.1$ but it increases $n_{s}\simeq 0.987$, see Fig.\ref{fig2}.

\begin{figure}[t!]
 \centering
\includegraphics[width= 8.0cm]{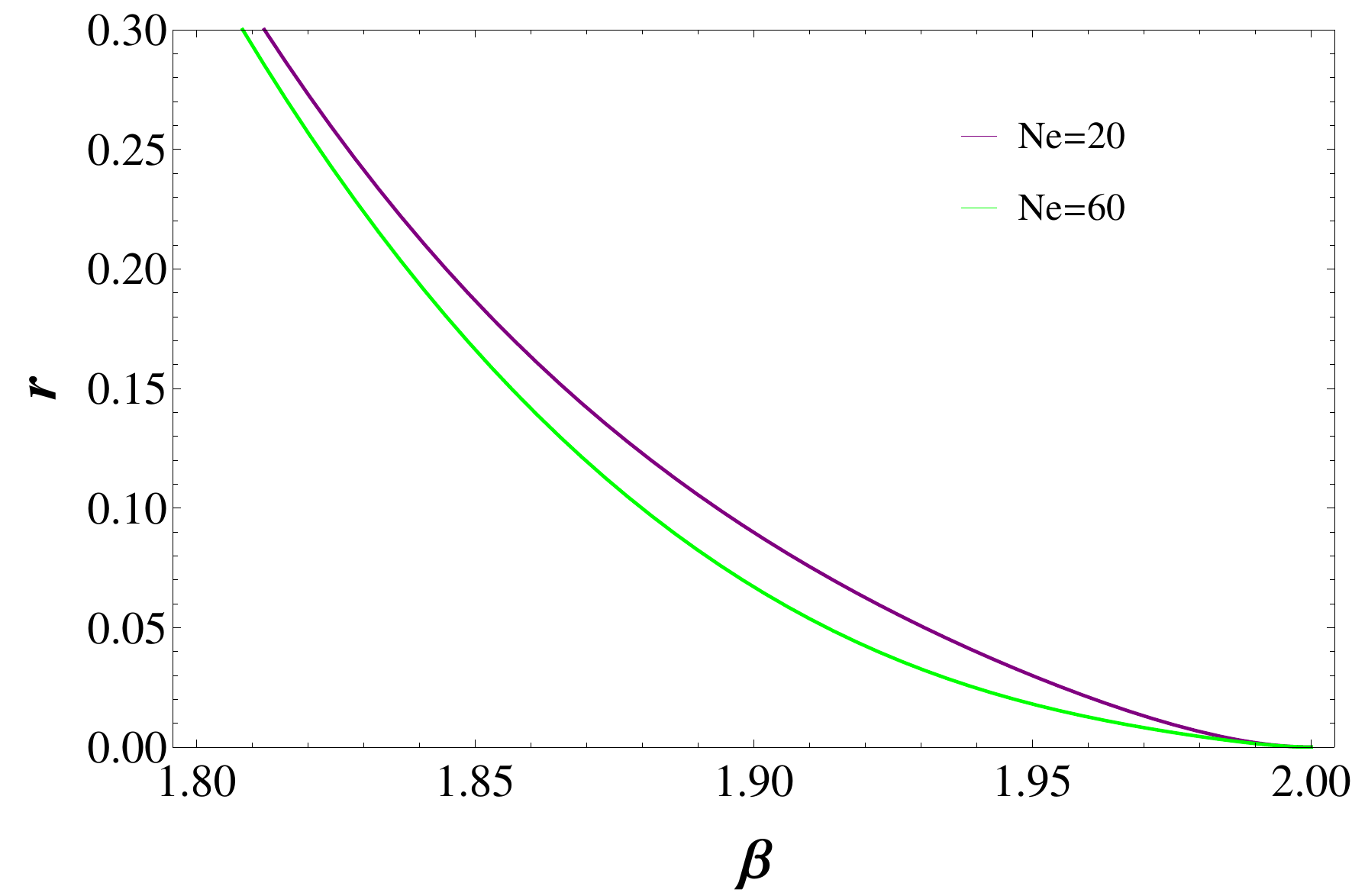}\hspace{0.0cm}
\caption[The variation of tensor-to-scalar ratio $r$ with parameter $\beta$ for running and no-running of spectral index $n_{s}$.]
   {The variation of $r$ with $\beta$ shown for two cases studied in our model: (i) for $N=20$ when running of $n_{s}$
   is considered and (ii) for $N=60$ when there is no running of $n_{s}$.}
\label{fig3gae}
\end{figure}
\subsubsection{Power Law Starobinsky Model from No-scale SUGRA}\label{SUGRA}

In this section we show that the power law Starobinsly model (higher order curvature theory) can be motivated from SUGRA as they arise naturally in this framework. The SUGRA embedding of the Higgs-inflation \cite{Bezrukov} does not produce a slow-roll
potential in MSSM  but a potential suitable for inflation is obtained in N(ext)MSSM \cite{Einhorn:2009bh}.
The potential in NMSSM however has a tachyonic instability in the direction orthogonal to the slow-roll
\cite{Ferrara:2010yw}. This instability can be cured by the addition of quartic terms of the fields  in the
K\"ahler potential \cite{Lee:2010hj,Ferrara:2010in} (see also \cite{Girish2016PLB}). In the context of a SUGRA embedding of the Starobinsky model, it was shown that quadratic Ricci  curvature terms  can be derived in SUGRA by adding two chiral superfields in the minimal SUGRA \cite{Cecotti:1987sa}. A no-scale SUGRA model \cite{Ellis:1984bm,Lahanas:1986uc} with a modulus field and the inflation field with a minimal Wess-Zumino superpotential gives the same F-term potential in the Einstein frame as the Starobinsky model \cite{Ellis:2013xoa}. The symmetry principle which can be invoked for the SUGRA generalization of the Starobinsky model is the spontaneous violation of superconformal symmetry. The quadratic
curvature can also arise from D-term in a minimal-SUGRA theory with the addition of a vector and chiral
supermultiplets \cite{Cecotti:1987qe}. The Starobinsky model has been derived from the D-term potential of
a SUGRA model \cite{Buchmuller:2013zfa}. Quartic powers of Ricci curvature in the bosonic Lagrangian can also be obtained in a SUGRA model by the D-term of higher order powers of the field strength superfield \cite{Ferrara:2013kca}.

More specifically, to get a no-scale SUGRA model corresponding to power law Starobinsky model which can give a larger $r$, we choose the minimal Wess-Zumino form of the superpotential (\ref{WZSP})
and a minimal no-scale K\"ahler potential with an added $(\phi+\phi^{*})^{n}$ term as
\be\label{KPgae}
K = -3 \ln \left[T+T^{*}-\frac{(\phi+\phi^{*})^{n}}{12}\right]
\ee
which can be motivated by a shift symmetry
 \[
 T\rightarrow T + i C\,, \quad  \phi \rightarrow \phi +i C\,,
 \]
with $ C $ real, on the K\"ahler potential. Here $T$ is a modulus field and $\phi$ is a matter filed which plays the role of inflaton.
We assume that the $T$ field gets a vev $\langle T + T^{*} \rangle = 2 \langle Re T \rangle = c >0$ and
 $\langle ImT \rangle =0 $. We write $\phi$ in terms of its real and imaginary parts $\phi=\phi_{1}+i\phi_{2}$.
If we fix the imaginary part of the inflaton field $\phi$ to be zero then $\phi=\phi^{*}=\phi_{1}$ and for simplicity we
replace $\phi_{1}$ by $\phi$, the effective Lagrangian in the Einstein frame is given by
 \bea
{\cal{L}}_{E} = \frac{n (2\phi)^{n-2} [c(n-1) + \frac{(2\phi)^{n}}{12}]}{4[c - \frac{(2\phi)^{n}}{12}]^{2}}
\left|\partial_{\mu}\phi \right|^{2} - \frac{4(2\phi)^{2-n}}{n(n-1)[c-\frac{(2\phi)^{n}}{12}]^{2}}
\left|\frac{\partial{W}}{\partial{\phi}}\right|^{2}. \label{LE1}
\eea
To make the kinetic term canonical in the ${\cal{L}}_{E}$, we redefine the field $\phi$ to $\chi$ with
\bea
\frac{\partial \chi}{\partial \phi} = -\frac{\sqrt{n(2\phi)^{n-2}[c(n-1)+\frac{(2\phi)^{n}}{12}]}}
{2[c - \frac{(2\phi)^{n}}{12}]} \label{chiphi}
\eea
Assuming that $n \sim \mathcal{O}(1)$ and the large field limit $ (2\phi)^{n} \gg 12c $ during inflation, integrating Eq. (\ref{chiphi}) gives
\bea
\phi \simeq \frac{1}{2} \exp\left(\frac{2\chi}{\sqrt{3n}}\right) \left[1 + \frac{6c(n+1)}{n}
\exp\left(\frac{-2 n\chi}{\sqrt{3n}}\right) \right] \label{phi1}
\eea
Substituting from (\ref{SP}) and (\ref{phi1}) into the potential term of (\ref{LE1}) and simplifying, one derives
the effective scalar potential in the EF
\begin{eqnarray}\label{V3}
V &=& \frac{144\mu^{2}}{n(n-1)} \left[1- \frac{2\mu}{\lambda}\exp\left(\frac{-2\chi}{\sqrt{3n}}\right) -\frac{9c(n^{2}-n-2)}{n}
\exp\left(\frac{-2n\chi}{\sqrt{3n}}\right) \right]^{2} \times \\
 & & \nonumber \\
 & &\times
\exp\left[\frac{2\chi}{\sqrt{6}}\left(\frac{3\sqrt{2}(2-n)}{\sqrt{n}}\right)\right]\,, \nonumber
\end{eqnarray}
which, assuming $1<n<2$, in the large field limit $\chi \gg \frac{\sqrt{3n}}{2}$ is equivalent to
\bea\label{V4}
V \simeq \frac{144\mu^{2}}{n(n-1)} \exp\left[\frac{2\chi}{\sqrt{6}}\left(\frac{3\sqrt{2}(2-n)}{\sqrt{n}}\right)\right]\,.
\eea
In the limit $n \rightarrow 2$ and with the specific choice  $\frac{\lambda}{\mu}=\frac{1}{2}$,
the potential (\ref{V3}) reduces to Starobinsky Model potential.
We can now compare the power law potential (\ref{U2}) and SUGRA potential (\ref{V4}) for inflaton to show the relation
between the parameters of the two model. Comparing the constant coefficient and exponent in the two potentials we get
\begin{eqnarray}\label{beta1}
\beta &=& \frac{2\sqrt{n}+3\sqrt{2}(2-n)}{\sqrt{n}+3\sqrt{2}(2-n)}\,, \\
 & & \nonumber \\
M^{2} &=& \frac{\beta^{\beta}}{6} \left[\frac{288 \mu^{2}}{n(n-1)(\beta-1)}\right]^{\beta-1}\,. \nonumber
\end{eqnarray}
Numerically we evaluate the SUGRA model parameter values (in $M_p = 1$ unit) for three values of $\beta$ corresponding to running and without running of spectral index $n_{s}$ as depicted in Fig.\ref{fig2} and for Starobinsky limit $\beta=2$. These values are shown in the TABLE \ref{sugrapara}.
\begin{table}
\centering
\caption[The SUGRA model parameters values for three different values of parameter $\beta$ corresponding to
    running and without running of spectral index $n_s$, and Starobinsky limit $\beta\rightarrow2$.]{The SUGRA model parameter values (in $M_p = 1$ unit) for three values of $\beta$ corresponding to
    running and without running of spectral index $n_{s}$ as depicted in Fig.\ref{fig2} and for Starobinsky limit $\beta=2$.}
    \begin{tabular}{|l|l|l|l|l|}
        \hline
        $ \beta $ & $ \ M$ & $ n  $ &  $ \mu=\frac{|\lambda|}{2} $ & $ \alpha_{s}= \frac{dn_{s}}{d\ln{k}}$ \\ \hline
        1.83 & $ 1.7\times 10^{-4} $ & $ 1.93 $ &  $ 3.13\times 10^{-6} $ & $ -9.16\times 10^{-6} $ \\ \hline
        1.88 & $ 1.7\times 10^{-4} $  & $ 1.96  $ & $ 5.54 \times 10^{-6} $ & $ -2.86\times 10^{-3} $\\ \hline
        2.00 & $ 1.1 \times 10^{-5}  $ & $ 2.00  $ &   $ 1.16 \times 10^{-6} $ & $ -5.23\times 10^{-4} $\\ \hline
   \end{tabular}
    \label{sugrapara}
\end{table}

To summarize, this power law model with $\frac{1}{M^2}R^{\beta}$ correction to Einstein gravity
is a generalisation of the Starobinsky model of inflation. The interesting feature of this form of generalization is that
small deviations from the Starobinsky limit $\beta=2$ can change the value of tensor-to-scalar ratio from
$r \sim \mathcal{O}(10^{-3})$ to $r\sim \mathcal{O}(0.1)$. We find that this model predicts large tensor-to-scale
$r\approx 0.22$ as indicated  by BICEP2 measurements, for the value of $\beta \approx 1.83$
and $M\sim10^{-4}$. Also we showed that the general $R^\beta$ model can be obtained from a SUGRA construction
with minimal Wess-Zumino form of superpotential and by adding a power law $(\phi +\bar \phi)^n$ term to the
minimal no-scale SUGRA K\"ahler potential. We further showed that this two parameter power law generalization  
of the Starobinsky model is equivalent to generalized non-minimal curvature coupled models with quantum 
corrected $\lambda\phi^{4}$ potentials $i.e.$ models of the form $\xi \phi^{a} R^{b} + \lambda \phi^{4(1+\gamma)}$,
and thus the power law Starobinsky model is the most economical parametrization of such models. Since
such a power law correction to Einstein gravity generates large amplitude of gravity waves,
therefore they are ruled out by the current status of the observations by Planck and
BKP Collaboration.

\subsection{The model with two scalar fields}

In this section we discuss a two-field inflationary model where the
inflaton field $\phi$ is assisted by a dilaton field $\sigma$ and has
a non-canonical kinetic term due to the presence of a dilaton
field. Supergravity theories which are low energy limit of string theory contains several scalar fields
which can be of cosmological interest. The action of the model can be generically written as
\bea
S &=&\frac12\int d^4 x\sqrt{-g}\left[R-\nabla^\mu\sigma\nabla_\mu\sigma-e^{-\gamma\sigma}\nabla^\mu \phi\nabla_\mu \phi-2e^{-\beta\sigma} V(\phi)\right],\label{action-sigma-einstein2}
\eea
where $\beta$ and 
$\gamma$ are arbitrary independent parameters. Brans-Dicke (BD) gravity in Einstein frame (EF) is a special case where $\beta=2\gamma$ 
\cite{Brans:1961sx,Starobinsky:1994mh,GarciaBellido:1995fz,DiMarco:2002eb,Gong:1998nf}.
However BD gravity predicts $r$ larger than the observed limit, therefore generalization of BD theory 
is necessary for application to inflation. In this paper we generalize the BD theory in EF to a
two-parameter scalar-tensor theory where we treat $\beta$ and $\gamma$ as two independent arbitrary
parameters. Addition of one extra parameter allows us to obtain viable inflation
with otherwise ruled out quadratic and quartic potentials as we can have tensor-to-scalar 
ratio $r$ in the range of interest for forthcoming experiments. 

To note, it was shown by Ellis et al.~\cite{Ellis:2013xoa} that the inflaton field accompanied by a
moduli field $T$, which appear in string theories and have a no-scale supergravity 
form, give a potential for inflation equivalent to the $R+R^2$ Starobinsky model, producing $r\sim10^{-3}$.
We show in this paper that the above mentioned two parameter 
scalar-tensor theory can be obtained from no-scale supergravity theories
~\cite{Cremmer:1983bf,Ellis:1983sf,Lahanas:1986uc} which now can produce $r$ much larger than $10^{-3}$
and thus are observationally falsifiable by future experiments. Also in contrast to 
the supergravity embedding of the Starobinsky model, studied in~\cite{Kallosh:2014qta,Hamaguchi:2014mza},
where the imaginary part of the superfield $T$ $(i.e.$ axion $\phi)$ decreases rapidly and its real 
part $(i.e.$ dilaton $\sigma)$ drives the inflation, we will see that in our model dilaton-axion
pair evolves sufficiently during inflation and the axion acts as the inflaton. 

 First we look at the background dynamics of our model. Starting from the action 
(\ref{action-sigma-einstein2}), the equations of motion of the fields $\phi$ and $\sigma$ and 
the Friedmann equations can be obtained as 
\begin{eqnarray}
&&\ddot{\sigma}+3H\dot{\sigma}+\frac{\gamma}{2}e^{-\gamma\sigma}\dot{\phi}^2-\beta e^{-\beta\sigma}V(\phi)=0, \label{fullsigmadot} \\
&&\ddot{\phi}+3H\dot{\phi}-\gamma\dot{\sigma}\dot{\phi}+e^{(\gamma-\beta)\sigma}V'(\phi)=0, \label{fullphiddot} \\
&& 3 H^2 =\frac12\dot\sigma^2+\frac12 e^{-\gamma\sigma}\dot{\phi}^2 +e^{-\beta\sigma}V(\phi), \label{fullHubble} \\
&&\dot{H}=-\frac{1}{2}\left(\dot\sigma^2+e^{-\gamma\sigma}\dot{\phi}^2\right), \label{fullHdot}
\end{eqnarray}
where an over dot represents derivatives w.r.t. time and prime denotes
derivative with respect to $\phi$.  In the slow-roll regime when both
the fields slow-roll, terms with double time derivatives
can be neglected and therefore the background equations reduce to
\begin{eqnarray}
\hspace{-.8cm}&&3H\dot{\sigma}=\beta e^{-\beta\sigma}V(\phi),~~~~~~~3H\dot{\phi}=-e^{(\gamma-\beta)\sigma}V'(\phi),\label{phidot}\\
\hspace{-.8cm}&&3H^2=e^{-\beta\sigma}V(\phi),~~~~~~~~~\dot{H}=-\frac{\dot\sigma^2+e^{-\gamma\sigma}\dot{\phi}^2}{2} \label{Hdot}.
\end{eqnarray}
Here the full potential $W(\sigma,\phi)\equiv e^{-\beta\sigma}V(\phi)$ can be regarded as the product of potentials 
of the individual fields,
$U(\sigma)\equiv e^{-\beta\sigma}$ and $V(\phi)$, and thus 
we define the slow-roll parameters for both the fields in a usual way (following \cite{GarciaBellido:1995fz}) :
\begin{eqnarray}
\epsilon_\phi&\equiv&\frac{1}{2}\left(\frac{V'(\phi)}{V(\phi)}\right)^2,~~~~~~~~~~~~
\eta_\phi \equiv \frac{V''(\phi)}{V(\phi)},\nonumber\\
\epsilon_\sigma&\equiv&\frac{1}{2}\left(\frac{U_{\sigma}}{U}\right)^2=\frac{\beta^2}{2},~~~~~~~~
\eta_\sigma \equiv \frac{U_{\sigma\sigma}}{U}=\beta^2, \label{slowparameters}
\end{eqnarray}
where $U_\sigma\equiv\partial U/\partial\sigma$. To ensure the smallness of the slow-roll 
parameters we demand that the Hubble slow-roll parameter $\epsilon_H\equiv-\frac{\dot H}{H^2}\ll1$ during inflation. We notice that 
\begin{eqnarray}
\epsilon_H=\epsilon_\sigma+e^{\gamma\sigma}\epsilon_\phi, 
\label{epsilonH}
\end{eqnarray}
which implies that $\epsilon_\sigma\ll1$ and $e^{\gamma\sigma}\epsilon_\phi\ll1$ during inflation. 
Again, taking a time-derivative of the second equation of (\ref{phidot}) to obtain $\dot H$, one 
obtains $\epsilon_H=\frac{\ddot\phi}{H\dot\phi}+e^{\gamma\sigma}\eta_\phi-(\gamma-\beta)\beta$, 
which implies that $e^{\gamma\sigma}\eta_\phi\ll1$ and $\gamma\ll\beta+\frac1\beta\sim \frac1\beta$
(as $\beta$ is to be taken smaller than unity). 
We would show later on that the dilaton field $\sigma$ evolve slower than the inflaton
field $\phi$ throughout the inflationary phase which would enable us to treat $\sigma$ as a background field.

Furthermore, we would require the initial field values to calculate the inflationary 
observables such as $n_s$, $r$ and $f_{\rm NL}$. From the first equation of (\ref{Hdot}),
we notice that
\begin{eqnarray}
\sigma&=&\sigma_0+\beta \ln\left(\frac{a}{a_0}\right), \\
\int d\phi\frac{V(\phi)}{V'(\phi)} &=& -\frac{e^{\gamma\sigma_0}}{\beta \gamma} \left[\left(\frac{a}{a_0}
\right)^{\beta\gamma}-1\right], \label{phi-evo}
\label{sigma}
\end{eqnarray}
where subscript $0$ indicates the values of the quantities 60 e-foldings prior to end of inflation.
Defining $f(\phi)\equiv \int d\phi\frac{V(\phi)}{V'(\phi)}$ and requiring 
$\frac{a_f}{a_0}\gtrsim e^{\Delta N}$ (subscript $f$ denoting the quantities at the end of inflation)
for sufficient inflation, one gets
\begin{eqnarray}
\frac{1}{\beta\gamma}\ln\left[1+ \beta\gamma e^{-\gamma\sigma_0}(f(\phi_0)-f(\phi_f))\right] \gtrsim \Delta N.
\label{efolds}
\end{eqnarray}


The perturbation analysis of such a model has been extensively discussed in 
\cite{Starobinsky:1994mh, GarciaBellido:1995fz,DiMarco:2002eb}, where the comoving curvature perturbation is defined as
\begin{eqnarray}
\hskip-0.3cm \mathcal{R} = \Phi-\frac{H}{\dot{H}}\left(\dot{\Phi}+H\Phi\right)=\Phi+H\frac{\dot{\sigma}\delta\sigma
+e^{-\gamma\sigma}\dot{\phi}\delta \phi}{\dot{\sigma}^2+e^{-\gamma\sigma}\dot{\phi}^2},
\label{comov-curv-pert}
\end{eqnarray}
where $\Phi$ is the scalar metric perturbation in the longitudinal gauge. As we are
dealing with a multi-field inflationay model, $\mathcal{R}$ is not a frozen quantity on superhorizon scales and
 its time evolution is given by
\bea
\mathcal{\dot R}=\frac{k^2}{a^2}\frac{H^2}{\dot H}\Phi + \mathcal{S},
\eea
where $\cal{S}$ represents isocurvature (or entropy) perturbations given by
\bea
\hskip-0.3cm\mathcal{S} = \frac{2H (\beta \dot\sigma \dot\phi^2
V(\phi) e^{-\gamma\sigma} + \dot\phi \dot\sigma^2 V'(\phi))}
{e^{\beta\sigma}(\dot\sigma^2 + e^{-\gamma\sigma} \dot\phi^2)^2} \left(\frac{\delta\sigma}{\dot\sigma}
-\frac{\delta\phi}{\dot\phi}\right). \label{iso-pert}
\eea

Under slow-roll approximation, one can solve for the scalar perturbations of the model,
$\delta\sigma$, $\delta\phi$ and $\Phi$, on superhorizon scales, which turn out to be \cite{Starobinsky:1994mh}
\begin{eqnarray}
\hskip-0.7cm\frac{\delta\sigma}{\dot\sigma}&=&\frac{c_1}{H}-\frac{c_3}{H}\,;~~~~~~~~~~~
\frac{\delta \phi}{\dot \phi}=\frac{c_1}{H}+\frac{c_3}{H}
\left(e^{-\gamma\sigma}-1\right),\label{super-horizon-eqns} \\
\hskip-0.7cm\Phi&=&-c_1\frac{\dot H}{H^2}+c_3\left[\frac{1}{2} \left(\frac{V'(\phi)}
{V(\phi)}\right)^2 \left(1-e^{\gamma\sigma}\right) -\frac{\beta^2}{2}\right],
\label{super-horizon-eqns3}
\end{eqnarray}
where $c_1$ and $c_3$ are the time independent integration constants and can be fixed using initial conditions. 
In the above expression (\ref{super-horizon-eqns3}), terms proportional to $c_1$ represent the adiabatic modes
while those proportional to $c_3$ represent the isocurvature modes. 
Using eq.s~(\ref{phidot})-(\ref{slowparameters}), the comoving curvature perturbation~(\ref{comov-curv-pert})
can be simplified to the form: 
${\mathcal R} \simeq\Phi + c_1-c_3+c_3 \epsilon_\phi(\epsilon_\sigma + e^{\gamma\sigma} \epsilon_\phi).$ 

Since from eq.(\ref{super-horizon-eqns3}), it is clear that all the terms in $\Phi$ are proportional to
$(c_1,c_3)\times$ slow-roll parameters, therefore we will ignore the potential $\Phi$ compared to 
$c_1$ and $c_3$ in ${\mathcal R}$. From eq.~(\ref{super-horizon-eqns}), 
we can calculate $c_1$ and $c_3$. Substituting $c_1$ and $c_3$, 
the comoving curvature perturbation on super horizon scales becomes
\begin{eqnarray}
{\mathcal R} =H \frac{\delta \phi}{\dot \phi} e^{\gamma\sigma} A + H \frac{\delta\sigma}{\dot\sigma} B,
\label{comov-curv}
\end{eqnarray}
where $A=\epsilon_\phi/\epsilon_H$ and $B=\epsilon_\sigma/\epsilon_H$. 

Now we look at the observables predicted by this model. The mode functions for
superhorizon fluctuations of $\sigma$ and $\phi$, evaluated
at horizon crossing $k = a(t_k) H(t_k)$, are $\langle|\delta\sigma(k)|^{2}\rangle =H^{2}(t_k)/2k^3$
and $\langle|\delta\phi(k)|^{2}\rangle =e^{\gamma\sigma(t_k)} H^{2}(t_k)/2k^3$.
Therefore, the power spectrum of comoving curvature perturbations becomes
\begin{eqnarray}
{\mathcal P}_{\mathcal R}\equiv\frac{k^3}{2\pi^2}\langle{\mathcal R}^2\rangle 
=\frac{H^2}{8\pi^2\epsilon_H}. \label{powerR1}
\end{eqnarray}

 The tensor power spectrum retains its generic form, because the action involves only minimal Einstein curvature term $R$, given by
$\mathcal{P}_{\mathcal T}=8H^2(t_k)/4\pi^2$,
which yields the tensor-to-scalar ratio as 
\begin{eqnarray}
r\equiv\frac{\mathcal{P}_{\mathcal T}}{\mathcal{P}_{\mathcal R}}=16\epsilon_H.
\label{tensor}
\end{eqnarray}
The scalar spectral index $n_s$ in this model is
\bea
\hskip-0.3cm n_s-1\simeq A \left[(2\eta_\phi-6\epsilon_\phi)e^{2\gamma\sigma} - \beta(2\beta+\gamma)e^{\gamma\sigma}\right]
-B \beta^{2}.\label{ns}
\eea
It can be noted that in the limit $\beta\rightarrow0$ and $\gamma\rightarrow0$ (i.e. $A=1$ and $B=0$)
the forms of power spectrum, tensor-to-scalar ratio and 
spectral index  reduce to their standard forms in the single field slow-roll inflation.

It is important to note that in our model the amplitude of the isocurvature perturbations, 
\begin{eqnarray}
\hskip-0.2cm{\mathcal P}_{\mathcal S} = \frac{k^3}{2\pi^2}\langle{\mathcal S}^2\rangle
= \frac{H^4}{\pi^2} \frac{\left[\beta e^{-\gamma\sigma}
\dot\phi V(\phi) + \dot\sigma V'(\phi)\right]^2}{e^{(2\beta-\gamma)\sigma}(\dot\sigma^{2}
+ e^{-\gamma\sigma} \dot\phi^2)^3}, \label{powerS}
\end{eqnarray}
vanishes in the slow-roll approximation (using (\ref{phidot})). 
The multifield models with non-canonical kinetic term posses a strong single-field attractor 
solution~\cite{Kaiser:2013sna,Schutz:2013fua}
as has also been observed in this case. But generally these multifield models produce isocurvature perturbations
which can also account for a large angular scale suppression of the power spectrum. But our case differs from such
multifield models as it does not produce any isocurvature perturbations upto slow-roll approximation.

\subsubsection{Analysis of the Model with $\lambda_n \phi^n$ Potentials}\label{chaotic-potential}

\begin{figure}[t!]
 \centering
\includegraphics[width= 8cm]{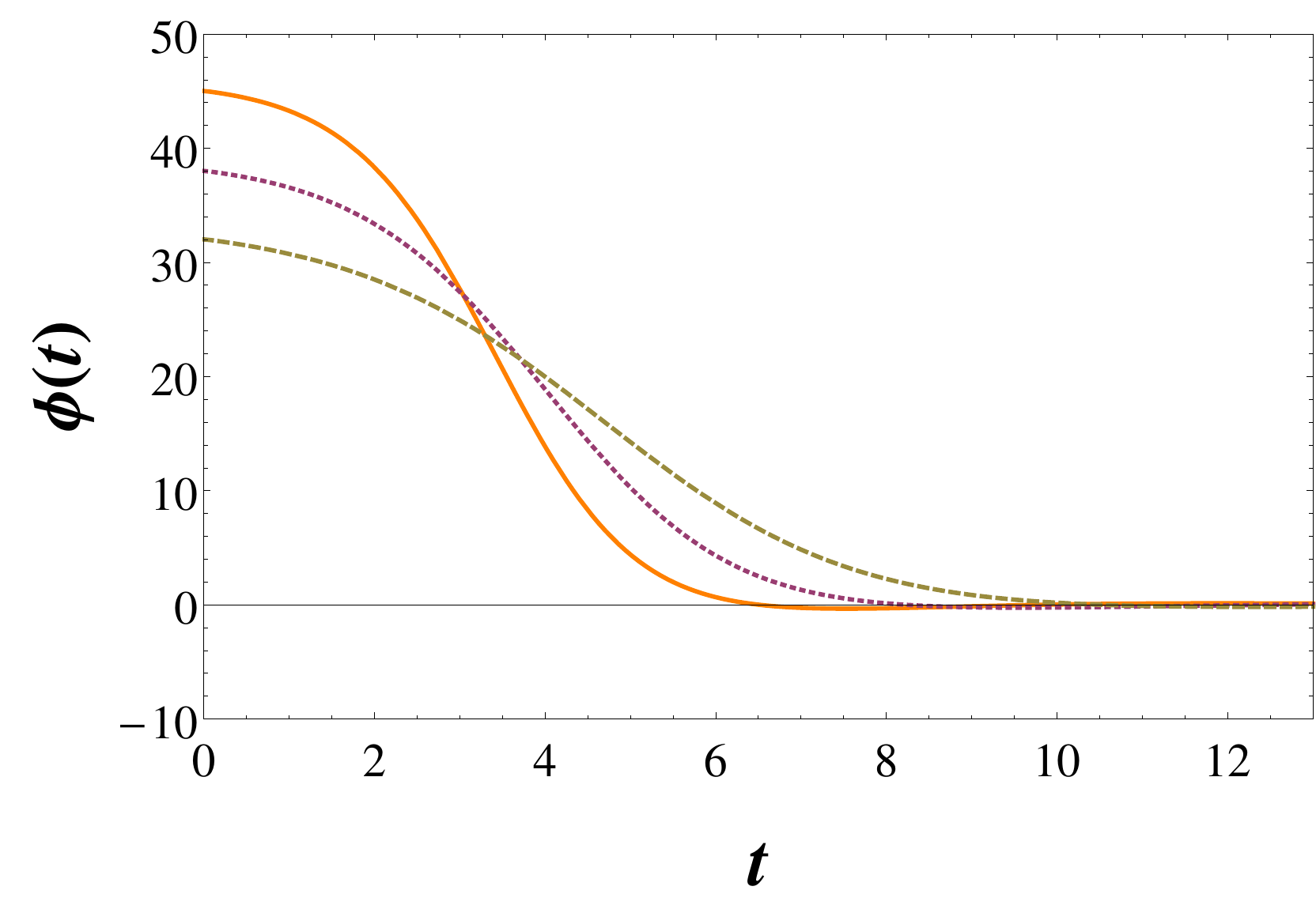}\vspace{0.3cm}
\includegraphics[width= 8cm]{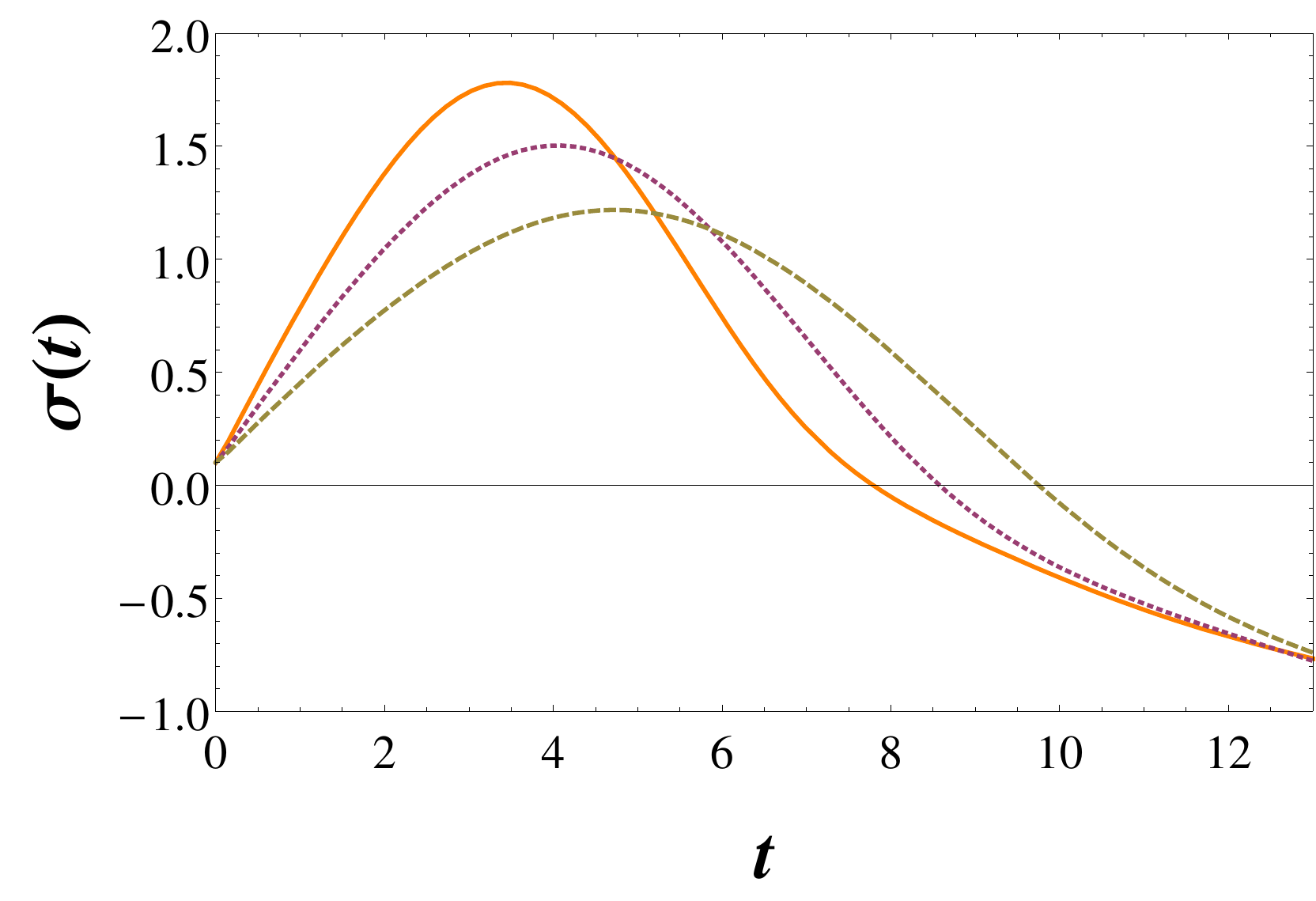}\hspace{0.0cm}
\caption{Evolution of the scalar fields w.r.t. time $t$ measured in the units of $m_{\phi}^{-1}$ is shown.}
 \label{fig1}
\end{figure}
\begin{figure}[t!]
 \centering
\includegraphics[width= 8cm]{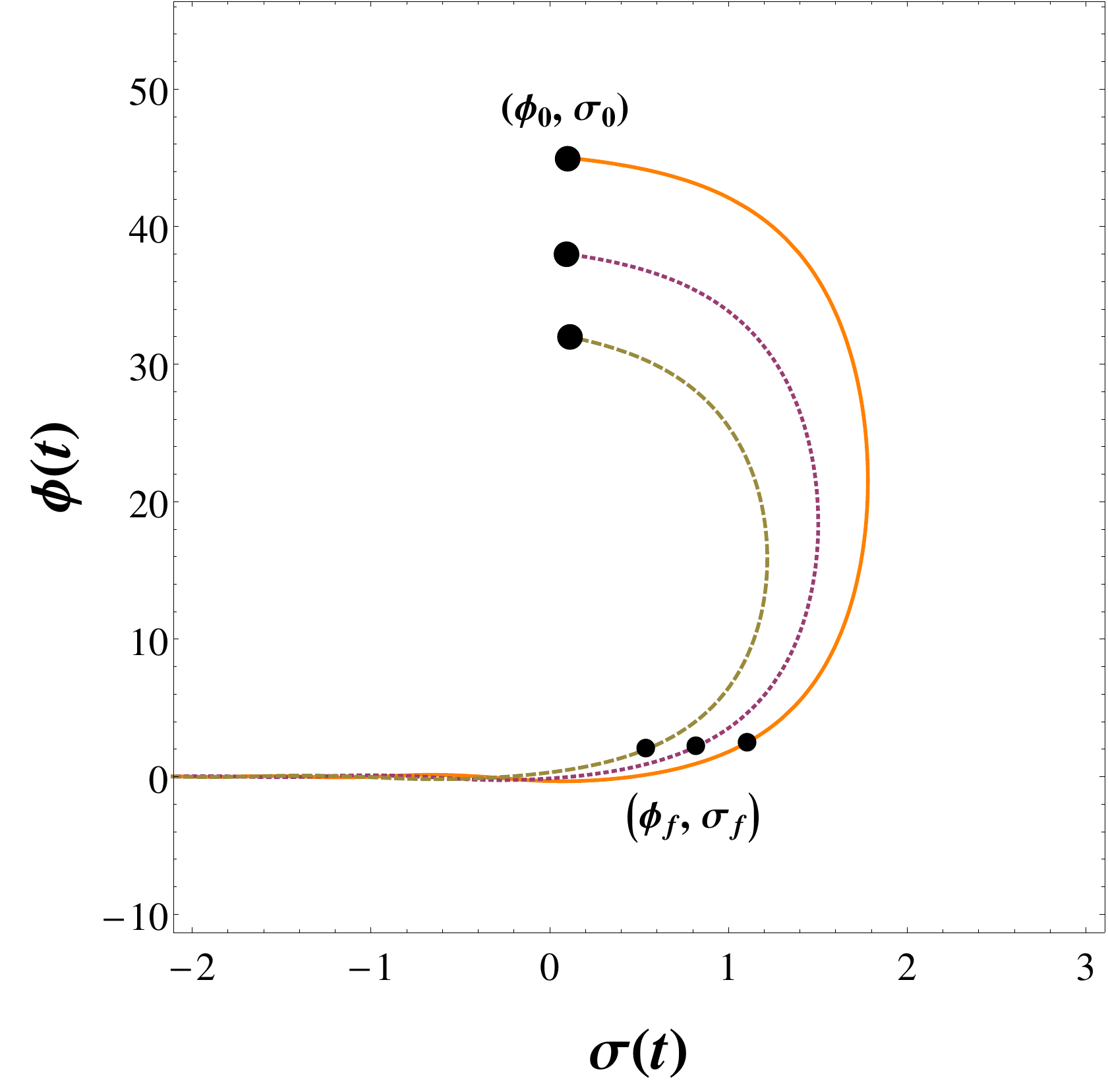}\vspace{0.3cm}
\includegraphics[width= 8cm]{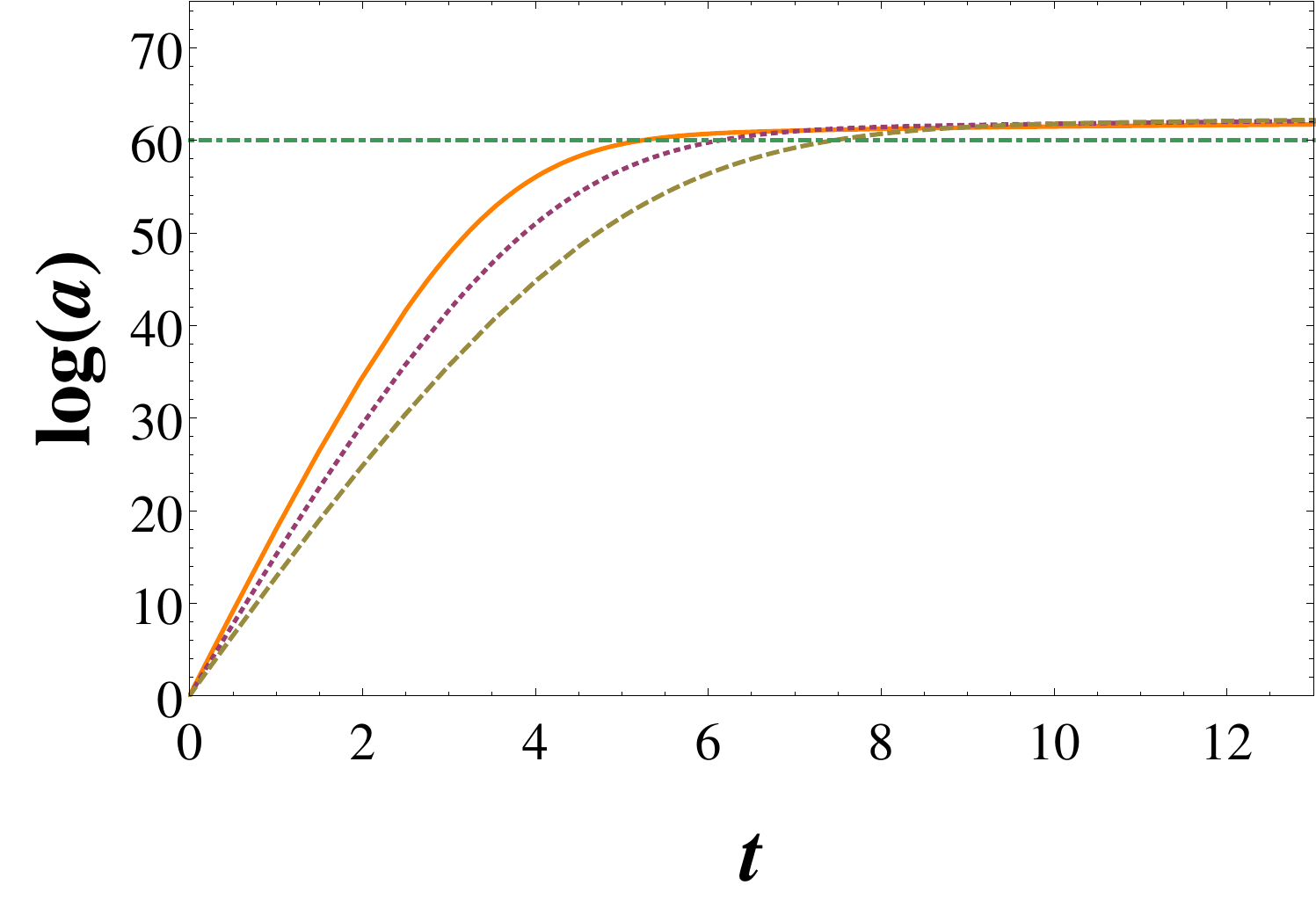}\hspace{0.0cm}
\caption{{\it Upper panel : } Evolution of the inflaton $\phi(t)$ with respect to dilaton $\sigma(t)$
is shown. Bigger black dots correspond to the field values when inflation starts and smaller dots correspond to the
field values at the end of inflation. We see that during inflation the evolution of dilaton
is much slower compared to inflaton. {\it Lower panel : } The cosmological evolution of the
scale factor is shown. The straight dot-dashed line represent the $\Delta N=60$ line where inflation ends.}
 \label{fig2}
\end{figure}

In this model, we treat the $\phi$ field as the inflaton field which is assisted by 
the dilaton field $\sigma$ during the inflationary evolution. This can only be ensured if we confirm 
that the $\sigma$ field evolves slower than the $\phi$ field during the entire inflationary epoch. To show this, we first
 study the background evolution of both the scalar fields $\phi$ and $\sigma$
  by numerically solving the field eq.s (\ref{fullsigmadot}-\ref{fullHdot}).  We first
  treat the case when the inflaton field has quadratic potential $V(\phi)=m_{\phi}^{2}\phi^{2}/2$.
As some representative initial conditions, we choose $\sigma_0=0.1$ and $\phi_0=45$ (Solid), 
$\phi_0=38$ (Dotted), $\phi_0=32$ (Dashed) corresponding to $\beta=0.04$, $\beta=0.035$,
$\beta=0.03$ respectively. Also we fix $\gamma=2\sqrt{2/3}$ for each case,
which is required for the SUGRA derivation of this model studied in the later part of this paper.
The initial conditions are chosen carefully such that we get correct $n_s$ and $r$ for $\Delta N\simeq60$ e-folds.
In FIG.~\ref{fig1}, we show the time evolution of the fields $\phi$ and $\sigma$ where time is given in the units 
of $m_{\phi}^{-1}$. The different colors in the figure correspond to different initial conditions as described above.
In FIG.~\ref{fig2} (upper panel) shows the evolution of the fields in $(\sigma,\phi)$ plane. 
This plot shows that during $60$-efolds inflation, dilaton $\sigma$ evolves much slower compared to inflaton 
$\phi$. After the end of inflation, inflaton goes to its minimum value  $\phi=0$.
 Such background evolution of the fields also ensure that the background spatial metric evolves 
(quasi-)exponentially during inflation which has been depicted in the lower panel of FIG.~\ref{fig2}.
Also we checked that for the case of quartic potential $V(\phi)=\lambda \phi^{4}/4$, the fields evolve in a similar way
ensuring that $\phi$ can be treated as an inflaton field.

We now analyze the observable parameters for inflaton potential $V(\phi)=\lambda_n \phi^{n}/n$. 
From eq.~(\ref{sigma}), we find $\sigma_f = \sigma_0 + \beta \Delta N$. 
Using $\epsilon_H=1$, which is the condition for the end of inflation, we obtain $\phi_f= n e^{\gamma \sigma_{f}/2}/\sqrt{2-\beta^2} $. 
From eq.(\ref{efolds}), the field value $\phi_0$ can be expressed in terms of $\phi_f$ and $\sigma_0$ as
$\phi_0^{2} \simeq \phi_f^{2}+ \frac{2n}{\beta\gamma}e^{\gamma\sigma_0}\left(e^{\beta\gamma \Delta N}-1\right).$
\begin{figure}[t!]
 \centering
\includegraphics[width= 8cm]{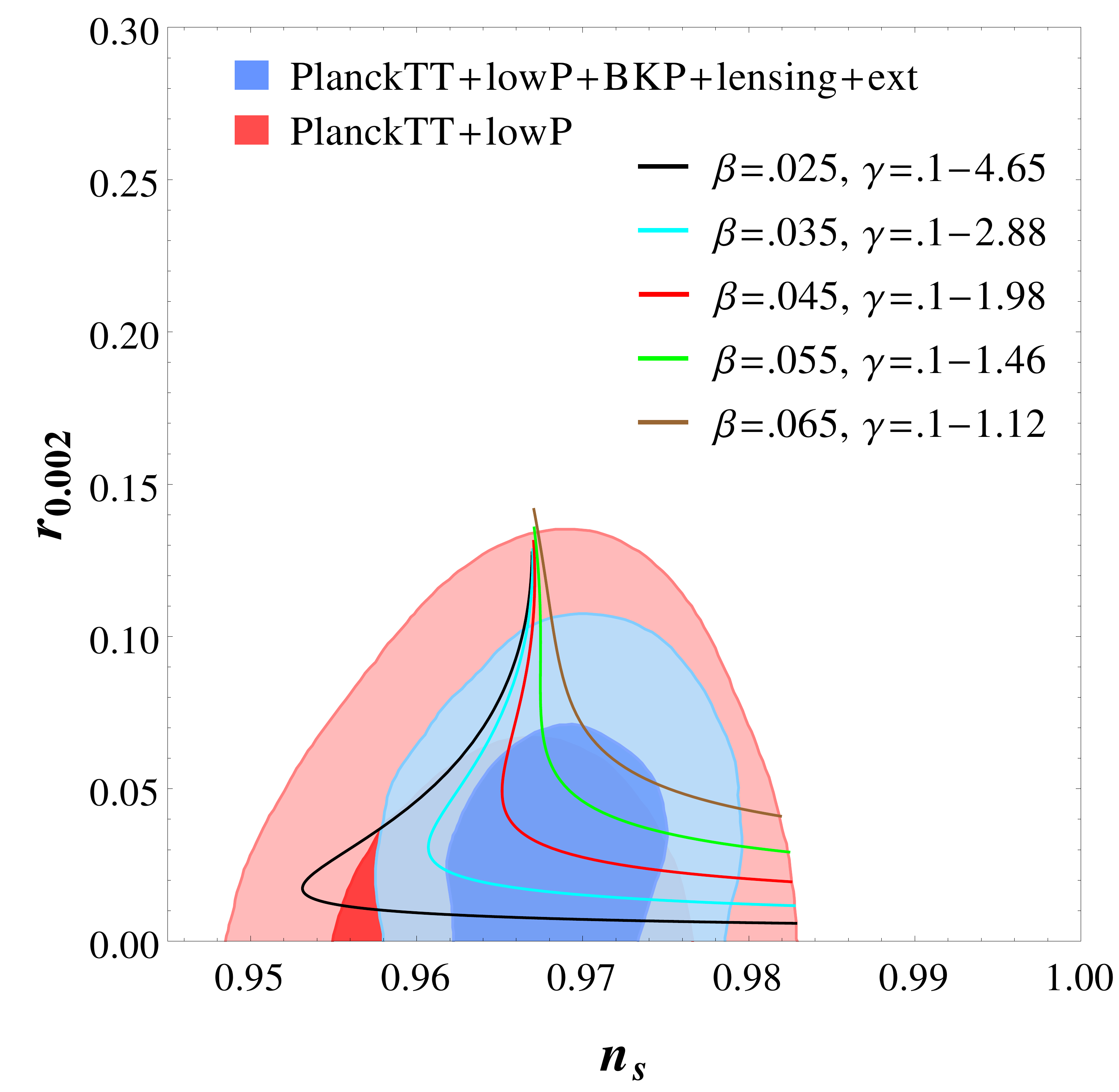}\vspace{0.3cm}
\includegraphics[width= 8cm]{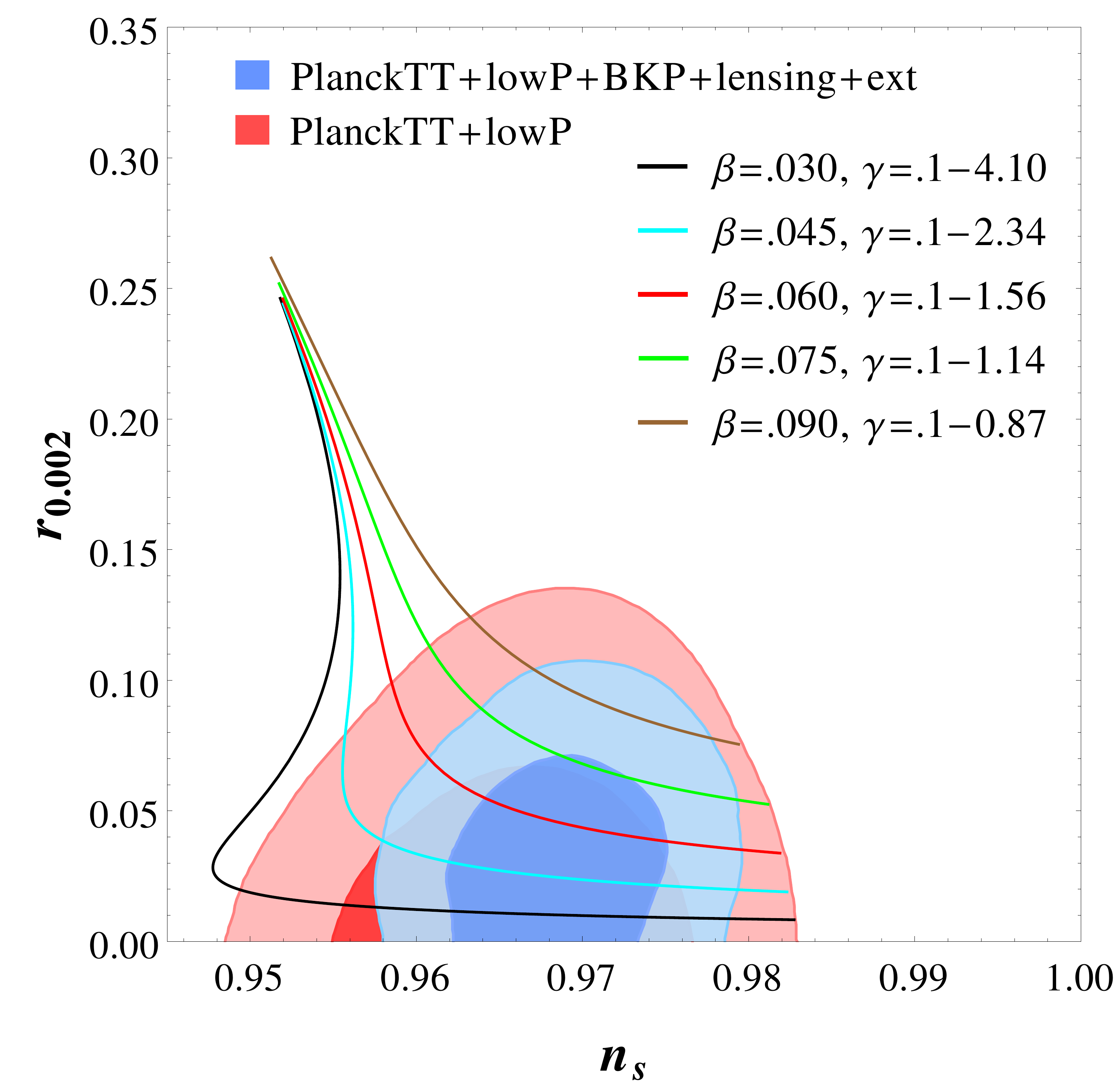}\hspace{0.0cm}
\caption{The $n_s - r$ predictions of the model for quadratic (upper panel) and quartic (lower panel) 
potentials are shown with various contour lines and compared with $1\sigma$ and $2\sigma$ contours of 
the Planck observations \cite{Ade:2015xua}. We take $\Delta N = 60$ and $\sigma_0=0.1$. In both the figures
   the range of values of $\gamma$ increases along the curves from top to bottom. It is also manifest that as the
   values of $\beta$ and $\gamma$ goes to zero, $n_s$ and $r$ values converges to standard slow-roll inflation 
   predictions.}
 \label{fig3}
\end{figure}
Now we substitute $\phi_0$ into eq.s~(\ref{powerR1}), (\ref{tensor}) and (\ref{ns}) to
give $n_s$, $r$ and ${\mathcal P}_{\mathcal R}$ in terms of $\sigma_0$, $n$, $\Delta N$, $\beta$ and $\gamma$.
For $\Delta N =60$ e-folds and for the choice $\sigma_0 = 0.1$ with various choices of the parameters $\beta$
and $\gamma$, the $n_s-r$ predictions for quadratic $(n=2)$ and quartic $(n=4)$ potentials are shown in the 
FIG.~\ref{fig3}. 
For $\sigma_0=0.1$, $\Delta N =60$ and for the range of the parameter values 
of ($\beta,\gamma$) as shown in FIG.~\ref{fig3}, we find inflaton mass in the range $\lambda_2=m_{\phi}^{2}
\sim 10^{-11}-10^{-14}$ and self-coupling in the range $\lambda_4=\lambda\sim 10^{-13}-10^{-17}$. 
E.g. for the choice $\beta=0.05$, $\gamma=0.7$, which can produce $n_s\simeq 0.9666$, $r\simeq0.06$,
gives $m_{\phi} \approx 2\times10^{-6}$. And for $\beta=0.06$, $\gamma= 1$, which produces $n_s\simeq 0.964$,
$r\simeq0.05$, gives $\lambda\approx 10^{-16}$. Therefore, in this model with quadratic and
quartic potentials, similar to the case of single-field slow-roll inflation, we require light inflaton mass
and fine-tuning of the inflaton self-coupling in order to fit the observed CMB amplitude. However, unlike the Higgs 
inflation which predicts very small $r\approx0.003$ and standard single-field inlation with quadratic and quartic potentials give
large $r$, the two field model can give $r$ close to the present bound $r<0.07$. For the above mentioned initial conditions, 
coupling constants and parameter values, the running of the spectral index
$\alpha_s \equiv \frac{dn_s}{d\ln k}\simeq \frac{1}{H}\frac{dn_s}{dt}$ comes out to be 
$\alpha_s\simeq -6.7\times 10^{-4}$ and $\alpha_s\simeq -1.2\times 10^{-3}$ for quadratic and quartic potentials, respectively, fully
consistant with the Planck observation $\alpha_s =−0.0084 \pm 0.0082 ~(68 \% CL, Planck TT+lowP)$~\cite{Ade:2015lrj}.

Finally, we notice that this model yields negligible non-Gaussianity (up to slow-roll approximation).
This can be seen as follows.
Let us first note that besides yielding $r$, $n_s$ and $\alpha_s$ within the observational bounds, a viable inflationary model should not produce
large Non-Gaussianity (NG) to remain in accordance with observations. NG in multifield
models where the fields have non-canonical kinetic terms has been calculated in Ref.s~\cite{Seery:2005gb,Choi:2007su}.
Following these Ref.s, we calculated the non-linearity parameter $f_{NL}$ which characterizes the amplitude of NG.
We find that for the range of parameters values as shown in FIG.~\ref{fig1},
$f_{NL}\sim \mathcal{O}(10^{-2})-\mathcal{O}(10^{-3})$ consistent with the observations.
Also we find that $f_{NL}$ does not depend on initial
value of the dilaton $\sigma_0$ and coupling constants for the considered chaotic form of potential.

The NG parameter $f_{NL}$ is given as
\be
 -\frac56 f_{NL}\equiv-\frac56f_{NL}^{(3)}-\frac56f_{NL}^{(4)}.
 \ee
If one writes the multifield action as
\begin{eqnarray}\label{multi-action}
S=\frac12\int d^4x\sqrt{-g}\left[R-G_{IJ}\partial_\mu\varphi^I\partial^\mu\varphi^J-2W(\varphi)\right],
\end{eqnarray}
then it is shown in \cite{Seery:2005gb} that the Bispectrum non-Gauassian parameter $f_{NL}$ is
 \begin{equation}
-\frac56 f_{NL} N_{,MN}{(N_{,I}N_{,J}G^{IJ})^2}= -\frac56f_{NL}^{(3)}-\frac56f_{NL}^{(4)}\,.
 \end{equation}
Following \cite{Choi:2007su} one gets
 \begin{equation}
 -\frac56f_{NL}^{(3)}=\frac{r}{16}(1+f)
 \end{equation}
 where
 \[
f\equiv-1-\frac1{2\displaystyle
 \sum_ik_i^3}\displaystyle\sum_{\rm perms}\left[-3\frac{k_2^2k_3^2}{k_t}-\frac{k_2^2k_3^2}{k_t^2}(k_1+2k_3)
 +\frac12k_1^3-k_1k_2^2\right],
 \]
 \[
 k_t=\displaystyle\sum_ik_i\,,
 \]
 \[
 -\frac56f_{NL}^{(4)}=\frac{2}{\left[e^{\gamma\sigma_*}\frac{v^2}{\epsilon_\phi^*}+\frac{u^2\alpha^2}{\epsilon_\sigma^*}
 \right]^2}\left[\frac{u^3\alpha^3}{\epsilon_\sigma^*}\left(1-\frac{\eta_\sigma^*}{2\epsilon_\sigma^*}\right)+
 e^{2\gamma\sigma^*}
 \frac{v^3}{\epsilon_\phi^*}\left(1-\frac{\eta_\phi^*}{2\epsilon_\phi^*}\right)- \right.
 \]
 \[ \left.
 - \frac{u^2\alpha^2v}{(2\epsilon_\sigma^*)^{\frac12} \epsilon_\sigma^*}\gamma e^{\gamma\sigma_*}-\left(\frac{u\alpha}{\epsilon_\sigma^{*}}-\frac{v}{\epsilon_\phi^*}\right)^2
 e^{2\gamma\sigma_*}\mathcal A_P\right],\nonumber\\\label{fnl3}
 \]
where $*$ denotes the quantities at horizon crossing,
 \[
 u\equiv \frac{\epsilon_\sigma^f}{\epsilon^f}\,, \quad v\equiv \frac{\epsilon_\phi^f}{\epsilon^f}\,,\quad \alpha\equiv1+\left(e^{\gamma\sigma_f}-e^{\gamma\sigma_*}\right)
 \frac{\epsilon_\phi^f}{\epsilon_\sigma^f}\,,
 \]
 \[
  \mathcal A_P=-\frac{\epsilon_\sigma^f\epsilon_\phi^f}
 {\epsilon^{f^2}}\left[\eta_{ss}^f+\frac{\gamma e^{\gamma\sigma_f}\epsilon_\phi^{f^2}}{\sqrt{2\epsilon_\sigma^f}
 \epsilon^f}-\frac{4\epsilon_\sigma^f\epsilon_\phi^f}{\epsilon_f}\right]\,,
 \]
 \[
 \eta_{ss}\equiv \frac{\eta_\phi\epsilon_\sigma+\eta_\sigma\epsilon_\phi}{\epsilon}\,.
 \]
This model is a multifield model where the inflaton has a non-canonical kinetic term. Multifield inflation with canonical kinetic terms generically yields a local NG which is peaked in the squeezed configuration ($k_3\ll k_1,\, k_2$), whereas multifield inflation with non-canonical kinetic term of inflaton yields NG of equilateral type ($k_1=k_2=k_3$) \cite{Ade:2015lrj}. We see that the momentum dependent factor $f$ in (\ref{fnl3}) yields 0 and 5/6 for squeezed and equilateral configuration respectively, and numerically find that for any choice of parameter values which give correct $n_s$ and $r$ produce the Non-Gaussianity peaking in the equilateral configuration. This indicates, that despite
having two fields, the model behaves like a single-field model with non-canonical kinetic term.

\subsubsection{Two-Field Model Action from No-scale Supergravity}\label{sugra}

In this section we show that, such a two-field inflationary model
can be realised in the realm of no-scale Supergravity.
The two-field models of inflation with string motivated tree-level no-scale K\"ahler
potential in no-scale supergravity framework are
analyzed in \cite{Casas:1998qx,Ellis:2014gxa,Ferrara:2014ima,Ellis:2014opa}. We consider the K\"ahler potential of the following form
\bea
K=-3 \ln[T+T^{\ast}]+ \frac{b \rho \rho^{\ast}}{(T+T^{\ast})^{\omega}},\label{KP}
\eea
here $T$ is the two component chiral superfield whose real part is the dilaton and imaginary part
is an axion. We identify axion as the inflaton of the model, and $\rho$ is an
additional matter field with modular weight $\omega$. In typical orbifold string compactifications with three
moduli fields, the modular weight $\omega$ has value $3$~\cite{Dixon:1989fj,Casas:1998qx,Ellis:2014gxa}.
 Here we shall treat $\omega$ as a
phenomenological parameter whose value can have small deviation from the canonical value $3$ which may 
be explained via string loop corrections to the effective supergravity action \cite{Derendinger:1991hq}.
In this model to obtain the correct CMB observables, the parameter $(3-\omega)$ has to be fine tuned
to the order of $10^{-2}$.

For the complete specification of supergravity, we assume the superpotential as
\be
W=\lambda_{m} ~\rho ~T^{m}\label{SP}.
\ee
We can decompose $T$ field in its real and imaginary parts parametrized by
two real fields $\sigma$ and $\phi$, respectively, as
\bea
T=e^{-\sqrt{2/3}\sigma} + i \sqrt{2/3}\phi.
\label{Tfield}
\eea
\begin{figure}[t!]
 \centering
\includegraphics[width= 8cm]{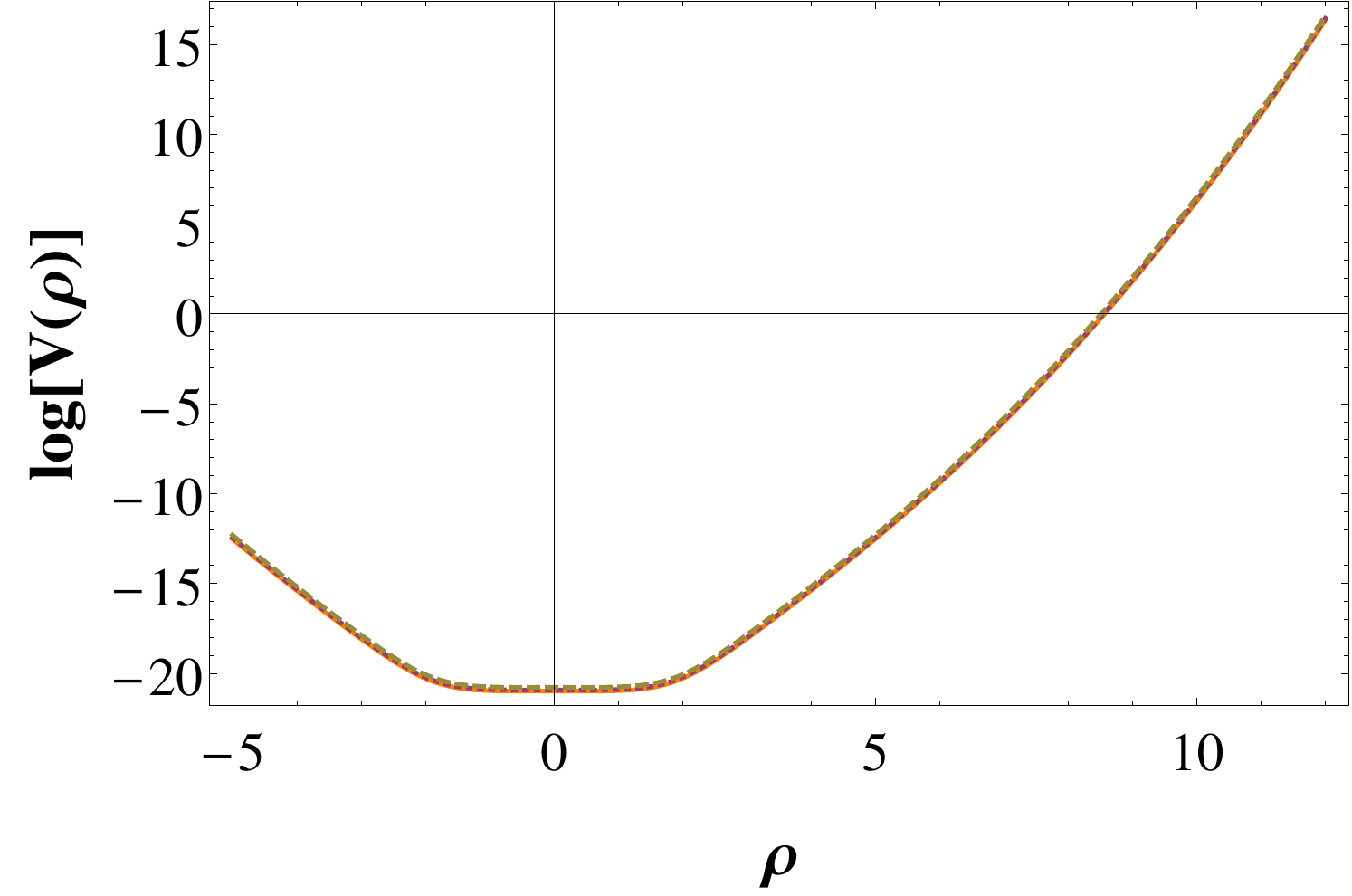}
\caption{Inflationary evolution of the potential $V(\rho)$ is shown for the three initial condition 
as discussed before. We see that during inflation the potential for the field $\rho$
is exponential steep and therefore the field $\rho$ rapidly falls towards the minima of the
potential and stabilizes at $\rho=0$.}
 \label{fig4}
\end{figure}

 The evolution of the matter field $\rho$ is constrained by the exponential factor $e^{K}$ via $e^{G}$ in 
the scalar potential (\ref{sugraVF1}) as $V\propto e^{\frac{b \rho\rho^{\ast}}{(T+T^{\ast})^{\omega}}}$.
Since $(T+T^{\ast})^{-\omega}=2 e^{\frac{2}{3}\omega \sigma}\gtrsim2$ for $\omega\approx3$ and 
$\sigma>0$ during inflation. Therefore field $\rho$, due to its exponentially steep 
potential, is rapidly driven to zero at the start of the inflation and stabilizes at $\rho=0$
~\cite{Ellis:2014opa}. In FIG.~\ref{fig4}, we show the stabilization of the field $\rho$ for 
different initial conditions as discussed before for $(2m=n=2)$. We also checked the evolution 
of $\rho$ for $(2m=n=4)$ and found that it stabilizes in the similar fashion.
Therefore for vanishing $\rho$, the scalar potential (\ref{sugraVF1}) and kinetic term (\ref{sugraLK}) takes the simple form
\be
V=\frac{\lambda_{m}^{2} T^{m}T^{\ast m}}{b (T+T^{\ast})^{3-\omega}}, ~~~~~~ \mathcal{L}_{K}=
\frac{3 \partial^{\mu}T \partial_{\mu}T^{\ast}}{(T+T^{\ast})^{2}},\label{VLK}
\ee
which upon using the decomposition of $T$ becomes
\bea
\mathcal{L}_{K}&=&\frac{1}{2} \partial^{\mu}\sigma \partial_{\mu}\sigma +\frac{1}{2} e^{-\gamma \sigma}
\partial^{\mu}\phi \partial_{\mu}\phi,\label{LK1} \\
V&=& \frac{2^{\omega-3}\lambda_{m}^{2}}{b}~ e^{-\beta\sigma} \left[e^{\gamma\sigma}+\frac{2}{3}\phi^{2} \right]^{m},\label{V1}
\eea
where $\gamma=2\sqrt{2/3}\simeq 1.633$ and $\beta=(3-\omega)\sqrt{2/3}$. Since during inflation, 
dilaton $\sigma$ evolves much slower compared to inflaton $\phi$, see FIG.~\ref{fig2}, therefore
$e^{\gamma\sigma}\ll \phi^{2}$ and hence the first term inside the bracket in (\ref{V1}) can be neglected compared to second term. 

\begin{figure}[t!]
 \centering
\includegraphics[width= 8cm]{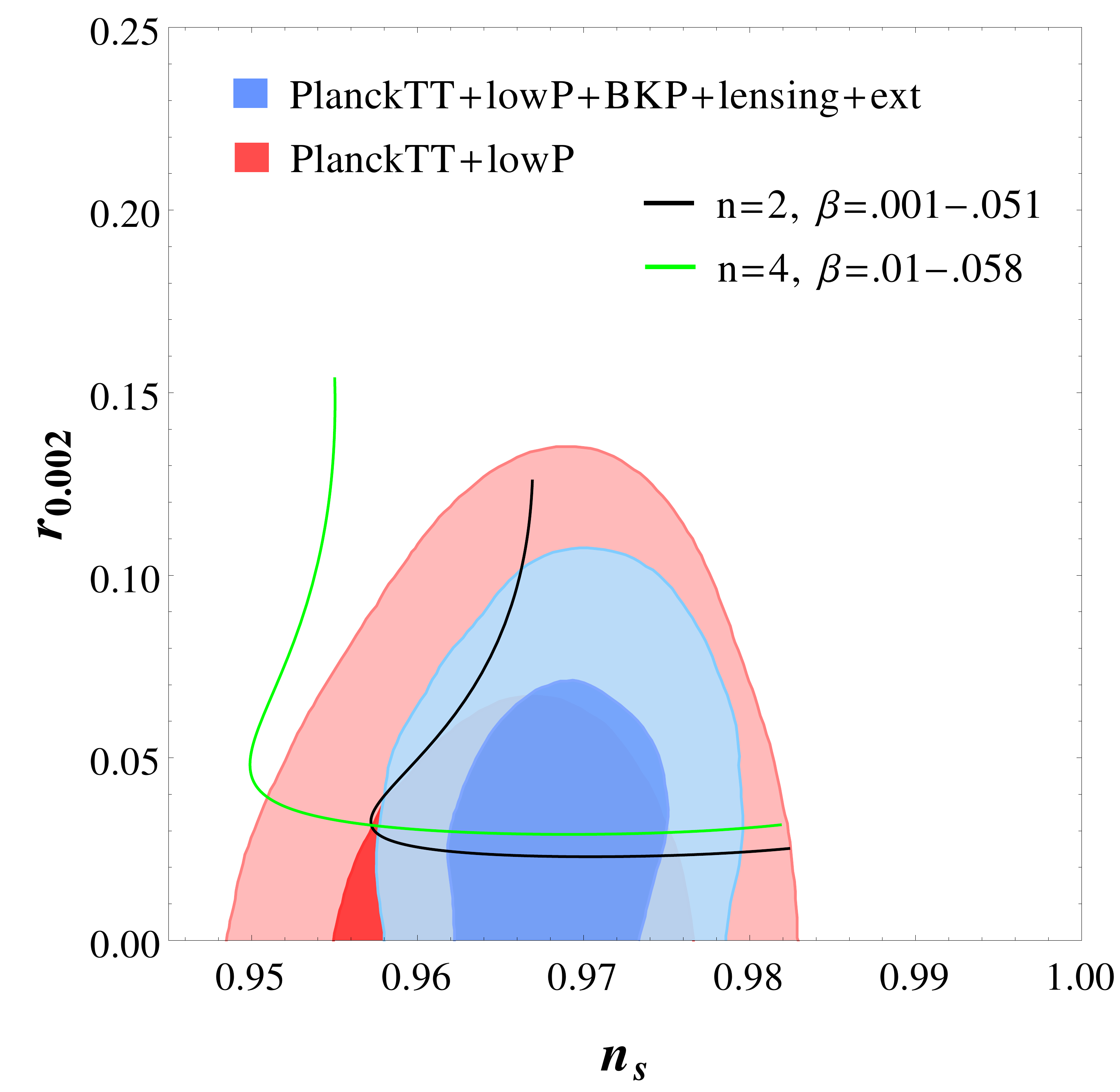}
\caption{The $n_{s}-r$ predictions for quadratic ($n=2$) and quartic ($n=4$) potentials,
with a fixed value of $\gamma=2\sqrt{2/3}$ are shown and compared with 
$1\sigma$ and $2\sigma$ contours of the Planck observations \cite{Ade:2015xua}. 
The range of values of $\beta$ increases along the curves from top to bottom.}
 \label{fig5}
\end{figure}

Therefore, from (\ref{LK1}) and (\ref{V1}), the Lagrangian in EF becomes
\be
\mathcal{L}_{M}=\frac{1}{2} \partial^{\mu}\sigma \partial_{\mu}\sigma +\frac{1}{2} e^{-\gamma \sigma} 
\partial^{\mu}\phi \partial_{\mu}\phi + e^{-\beta\sigma} V(\phi),
\ee
where $V(\phi)=\lambda_{m}^{2}\phi^{2m}/2m$ and we set $b=2^{\omega}/6$ for quadratic
potential ($2m=n=2$) and $b=2\times2^{\omega}/9$ for quartic potential ($2m=n=4$). 
We see that the parameter $b$ is no new parameter and can be given in terms of $\omega$. 
For $\Delta N =60$ and $\sigma_0 = 0.1$, the $n_{s}-r$ predictions for a fixed value of
$\gamma=2\sqrt{2/3}$ and with varying $\beta$ are shown in FIG.~\ref{fig5}. 

To summarize, this two-field two-parameter inflationary model, where the inflaton field has
a non-canonical kinetic term due to the presence of the dilaton field, renders quartic and
quadratic potentials of the inflaton field viable with current observations. Unlike Higgs-inflationary 
scenario which predicts very small tensor-to-scalar ratio $r\approx0.003$, this model can produce large
$r$ in the range $r\sim10^{-1}-10^{-2}$ which would definitely be probed by future $B-$mode experiments 
and thus such a model can be put to test with the future observations. Also this model produces no 
isocurvature perturbations upto slow-roll approximation and predicts negligible non-Gaussianity consistent
with the observations. In addition, we showed that this model can be obtained from a no-scale SUGRA model
which makes this model of inflation phenomenologically interesting from the particle physics perspective.

\section{Conclusions}
\setcounter{equation}{0}

The idea, that the universe through a period of exponential expansion, called inflation, has proved useful for solving the horizon and flatness
problems of standard cosmology and in addition providing an explanation for the scale invariant super-horizon
perturbations which are responsible for generating the CMB anisotropies and formation of structures in the universe.
A successful theory of inflation requires a flat potential where a scalar field acquires a slow-roll over a sufficiently
long period to enable the universe to expand by at least $60$ e-foldings during the period of inflation.

There are a wide variety
of particle physics models which can provide the slow-roll scalar field 'inflaton' for inflation \cite{Lyth:1998xn,Martin:2013tda}.
From the observations of CMB anisotropy spectrum by COBE, WMAP and Planck \cite{Smoot:1998jt,WMAP9,Planck:2015},
it is not yet possible to pin down a specific particle physics model as the one responsible for inflation.
Though all of the above experiments gave tighter and tighter constraints on inflationary observables, $e.g.$
power spectrum and spectral index, which allowed several models of inflation to be ruled out but still there
is a large degeneracy in inflation models. The 2015 data from Planck observation gives the amplitude, spectral
index and tensor-to-scalar ratio as $10^{10}\ln(\Delta_{\mathcal R}^{2}) = 3.089\pm 0.036$, $n_{s}= 0.9666
\pm 0.0062$ at ($68 \%$ CL) and $r_{0.002}<0.11$ at ($95\%$CL), respectively~\cite{Ade:2015lrj}. Also the latest
results by BKP collaboration put $r$ at $r_{0.05}<0.07$ at ($95\%$CL)\cite{bicep2keck2015}.
Therefore, all those models which can produce the correct amplitude of CMB power spectrum and its spectral tilt along
with producing $r<0.07$ are allowed. The future $B-$mode observations are expected to fix $r$ which
will allow many existing inflation models to be ruled out.

In this review we have presented some recent results of a single-field generalized
non-minimal model of inflation, a power law model of inflation and a two-field model of inflation with a
non-canonical kinetic term. We have calculated the key inflationary observables : amplitude of the power spectrum of
curvature perturbations, spectral index and its running, tensor-to-scalar ratio and amplitude of isocurvature
perturbations. We fix the parameters of the models using the measured values of these observables. Also we
motivate these models from a fundamental theory called no-scale supergravity. From the analysis of these models we find
that the power law model $R+R^{\beta}/M^{2}$ of inflation and therefore all the models of inflation with the Lagrangian
$R + \xi \phi^{a} R^{b} + \lambda \phi^{4(1+\gamma)}$ are ruled form the current bound on $r_{0.05}<0.07$ at ($95\%$CL).
However for $\beta=2$ which corresponds to Starobinsky model $R+R^{2}/M^{2}$ of inflation produces $r\simeq0.0033$. This 
prediction of $r\simeq0.0033$ however very small but is currently favored by the observations.
Contrary to the large $r$ prediction of power law model, the two field model predicts a range of $r\sim 10^{-1}-10^{-2}$ 
values close the present bound on $r$.

Currently tensor-to-scalar ratio is an extremely important inflationary parameter
in view of validating and ruling out models of inflation, and therefore many models that can be
classified as extended theories of gravity. The observation of primordial $B$-modes
(CMB polarization) will provide the constraint on $r$. The $B$-modes are the signature of inflationary tensor modes,
precise observations of which will provide a most distinctive confirmation of occurrence of an inflationary era
in the early universe. There are several ongoing experiments for $B-$mode detection and all hope to
observe these signals from the inflationary era. There are several experiments, $e.g.$ ground based (Keck/BICEP3,
SPT-3G, AdvACT, CLASS, Simons Array), balloons based (Spider,EBEX) and satellites based (CMBPol, LiteBIRD and COrE).
These observational experiments will be taking into account the recent Planck data on polarized dust. They aim to probe the tensor-to-scale ratio at the level of $r\sim10^{-3}$ which is a theoretically motivated limit~\cite{Creminelli:2015oda}. High precision measurements of small-scale temperature anisotropies along with the observations of $B-$mode will not only test the inflationary hypothesis but allow to remove a large degeneracy in the models of inflation.

\begin{acknowledgement}
G.L. thanks the project COST project CA15117 CANTATA.
\end{acknowledgement}

\end{document}